\documentclass[prl,twocolumn,amsmath,amssymb,10pt,aps,longbibliography,superscriptaddress,citeautoscript,bibnotes,footinbib]{revtex4-1}

\usepackage{feynmp}
\usepackage{amsmath}
\usepackage{graphicx}
\usepackage{multirow}
\usepackage{latexsym}
\usepackage{textcomp}
\usepackage{verbatim}
\usepackage{color}
\usepackage{bm}
\usepackage{subfigure}
\usepackage{everysel}
\usepackage{keyval}
\usepackage{ragged2e}
\usepackage{dsfont}
\usepackage{amssymb}
\usepackage{enumitem}
\usepackage{pstricks}
\usepackage{mathtools}
\usepackage{braket}
\usepackage{slashed} 
\usepackage[colorlinks=true]{hyperref}

\usepackage{graphicx}
\usepackage{xcolor}
\usepackage{amsmath}
\usepackage{amsfonts}
\usepackage{bbm}

\usepackage{ulem}

\newcommand{\osum}{{%
    \setbox0\hbox{\circ}%
    \rlap{\hbox to \wd0{\hss\sum\hss}}\box0
}}

\begin{document}

\title{Metrology of Band Topology via Resonant Inelastic X-ray Scattering}

\author{Sangjin Lee}
\thanks{Electronic Address: sangjinlee519@ibs.re.kr}
\affiliation{Asia Pacific Center for Theoretical Physics, Pohang 37673, Republic of Korea}
\affiliation{Center for Artificial Low Dimensional Electronic Systems, Institute for Basic Science (IBS), Pohang 37673, Republic of Korea}

\author{Kyung-Hwan Jin}
\affiliation{Center for Artificial Low Dimensional Electronic Systems, Institute for Basic Science (IBS), Pohang 37673, Republic of Korea}

\author{Byungmin Kang}
\affiliation{School of Physics, Korea Institute for Advanced Study, Seoul 02455, Korea}

\author{B. J. Kim}
\thanks{Electronic Address: bjkim6@postech.ac.kr }
\affiliation{Department of Physics, Pohang University of Science and Technology (POSTECH), Pohang 37673, Republic of Korea}
\affiliation{Center for Artificial Low Dimensional Electronic Systems, Institute for Basic Science (IBS), Pohang 37673, Republic of Korea}

\author{Gil Young Cho}
\thanks{Electronic Address: gilyoungcho@postech.ac.kr}
\affiliation{Department of Physics, Pohang University of Science and Technology (POSTECH), Pohang 37673, Republic of Korea}
\affiliation{Center for Artificial Low Dimensional Electronic Systems, Institute for Basic Science (IBS), Pohang 37673, Republic of Korea}
\affiliation{Asia Pacific Center for Theoretical Physics, Pohang 37673, Republic of Korea}

\date{\today}
 
\begin{abstract} 
Topology is a central notion in the classification of band insulators and characterization of entangled many-body quantum states. In some cases, it manifests as quantized observables such as quantum Hall conductance. However, being inherently a global property depending on the entirety of the system, its direct measurement has remained elusive to local experimental probes in many cases. Here, we demonstrate that various topological band indices can be directly probed by resonant inelastic x-ray scattering. Specifically, we show that the crystalline symmetry eigenvalues at the high-symmetry momentum points, which determine the band topology, leads to distinct scattering intensity for particular momentum and energy. Our approach can be explicitly demonstrated in several examples such as 1D Su-Schrieffer-Heeger chain, 2D quadrupole insulator, 3D topological band insulator and chiral hinge insulator. Our result establishes an incisive bulk probe for the measurement of band topology. 
\end{abstract}


\maketitle
{\color{green}\textit{\textbf{1.Introduction}}}: In condensed matter system, scattering techniques generally measure two-point correlators of physical quantities that determine the transport and dissipation of energy and momentum through collective excitations of constituent nuclei and electrons. In resonant inelastic x-ray scattering (RIXS)\cite{rixs_rmp,rixs_rmp2}, sensitivity to valence electrons is greatly amplified by resonant processes when the incident x-ray is tuned to the binding energy of a core electron. Besides detection of otherwise vanishingly small signal associated with subtle electronic orders\cite{peter_2014,peter_1999}, RIXS sometimes provides additional valuable information on the electronic wave functions; prominent examples include the detection of pseudospin structure in spin-orbit Mott insulators and the bonding/anti-bonding nature of quasi-molecular\cite{Kim1329} orbitals in quantum dimers.\cite{BJ2017,BJKim2014,Revelli2019,lucile_RIXS,rixs_method7,RIXS_SC,kourtis_2016} The method relies on the coherent interference\cite{Revelli2019} among different intermediate states, whose phase differences encode information beyond that contained in the correlation functions.

In this Letter, we investigate the possibility of exploiting RIXS for probing the topology of an electronic band structure, which is classified by a topological band index.\cite{RevModPhys.82.3045, Qi_rmp, chiu_rmp} We note that the band topology is difficult to directly access through bulk energy spectrum as two topologically distinct phases can share the same spectrum, but its change can be detected at their interface. Hence, much of the previous experimental efforts focused on detecting the topological boundary modes via, e.g. photoemission spectroscopy\cite{Chen_ARPES2009,Xia_ARPES,    ARPES_hsieh2009,Nishide_ARPES2010,ARPES_STM_Roushan:2009} and tunnelling probes\cite{STM_Seo2010,STM_Beidenkopf2011,STM_Alpichshev2010,     STM_majorana, STM_majroana2015, STM_majorana_edge2019}. However, the topological boundary states are not always easily accessible in experiment, for example when the topological boundary states are buried deep inside the bulk states. Thus, it is desirable to establish a \textit{bulk probe} that can directly diagnose the non-trivial band topology.

To this end, we will show that the band topology manifests in RIXS spectral intensity at high-symmetry momentum. More precisely, we will prove that from the momentum dependence of the RIXS intensity, one can directly infer their crystalline symmetry eigenvalues of the reflection, inversion, and rotational symmetries, which constitute several important topological band indices.\cite{Benalcazar2017, hughes2017,Fang_Chern_number,inv_TI,hinge_HOTI} We will explicitly demonstrate our protocol on a few, most important examples of topological insulators: 1D Su-Schrieffer-Heeger (SSH) chain \cite{PhysRevLett.42.1698} relevant for A$_2$W$_6$X$_6$ (A$=$Rb,Cs, X$=$S,Se)\cite{KHJIN2020,SSH_mat_1,SSH_mat_2,SSH_mat_3}, 3D topological band insulators (TBI)  for Bi$_{1-x}$Sb$_x$ \cite{inv_TI, BiSb2009,BiSb2009a}, and also 2,3D higher-order topological insulators (HOTI) \cite{Benalcazar2017,hughes2017,hinge_HOTI} with material candidates XY (X$=$Ge, Y$=$S,Se)\cite{C4_TI_material}. 


{\color{green}\textit{\textbf{2.RIXS}}}: We first review the formalism for RIXS in the fast collision approximation \cite{rixs_rmp,rixs_method6,rixs_method4,haverkort_RIXS_NSF}. The RIXS spectral intensity is $\mathcal{I} (\bm{q}, \omega) = \sum_f |\mathcal{A}_f (\bm{q}, \omega) |^2\delta(E_{f}- E_{\text{GS}} - \omega) $ with the quantum amplitude  
\begin{align}
\mathcal{A}_f (\bm{q}, \omega)  =  \langle f| \hat{T} (\bm{q}, \omega) |\text{GS} \rangle.
\label{RIXS}
\end{align} 
Here $|\text{GS}\rangle$ and $|f \rangle$ are the many-electron states (the ground state and an excited, final state) and $\hat{T} (\bm{q}, \omega)$ is the scattering matrix describing the RIXS process. 

In the fast collision limit, $\hat{T} (\bm{q}, \omega)$ describes the creation of a local electron-hole pair at each atomic sites.\cite{rixs_rmp,rixs_method6,rixs_method4,haverkort_RIXS_NSF} 
In this Letter, we focus on the non-spin-flip RIXS \cite{non_SF_RIXS,SI}
\begin{align}
\hat{T} (\bm{q}, \omega) = \mathcal{C}(\omega) \sum_{\bm{r}, \sigma} c_{\bm{r}, \sigma}c_{\bm{r},\sigma}^{\dagger} e^{i\bm{q}\cdot \bm{r}}, 
\label{T}
\end{align} 
which can be achieved by appropriately selecting edges (e.g. Bi $2s\to6p$ transition) or polarizations of photons. Here, $\mathcal{C}(\omega)$ is a constant containing the information of atomic details, e.g., dipole matrix elements, polarization of (incident and outgoing) photons, and lifetime of the intermediate virtual states. $c_{\bm{r},\sigma}^{\dagger}$ ($c_{\bm{r}, \sigma}$) is a creation (annihilation) operator of low-energy spin-$\sigma$ electrons of an atom at $\bm{r}$. More discussion on $\hat{T} (\bm{q}, \omega)$ is in \cite{SI}.



\begin{figure}[t!]
	\includegraphics[width=.5\textwidth]{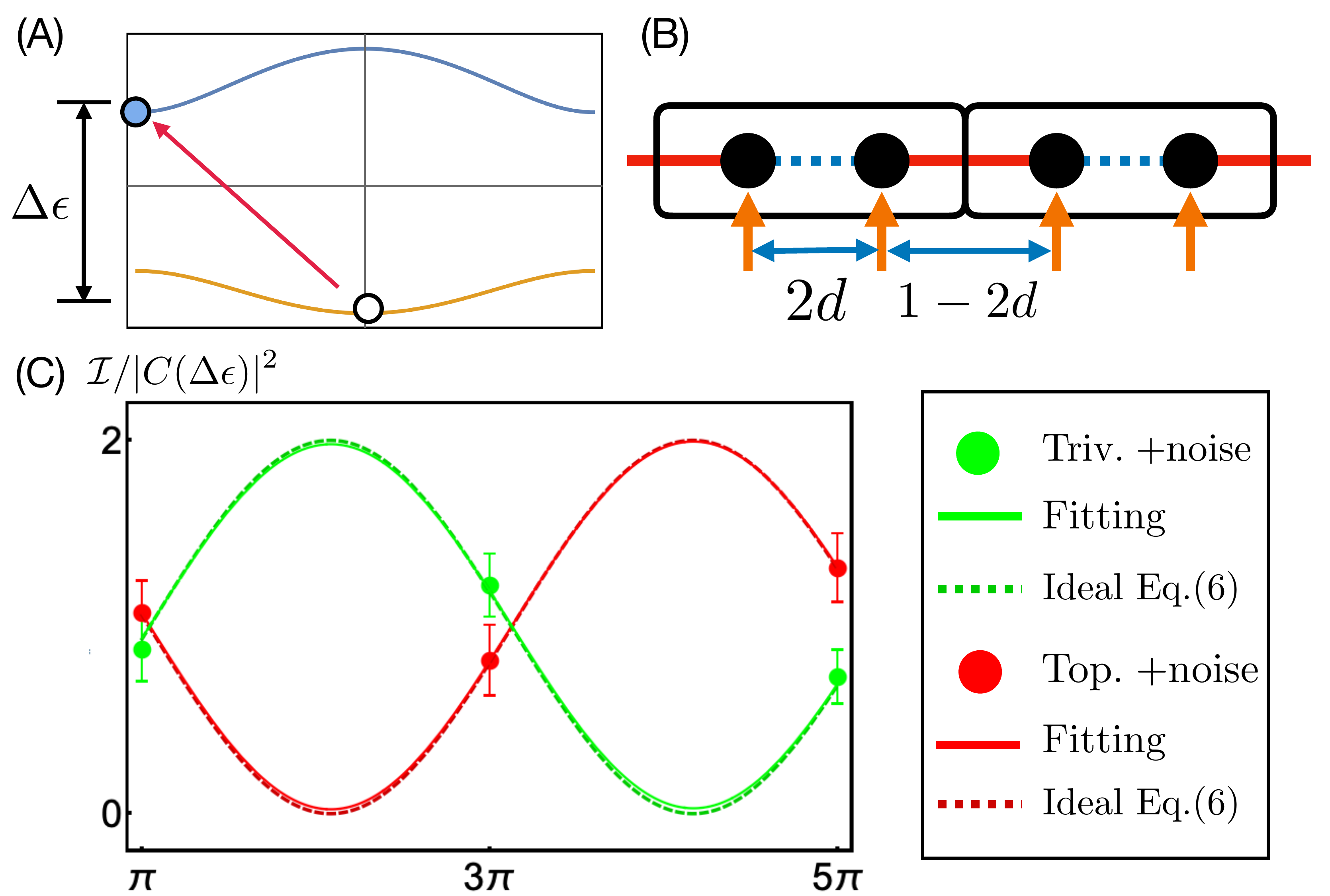}
	\caption{(A)RIXS process Eq.\eqref{BandRIXS}. RIXS induces a transition from the ground state to an excited state with an electron-hole pair. Here, an empty circles represents the hole in the valence band and the filled circle represents the created electron in the conduction band. (B)SSH model. The black box represents a unit cell, which consists of the two sublattices $\{1, 2\}$ (black circles). They are physically separated each other by ``$2d$" within the unit cell and ``$1-2d$" between the unit cell. (C)RIXS intensity of SSH model. Here, the error bars for the data (filled circles) are due to the noise that we included in our simulation. See \cite{SI} for numerical details. 
}
	\label{fig:1}
\end{figure}

Inserting Eq.\eqref{T} to Eq.\eqref{RIXS}, we can recast Eq.\eqref{RIXS} in terms of Bloch functions $\psi_{c/v}(\bm{k})$ (which depend on spin for strongly spin-orbit coupled systems): 
\begin{align}
\mathcal{A}_f (\bm{q}, \omega)  = \mathcal{C}(\omega) ~\psi^{\dagger}_{c}(\bm{k}+\bm{q}) \cdot \hat{\mathcal{M}}_{\bm{k}+\bm{q}, \bm{k}} \cdot \psi_{v}(\bm{k}), \label{BandRIXS}
\end{align} 
where $E_{c}(\bm{k}+\bm{q}) - E_{v}(\bm{k}) = \omega$. Here ``$c$" and ``$v$" represent the conduction and valence bands. $\hat{\mathcal{M}}_{\bm{q}+\bm{k}, \bm{k}}$ contains the information of the sublattice structure\cite{SI}. The final expression Eq.\eqref{BandRIXS} can be summarized as [Fig.\ref{fig:1}(A)], where an electron is pulled from the valence band to the conduction band. The momentum and energy of this excited state is fixed by those of the photons involved in RIXS. 
 



{\color{green}\textit{\textbf{3.Main Results}}}: We now present our central results by applying Eq.\eqref{BandRIXS} to various topological insulators: 1D SSH chain, 3D TBI, and 2,3D HOTI \cite{PhysRevLett.42.1698,inv_TI,hughes2017,hinge_HOTI}.

\textit{A. SSH chain}: The SSH model\cite{PhysRevLett.42.1698} is the simplest reflection-symmetric 1D band insulator, which exhibits a quantized electric polarization. Within each unit cell labelled by $x \in \mathbb{Z}$, there are two sublattices $a=1,2$, at the position $x - d$ for $a=1$ and $x+d$ for $a=2$ [Fig.\ref{fig:1}(B)].  We assume $d<1/4$ without losing generality, because the dimerization of the SSH model accompanies structural deformation in real materials \cite{KHJIN2020,SSH_mat_1,SSH_mat_2,SSH_mat_3}, which pushes $d$ away from $1/4$. 
The Hamiltonian is  
\begin{align}
H(k) = -\left( t_1 + t_2 \cos k \right)\sigma^{x}\otimes s^{0}  - t_2 \sin k \sigma^{y} \otimes s^{0}. \label{SSH-Hamiltonian}
\end{align}
Here, $\sigma^{a}$ is the Pauli matrix acting on the sublattice index and $s^{b}$ is that on the spin index. We set $t_1 >0, t_2 >0$. The model is spin-rotational symmetric and hence only the identity operator $s^{0} = \text{Id}_{2\times 2}$ appears in Eq.\eqref{SSH-Hamiltonian}. 
The topological invariant, i.e. polarization $\mathcal{P}$, can be written as\cite{PhysRevB.47.1651,PhysRevLett.80.1800}
\begin{align}
\exp \Big( 2\pi i \mathcal{P} \Big) = \mathcal{R}_v(\Gamma) \mathcal{R}_v(X),  \label{Pol}
\end{align}
where $\mathcal{R}_v (k)$ is the reflection eigenvalue of the valence band at $k$. $\Gamma$ ($X$) represents $k =0$ ($k= \pi$). The reflection symmetry is $\hat{\mathcal{R}} = \sigma^{x}$ with $k \to -k$. Many aspects of the model have been well known in literature \cite{SI,  hughes2017}. Here we only need to know that there are two phases: the non-trivial phase $\mathcal{P}=1/2$ mod 1 for $t_1<t_2$ and the trivial one $\mathcal{P}=0$ mod 1 for $t_1>t_2$.




To grasp $\mathcal{P}$ via RIXS, one would naturally consider the scattering between $X$ and $\Gamma$ because of Eq.\eqref{Pol}, which involves the symmetry data at those momentum points. Indeed, $\mathcal{P}$ directly manifests in the intensity
\begin{align}
\mathcal{I} \left[ q_n, \Delta \epsilon \right] = 2  |\mathcal{C} (\Delta \epsilon)|^2   \sin^2 (q_nd + \mathcal{P} \pi ),
\label{SSH-Intensity}
\end{align}
where $q_n = (2n+1)\pi, n \in \mathbb{Z}$ is precisely the momentum connecting $X$ and $\Gamma$. $\Delta \epsilon$ is the energy difference between the valence band at $\Gamma$ and the conduction band at $X$. These two fix the scattering process as [Fig.\ref{fig:1}(A)]. Remarkably, the intensity Eq.\eqref{SSH-Intensity} is largely independent of the hopping parameters of the SSH model (except the overall coefficient $\mathcal{C} (\Delta \epsilon)$). Even more, its oscillation in momentum is entirely fixed by the topology $\mathcal{P}$. This allows us to apply Eq.\eqref{SSH-Intensity} irrespective of the detailed parameters of the Hamiltonian.

Below we will first show that we can correctly infer the band topology by fitting the numerically-simulated RIXS intensity data with Eq.\eqref{SSH-Intensity}. Then, we will present our mathematical proof of Eq.\eqref{SSH-Intensity}.

- \textit{Numerics}: We want to demonstrate that Eq.\eqref{SSH-Intensity} allows us to decipher $\mathcal{P}$ from the RIXS intensity. For this, we first numerically generated the RIXS intensity data for both the topological and trivial states of the SSH model (with $d=0.24$), and we attempted to fit the data with Eq.\eqref{SSH-Intensity}. The data, which are represented as the filled circles in [Fig.\ref{fig:1}(C)], consist of the ideal signals (obtained by applying Eq.\eqref{BandRIXS} to the SSH model) and white noises (added to mimic the experimental situations). See the numerical details in \cite{SI}. The green (red) circles are the data points generated from the trivial (topological) states. Then, we attempted to fit these data with $\mathcal{I}(q_n)/|\mathcal{C}(\Delta \epsilon)|^2 = A \sin^{2} (q_n B+P) +C$, where $(A, B, C, P)$ are the fitting parameters (we have restricted $B \in [0, 1/4]$ without losing the generality). See the solid lines in [Fig.\ref{fig:1}(C)], which are obtained from the fitting. The results excellently agree with the dotted lines, which are the ideal ones Eq.\eqref{SSH-Intensity}. When the noises are removed, then the simulated data exactly agree with the dotted lines Eq.\eqref{SSH-Intensity}. In particular, the fitting (solid lines) correctly determines the polarization $\mathcal{P} = P/\pi$ to be either $0$ or $1/2$. Hence, we conclude that the fitting of the RIXS data against Eq.\eqref{SSH-Intensity} can successfully diagnose $\mathcal{P}$. 

Our theory can be tested in 3D bulk materials A$_2$W$_6$X$_6$ (A=Rb,Cs, X=S,Se), which realize the quasi-1D SSH chains  \cite{KHJIN2020,SSH_mat_1,SSH_mat_2,SSH_mat_3}. We append their DFT band structures and some information on these materials in \cite{SI}, showing that they indeed have $\mathcal{P}=1/2$. In these materials, one can use W $2p\to5d$ transition to perform RIXS experiments to test our theory.

- \textit{Proof}: Here we sketch our proof of Eq.\eqref{SSH-Intensity}, while deferring minor details to \cite{SI}. Our starting point is Eq.\eqref{BandRIXS} 
\begin{align}
\mathcal{A}_f \left(q_n , \Delta \epsilon \right) = \mathcal{C}(\Delta \epsilon) \psi_c^{\dagger}(X) \hat{\mathcal{M}}_{q_n} \psi_v(\Gamma),  \nonumber
\end{align}
where $\hat{\mathcal{M}}_{q_n}= \cos (q_n d) + i \sin (q_n d) \cdot \hat{\Pi}
$.\cite{SI} 
Here $\hat{\Pi}=\sigma_z$ satisfies $\{\hat{\mathcal{R}}, \hat{\Pi}\} =0$. To proceed, we separate $\mathcal{A}_f$ into the two parts, which have distinct oscillations in momentum 
\begin{align}
\mathcal{A}_f = \mathcal{C}(\Delta \epsilon)\cdot\Big[ a_c \cos (q_n d) + i a_s  \sin (q_n d) \Big] \label{SSH-Amplitude}
\end{align} 
with 
$
a_c = \psi_c^{\dagger}(X) \psi_v(\Gamma)
$
and 
$
a_s = \psi_c^{\dagger}(X) \hat{\Pi} \psi_v(\Gamma)
$.

Our key observation is that the reflection symmetry enforces $a_s = 0$ for $\mathcal{P} = 1/2$  and $a_c = 0$ for $\mathcal{P} = 0$. For example, for $\mathcal{P}=0$
\begin{align}
a_c &=  \psi_c^{\dagger}(X) \hat{\mathcal{R}}^{-1} \hat{\mathcal{R}} \psi_v(\Gamma),  \nonumber\\ 
&=   \mathcal{R}_c(X) \mathcal{R}_v(\Gamma) \psi_c^{\dagger}(X) \psi_v(\Gamma) = - e^{2\pi i \mathcal{P}} a_c,   \nonumber
\end{align}
where we used $\mathcal{R}_c(X) = - \mathcal{R}_v(X)$\cite{SI} and Eq.\eqref{Pol}. Hence, $a_c =0$ for $\mathcal{P}=0$. On the other hand, the value of $a_s$ has no symmetry constraint and can be generically non-zero. Thus, we find $\mathcal{A}_f \sim \sin(q_n d)$ and $\mathcal{I}\sim \sin^{2} (q_n d)$ for $\mathcal{P}=0$. This agrees with Eq.\eqref{SSH-Intensity} up to an overall coefficient. In fact, a bit more of algebra \cite{SI} allows us to fix the non-zero coefficient $a_s$, which leads to Eq.\eqref{SSH-Intensity}. The similar discussion applies to the $\mathcal{P}=1/2$ case, see \cite{SI}, which completes our proof. Note that in deriving the above, we used only the symmetries. One obvious consequence of this is that the precise positions of the states in energy is not important in obtaining the above formula Eq.\eqref{SSH-Intensity}.

Finally, we remark that the above can be easily generalized to the multi-band models, which explicitly lack the chiral symmetry. For example, in \cite{SI}, we consider the four-band model which explicitly breaks the chiral symmetry and demonstrate that we can read off the polarization per each filled band.

\textit{B. 3D TBI}: Our theory is not restricted to 1D and applies equally well to the 3D TBI\cite{Fu-Kane-Mele2007,inv_TI}. We will show that the $\mathbb{Z}_2$ topological band index $\nu_0$\cite{Fu-Kane-Mele2007} can be completely determined from RIXS intensity.

For concreteness, we focus on the realistic, 4-band model $H_{TBI}({\bm{k}})= \sum_{i=0}^5 d_i ({\bm{k}})\Gamma_i$\cite{SI, inv_TI} describing Bi$_{1-x}$ Sb$_x$.\cite{inv_TI,BiSb2009,BiSb2009a} The specific form of $H_{TBI}({\bm{k}})$ is not important for now and can be found in \cite{SI}. The $\mathbb{Z}_2$ index $\nu_0$ is determined by a product of the inversion eigenvalues $\mathcal{I}_v ({\bm{k_*}})$ of the valence bands at the time-reversal invariant momentum (TRIM) points $\bm{k}_{*}$,\cite{inv_TI}  
\begin{align}
(-1)^{\nu_0} = \prod_{\bm{k}_{*}\in \text{TRIM}} \mathcal{I}_v ({\bm{k_*}}). \nonumber 
\end{align}
If $\nu_0 = 1$ mod $2$, then the insulator is topological with a single surface Dirac cone\cite{Fu-Kane-Mele2007}. Otherwise, it is trivial. 

\begin{figure}[t!]
	\includegraphics[width=.5\textwidth]{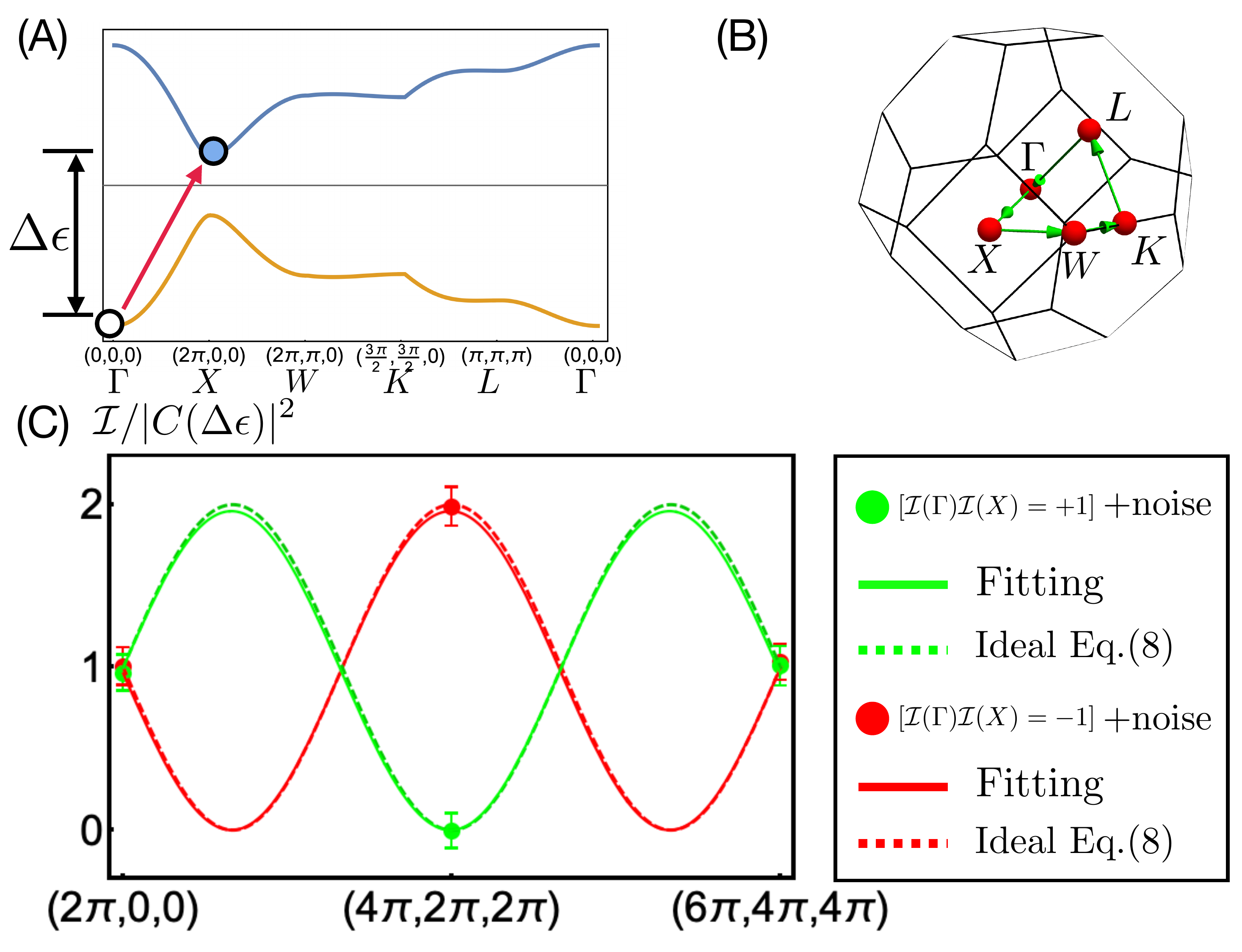}
	\caption{(A)Band structure of the 3D TBI along a momentum cut [Fig.\ref{fig:2}(B)] connecting the high symmetry points in the diamond lattice. The example in the main text considers the momentum transfer $\bm{q}_n= X-\Gamma$. This excites a valence electron at $\Gamma$ to the conduction band at $X$. (B)Momentum space of the diamond lattice. Here, the arrows represent the momentum cut along which the band structure is obtained in [Fig.\ref{fig:2}(A)]. (C)RIXS intensity of 3D TBI for the momentum transfer $\bm{q}_n=X-\Gamma$. As in the SSH case, the error bars for the data (filled circles) are due to the noise that we included in our simulation. See \cite{SI} for details. }
	\label{fig:2}
\end{figure}

Our key observation here is that the RIXS intensity connecting arbitrary two TRIM points, say $\bm{k}_*$ and $\Gamma$, allows us to determine the product $\mathcal{I}_v ({\bm{k_*}}) \mathcal{I}_v (\Gamma)$. This is because $\mathcal{I}_v ({\bm{k_*}}) \mathcal{I}_v (\Gamma)$ manifests in the intensity:
\begin{align}
\mathcal{I}(\bm{q_n}, \Delta\epsilon_{\bm{k}_*}) =  2|\mathcal{C}(\Delta\epsilon_{\bm{k}_*})|^2 \sin^2\Big( {\bm{q_n}\cdot \bm{d}}  + \frac{1}{2}\nu_{k_*}  \pi\Big),
\label{TBI-Intensity}
\end{align}
with ${\bm{d}} = \frac{1}{8}(1,1,1)$ and $\nu_{k_*} = \frac{1}{2} \left(\mathcal{I}_v ({\bm{k_*}})\mathcal{I}_v (\Gamma)-1\right)$. Here, we have tuned the momentum transfer as $\bm{q}_{\bm{n}} = {\bm{k_*}} + \sum_{i=1,2,3} n_i \bm{G}_i, n_i \in \mathbb{Z}$ where $\bm{G}_i$ is a primitive reciprocal vector. The energy transfer $\Delta \epsilon_{\bm{k}_*}$ is tuned to the energy difference between the valence band at $\Gamma$ and the conduction band at $\bm{k}_*$. See [Fig.\ref{fig:2}(A)] for the RIXS process, when $\bm{k}_* = X$. We note that the intensity Eq.\eqref{TBI-Intensity} is largely independent of the hopping parameters of the model (except the overall coefficient $\mathcal{C} (\Delta \epsilon)$). Moreover, its oscillation in momentum is fixed solely by $\mathcal{I}_v ({\bm{k_*}}) \mathcal{I}_v (\Gamma)$. We will see later that this allows us to deduce $\mathcal{I}_v ({\bm{k_*}}) \mathcal{I}_v (\Gamma)$ from the intensity. 

Hence, if we repeat the RIXS experiment on all the TRIM $\bm{k}_*$, we can derive $\mathcal{I}_v ({\bm{k_*}}) \mathcal{I}_v (\Gamma)$ for all the $\bm{k}_*$. This allows us to determine $\nu_0$, because  
\begin{align}
(-1)^{\nu_0} =  \prod_{\bm{k}_{*} \neq \Gamma} \Big[\mathcal{I}_v ({\bm{k_*}})\mathcal{I}_v (\Gamma)\Big]. \label{Z2index}
\end{align} 

Below, we will first discuss an example to demonstrate that Eq.\eqref{TBI-Intensity} allows us to obtain $\mathcal{I}_v ({\bm{k_*}}) \mathcal{I}_v (\Gamma)$ from the RIXS intensity. Then, we sketch our proof of Eq.\eqref{TBI-Intensity}.


- \textit{Numerics}: Let us illustrate the utility of Eq.\eqref{TBI-Intensity} in determining the topology. For this, we remind that to correctly infer the band topology of $H_{TBI}(\bm{k})$, the inversion eigenvalue at $\bm{k}_{*} = X$, i.e., $\mathcal{I}_v (X)$, needs to be carefully determined. (There are $3$ symmetrically-equivalent $X$'s [Fig.\ref{fig:2}(B)], but we will take one below.) This is because the band inversion occurs at $\bm{k}_{*} = X$, which controls the phase transition between $\nu_0 = 0$ and $\nu_0 = 1$.\cite{Fu-Kane-Mele2007,inv_TI} Hence, below we will focus on $\bm{k}_{*}=X$ [Fig.\ref{fig:2}(A)] and demonstrate that we can deduce $\mathcal{I}_v (X)\mathcal{I}_v (\Gamma)$ via Eq.\eqref{TBI-Intensity} from the RIXS intensity. 

For this, we first numerically generate the RIXS intensity data  by applying Eq.\eqref{BandRIXS} to $H_{TBI}(\bm{k})$. The parameters in $H_{TBI}(\bm{k})$ allow us to access both the cases $\mathcal{I}_v (X)\mathcal{I}_v (\Gamma) = \pm 1$. The RIXS intensity data of these two cases are represented as the filled circles in [Fig.\ref{fig:2}(C)]. Green (red) circles represent the data of the system with $\mathcal{I}_v (X)\mathcal{I}_v (\Gamma) = 1$ ($\mathcal{I}_v (X)\mathcal{I}_v (\Gamma) = -1$). As in the SSH chain, we have included white noises to the data to mimic the experimental situations.\cite{SI} Next, we attempted to fit these simulated data with $\mathcal{I}(\bm{q}_n)/|\mathcal{C}(\Delta\epsilon_{\bm{k}_*})|^2 = A \sin^{2} (\bm{q}_n \cdot \bm{d} + B) + C$, where $(A, B, C)$ are the fitting parameters. The solid lines in [Fig.\ref{fig:2}(C)] are the results of the fitting. They agree well with the dotted lines, which are the ideal ones Eq.\eqref{TBI-Intensity}. These dotted lines exactly agree with the simulated data when the noise is removed, and the small discrepency between the ideal ones (dotted lines) and the fitting (solid lines) is due to the white noise. In particular, we found that the fitted parameter $B$ correctly determines $\mathcal{I}_v (X)\mathcal{I}_v (\Gamma)$. See \cite{SI} for the numerical details. Hence, we conclude that we can successfully diagnose $\mathcal{I}_v (X)\mathcal{I}_v (\Gamma)$ via RIXS. Here, although we have focused only on $X$ points, the procedure can be repeated for all the TRIM points. This will determine all the $\mathcal{I}_v ({\bm{k_*}}) \mathcal{I}_v (\Gamma)$ for any TRIM $\bm{k}_*$ and thus $\nu_0$ via Eq.\eqref{Z2index}.  
 
Finally, we note that our theory can be tested in RIXS experiments on the standard topological insulator Bi$_{1-x}$Sb$_x$, where Bi $2s\to6p$ transition can be used. 

- \textit{Proof}: The technical proof of Eq.\eqref{TBI-Intensity} is similar to that of the SSH case, and we will sketch only the key steps here. Minor details can be found in \cite{SI}. Our starting point is the quantum amplitude Eq.\eqref{BandRIXS} between the reference point $\Gamma$ and TRIM ${\bm{k_*}}$
\begin{align}
\mathcal{A}_f \left(      {\bm{q}}_{\bm{n}} , \Delta \epsilon_{\bm{k}_*} \right) = \mathcal{C}(\Delta \epsilon_{\bm{k}_*}) \psi_{c}^{\dagger}({\bm{k_*}}) \hat{\mathcal{M}}_{ {\bm{q}}_{\bm{n}}   } \psi_{v}(\Gamma),  \nonumber
\end{align}
where $\bm{q}_{\bm{n}} = {\bm{k_*}} + \sum_{i=1,2,3} n_i \bm{G}_i, n_i \in \mathbb{Z}$ and we tuned the energy transfers properly to induce the transition from $\Gamma$ and ${\bm{k_*}}$. In the diamond lattice, 
\begin{align}
 \hat{\mathcal{M}}_{{\bm{q_{\bm{n}}}} }=  \cos ({\bm{q_{\bm{n}}}\cdot \bm{d}}) + i \sin ({\bm{q_{\bm{n}}}\cdot \bm{d}})  \cdot\hat{\Pi}, \nonumber 
\end{align}
with $\hat{\Pi}= \Gamma^{12}$.\cite{SI} Similar to the SSH case, we decompose $\mathcal{A}_f$ into the two parts 
\begin{align}
\mathcal{A}_f =\mathcal{C}(\Delta\epsilon_{k_*}) \cdot\Big[ a_c  \cos ({\bm{q_{\bm{n}}}\cdot \bm{d}}) + i a_s \sin ({\bm{q_{\bm{n}}}\cdot \bm{d}}) \Big], \nonumber 
\end{align} 
with 
$
a_c = \psi_c^{\dagger}(\bm{k}_*) \psi_v(\Gamma)
$
and 
$
a_s = \psi_c^{\dagger}(\bm{k}_*) \hat{\Pi} \psi_v(\Gamma)
$.

Similar to the SSH chain, we note that the inversion symmetry enforces $a_s = 0$ for $\mathcal{I}_v (\bm{k_*})\mathcal{I}_v (\Gamma) =-1$  and $a_c = 0$ for $\mathcal{I}_v (\bm{k_*})\mathcal{I}_v (\Gamma) =1$. For example, for $\mathcal{I}_v (\bm{k_*})\mathcal{I}_v (\Gamma) =-1$
\begin{align}
a_s &=  \psi_c^{\dagger}(\bm{k_*}) \cdot \hat{\mathcal{I}}^{-1} \hat{\mathcal{I}} \cdot \hat{\Pi}  \psi_v(\Gamma) =  -\psi_c^{\dagger}(\bm{k_*}) \cdot \hat{\mathcal{I}}^{-1} \hat{\Pi} \hat{\mathcal{I}} \cdot \psi_v(\Gamma) \nonumber\\
&=   -\mathcal{I}_c(\bm{k_*}) \mathcal{I}_v(\Gamma) \psi_c^{\dagger}(\bm{k_*})\hat{\Pi} \psi_v(\Gamma)=  \mathcal{I}_v(\bm{k_*})\mathcal{I}_v(\Gamma)  a_s,   \nonumber
\end{align}
where we have used $\{\hat{\mathcal{I}}, \hat{\Pi}\}=0$ in the first line and $\mathcal{I}_c(\bm{k_*}) = - \mathcal{I}_v(\bm{k_*})$ in the second line.\cite{SI} Hence, $a_s =0$ for $\mathcal{I}_v (\bm{k_*})\mathcal{I}_v (\Gamma) =-1$. On the other hand, the value of $a_c$ has no symmetry constraint and thus is generically non-zero. This leads to $\mathcal{A}_f \sim \cos(\bm{q}_n \cdot \bm{d})$ and $\mathcal{I}\sim \cos^{2} (\bm{q}_n\cdot \bm{d})$. This agrees with Eq.\eqref{TBI-Intensity} up to an overall coefficient. In fact, a bit more of algebra \cite{SI} allows us to fix the non-zero coefficient $a_c$, which precisely leads to Eq.\eqref{TBI-Intensity}. The similar discussion applies to the other case, see \cite{SI}, which completes our proof.

\textit{C. HOTIs}: Finally, we will apply our theory to the higher-order topological insulators, i.e. quadrupole insulator (QI) \cite{Benalcazar2017, hughes2017} and chiral hinge insulator (CHI)\cite{hinge_HOTI}. The discussion here are largely parallel to those of the SSH and TBI cases. The only difference is that the relevant crystal symmetries are $C_4$ and $C_4 \mathcal{I}$. Hence, we will focus on stating the main results, while deferring the details and mathematical proofs to \cite{SI}. 


\textit{- 2D QI}: QI is a $C_4$-symmetric, 4-band insulator, which can support the quantized quadrupolar moment $\mathcal{Q}_{xy}$ mod 1.\cite{Benalcazar2017, hughes2017} When $C_4$ symmetry is imposed, $\mathcal{Q}_{xy}$ can take only the discrete values, either $\mathcal{Q}_{xy} = 1/2$ (topological) or $\mathcal{Q}_{xy} =0$ (trivial). This quadrupole moment can be related with the $C_4$-symmetry eigenvalues of valence bands at $\Gamma$ and $M = (\pi, \pi)$\cite{Benalcazar2017, hughes2017}   
\begin{align}
\exp \Big( 2\pi i \mathcal{Q}_{xy} \Big) = r_{4}(M) r_{4}^{*}(\Gamma), \label{Quadrupole}
\end{align}
where $r_{4} (\bm{k}_*)$ reperesents the eigenvalue of the $C_4$ symmetry [\onlinecite{SI}] at the $C_4$-symmetric $\bm{k}_*$ such that $(r_{4} (\bm{k}))^2 = i$. Some details, which are not crucial for the discussion below, can be found in \cite{SI}. 

From Eq.\eqref{Quadrupole}, we are naturally led to consider the RIXS intensity between $\Gamma$ and $M$ to grasp $\mathcal{Q}_{xy}$. Indeed, we find \cite{SI} that the intensity depends on $\mathcal{Q}_{xy}$
\begin{align}
I[{\bm{q}_n} ,\Delta \epsilon]= 4  |\mathcal{C} (\Delta \epsilon)|^2   \sin^2 \Big[ \left(2n+1\right)\pi d   + \mathcal{Q}_{xy}\pi \Big], \nonumber 
\end{align}
with ${\bm{q}_n} =(2n+1)  (\pi, \pi)$, $n\in \mathbb{Z}$. Here, ``$d$" is the half of the real-space distance (in unit of the lattice constant) between the sublattices within a unit cell.\cite{SI} Thus, the topological index $\mathcal{Q}_{xy}$ can be deduced from the RIXS intensity. This can be tested in candidate materials for QI such as XY (X=Ge, Y=S,Se)\cite{C4_TI_material} where Ge $1s\to4p$ transition is available for the RIXS experiment. In \cite{SI}, we generalize the above to another class of non-degnerate multi-band $C_4$-symmetric insulator $H_{1b}^{(4)}$ of \cite{hughes2017,hughes2019}.

\textit{- 3D CHI}: The above can be generalized to 3D CHI \cite{hinge_HOTI,SI}, which hosts a chiral 1D mode at the hinges. The insulator is classified by another $\mathbb{Z}_2$ topological band index $\nu' \in\{0,1\}$, which can be determined by the inversion eigenvalues $\mathcal{I}(\bm{k}_*)$ at $C_{4z} \mathcal{I}$-symmetric points $\bm{k}_*$  
\begin{align}
(-1)^{\nu'} = \prod_{C_{4z} \mathcal{I}\cdot \bm{k}_* = \bm{k}_*} \mathcal{I}(\bm{k}_*). \nonumber 
\end{align}
Similar to the 3D TBI, this $\mathbb{Z}_2$ index for the CHI can be reformulated in terms of the product of $\mathcal{I}(\bm{k}_*)$  between the two high-symmetric points, e.g., $\mathcal{I}(\bm{k}_*)\mathcal{I}(\Gamma)$.\cite{SI} Hence, the topology of the CHI can be diagnosed in the exactly same fashion as we did for the 3D TBI.

{\color{green}\textit{\textbf{4.Conclusion}}}: In this Letter, we have shown that the RIXS intensity directly reflects the band topology of various insulators in different spatial dimensions. Essentially, the RIXS intensity is strongly constrained by the crystalline symmetry eigenvalues and this allows one to reconstruct the topological band indices. We have explicitly demonstrated this strategy in 1D SSH chain, 2D quadrupolar insulator, 3D topological band insulator and also higher-order chiral hinge insulators. From these examples, it is not difficult to anticipate that a broader class of topology\cite{bradlyn2017topological, Kitaev_periodic, hughes2019, PhysRevX.7.041069,bernevig_2019,Choi:2020wa} of band insulators and superconductors (beyond the examples discussed in this Letter) may as well be manifested in RIXS. We leave the generalization to the future research. 





\acknowledgements
We thank J. Ahn and Y.K Kim for the helpful discussions. G.Y.C. acknowledges the support of the National Research Foundation of Korea (NRF) funded by the Korean Government (Grant No. 2020R1C1C1006048 and 2020R1A4A3079707), as well as Grant No. IBS-R014-D1. G.Y.C. is also supported by the Air Force Office of Scientific Research under Award No. FA2386-20-1-4029. BJK was supported by IBS-R014-A2. SJL was supported by an appointment to the JRG/YST Program at the APCTP through the Science and Technology Promotion Fund and Lottery Fund of the Korean Government. This was also supported by the Korean Local Governments - Gyeongsangbuk-do Province and Pohang City. SJL is also supported by IBS-R014-D1. BK is supported by KIAS individual Grant PG069402 at Korea Institute for Advanced Study and the National Research Foundation of Korea (NRF) grant funded by the Korea government (MSIT) (No. 2020R1F1A1075569). K.-H. J. is supported by the Institute for Basic Science (Grant No. IBS-R014-Y1).

\bibliography{RIXSref}

\begin{thebibliography}{53}%
\makeatletter
\providecommand \@ifxundefined [1]{%
 \@ifx{#1\undefined}
}%
\providecommand \@ifnum [1]{%
 \ifnum #1\expandafter \@firstoftwo
 \else \expandafter \@secondoftwo
 \fi
}%
\providecommand \@ifx [1]{%
 \ifx #1\expandafter \@firstoftwo
 \else \expandafter \@secondoftwo
 \fi
}%
\providecommand \natexlab [1]{#1}%
\providecommand \enquote  [1]{``#1''}%
\providecommand \bibnamefont  [1]{#1}%
\providecommand \bibfnamefont [1]{#1}%
\providecommand \citenamefont [1]{#1}%
\providecommand \href@noop [0]{\@secondoftwo}%
\providecommand \href [0]{\begingroup \@sanitize@url \@href}%
\providecommand \@href[1]{\@@startlink{#1}\@@href}%
\providecommand \@@href[1]{\endgroup#1\@@endlink}%
\providecommand \@sanitize@url [0]{\catcode `\\12\catcode `\$12\catcode
  `\&12\catcode `\#12\catcode `\^12\catcode `\_12\catcode `\%12\relax}%
\providecommand \@@startlink[1]{}%
\providecommand \@@endlink[0]{}%
\providecommand \url  [0]{\begingroup\@sanitize@url \@url }%
\providecommand \@url [1]{\endgroup\@href {#1}{\urlprefix }}%
\providecommand \urlprefix  [0]{URL }%
\providecommand \Eprint [0]{\href }%
\providecommand \doibase [0]{http://dx.doi.org/}%
\providecommand \selectlanguage [0]{\@gobble}%
\providecommand \bibinfo  [0]{\@secondoftwo}%
\providecommand \bibfield  [0]{\@secondoftwo}%
\providecommand \translation [1]{[#1]}%
\providecommand \BibitemOpen [0]{}%
\providecommand \bibitemStop [0]{}%
\providecommand \bibitemNoStop [0]{.\EOS\space}%
\providecommand \EOS [0]{\spacefactor3000\relax}%
\providecommand \BibitemShut  [1]{\csname bibitem#1\endcsname}%
\let\auto@bib@innerbib\@empty
\bibitem [{\citenamefont {Ament}\ \emph
  {et~al.}(2011{\natexlab{a}})\citenamefont {Ament}, \citenamefont {van
  Veenendaal}, \citenamefont {Devereaux}, \citenamefont {Hill},\ and\
  \citenamefont {van~den Brink}}]{rixs_rmp}%
  \BibitemOpen
  \bibfield  {author} {\bibinfo {author} {\bibfnamefont {Luuk J.~P.}\
  \bibnamefont {Ament}}, \bibinfo {author} {\bibfnamefont {Michel}\
  \bibnamefont {van Veenendaal}}, \bibinfo {author} {\bibfnamefont {Thomas~P.}\
  \bibnamefont {Devereaux}}, \bibinfo {author} {\bibfnamefont {John~P.}\
  \bibnamefont {Hill}}, \ and\ \bibinfo {author} {\bibfnamefont {Jeroen}\
  \bibnamefont {van~den Brink}},\ }\bibfield  {title} {\enquote {\bibinfo
  {title} {Resonant inelastic x-ray scattering studies of elementary
  excitations},}\ }\href {\doibase 10.1103/RevModPhys.83.705} {\bibfield
  {journal} {\bibinfo  {journal} {Rev. Mod. Phys.}\ }\textbf {\bibinfo {volume}
  {83}},\ \bibinfo {pages} {705--767} (\bibinfo {year}
  {2011}{\natexlab{a}})}\BibitemShut {NoStop}%
\bibitem [{\citenamefont {Kotani}\ and\ \citenamefont
  {Shin}(2001)}]{rixs_rmp2}%
  \BibitemOpen
  \bibfield  {author} {\bibinfo {author} {\bibfnamefont {Akio}\ \bibnamefont
  {Kotani}}\ and\ \bibinfo {author} {\bibfnamefont {Shik}\ \bibnamefont
  {Shin}},\ }\bibfield  {title} {\enquote {\bibinfo {title} {Resonant inelastic
  x-ray scattering spectra for electrons in solids},}\ }\href {\doibase
  10.1103/RevModPhys.73.203} {\bibfield  {journal} {\bibinfo  {journal} {Rev.
  Mod. Phys.}\ }\textbf {\bibinfo {volume} {73}},\ \bibinfo {pages} {203--246}
  (\bibinfo {year} {2001})}\BibitemShut {NoStop}%
\bibitem [{\citenamefont {Rusydi}\ \emph {et~al.}(2014)\citenamefont {Rusydi},
  \citenamefont {Goos}, \citenamefont {Binder}, \citenamefont {Eich},
  \citenamefont {Botril}, \citenamefont {Abbamonte}, \citenamefont {Yu},
  \citenamefont {Breese}, \citenamefont {Eisaki}, \citenamefont {Fujimaki},
  \citenamefont {Uchida}, \citenamefont {Guerassimova}, \citenamefont
  {Treusch}, \citenamefont {Feldhaus}, \citenamefont {Reininger}, \citenamefont
  {Klein},\ and\ \citenamefont {R\"ubhausen}}]{peter_2014}%
  \BibitemOpen
  \bibfield  {author} {\bibinfo {author} {\bibfnamefont {A.}~\bibnamefont
  {Rusydi}}, \bibinfo {author} {\bibfnamefont {A.}~\bibnamefont {Goos}},
  \bibinfo {author} {\bibfnamefont {S.}~\bibnamefont {Binder}}, \bibinfo
  {author} {\bibfnamefont {A.}~\bibnamefont {Eich}}, \bibinfo {author}
  {\bibfnamefont {K.}~\bibnamefont {Botril}}, \bibinfo {author} {\bibfnamefont
  {P.}~\bibnamefont {Abbamonte}}, \bibinfo {author} {\bibfnamefont
  {X.}~\bibnamefont {Yu}}, \bibinfo {author} {\bibfnamefont {M.~B.~H.}\
  \bibnamefont {Breese}}, \bibinfo {author} {\bibfnamefont {H.}~\bibnamefont
  {Eisaki}}, \bibinfo {author} {\bibfnamefont {Y.}~\bibnamefont {Fujimaki}},
  \bibinfo {author} {\bibfnamefont {S.}~\bibnamefont {Uchida}}, \bibinfo
  {author} {\bibfnamefont {N.}~\bibnamefont {Guerassimova}}, \bibinfo {author}
  {\bibfnamefont {R.}~\bibnamefont {Treusch}}, \bibinfo {author} {\bibfnamefont
  {J.}~\bibnamefont {Feldhaus}}, \bibinfo {author} {\bibfnamefont
  {R.}~\bibnamefont {Reininger}}, \bibinfo {author} {\bibfnamefont {M.~V.}\
  \bibnamefont {Klein}}, \ and\ \bibinfo {author} {\bibfnamefont
  {M.}~\bibnamefont {R\"ubhausen}},\ }\bibfield  {title} {\enquote {\bibinfo
  {title} {Electronic screening-enhanced hole pairing in two-leg spin ladders
  studied by high-resolution resonant inelastic x-ray scattering at cu $m$
  edges},}\ }\href {\doibase 10.1103/PhysRevLett.113.067001} {\bibfield
  {journal} {\bibinfo  {journal} {Phys. Rev. Lett.}\ }\textbf {\bibinfo
  {volume} {113}},\ \bibinfo {pages} {067001} (\bibinfo {year}
  {2014})}\BibitemShut {NoStop}%
\bibitem [{\citenamefont {Abbamonte}\ \emph {et~al.}(1999)\citenamefont
  {Abbamonte}, \citenamefont {Burns}, \citenamefont {Isaacs}, \citenamefont
  {Platzman}, \citenamefont {Miller}, \citenamefont {Cheong},\ and\
  \citenamefont {Klein}}]{peter_1999}%
  \BibitemOpen
  \bibfield  {author} {\bibinfo {author} {\bibfnamefont {P.}~\bibnamefont
  {Abbamonte}}, \bibinfo {author} {\bibfnamefont {C.~A.}\ \bibnamefont
  {Burns}}, \bibinfo {author} {\bibfnamefont {E.~D.}\ \bibnamefont {Isaacs}},
  \bibinfo {author} {\bibfnamefont {P.~M.}\ \bibnamefont {Platzman}}, \bibinfo
  {author} {\bibfnamefont {L.~L.}\ \bibnamefont {Miller}}, \bibinfo {author}
  {\bibfnamefont {S.~W.}\ \bibnamefont {Cheong}}, \ and\ \bibinfo {author}
  {\bibfnamefont {M.~V.}\ \bibnamefont {Klein}},\ }\bibfield  {title} {\enquote
  {\bibinfo {title} {Resonant inelastic x-ray scattering from valence
  excitations in insulating copper oxides},}\ }\href {\doibase
  10.1103/PhysRevLett.83.860} {\bibfield  {journal} {\bibinfo  {journal} {Phys.
  Rev. Lett.}\ }\textbf {\bibinfo {volume} {83}},\ \bibinfo {pages} {860--863}
  (\bibinfo {year} {1999})}\BibitemShut {NoStop}%
\bibitem [{\citenamefont {Kim}\ \emph {et~al.}(2009)\citenamefont {Kim},
  \citenamefont {Ohsumi}, \citenamefont {Komesu}, \citenamefont {Sakai},
  \citenamefont {Morita}, \citenamefont {Takagi},\ and\ \citenamefont
  {Arima}}]{Kim1329}%
  \BibitemOpen
  \bibfield  {author} {\bibinfo {author} {\bibfnamefont {B.~J.}\ \bibnamefont
  {Kim}}, \bibinfo {author} {\bibfnamefont {H.}~\bibnamefont {Ohsumi}},
  \bibinfo {author} {\bibfnamefont {T.}~\bibnamefont {Komesu}}, \bibinfo
  {author} {\bibfnamefont {S.}~\bibnamefont {Sakai}}, \bibinfo {author}
  {\bibfnamefont {T.}~\bibnamefont {Morita}}, \bibinfo {author} {\bibfnamefont
  {H.}~\bibnamefont {Takagi}}, \ and\ \bibinfo {author} {\bibfnamefont
  {T.}~\bibnamefont {Arima}},\ }\bibfield  {title} {\enquote {\bibinfo {title}
  {Phase-sensitive observation of a spin-orbital mott state in sr2iro4},}\
  }\href {\doibase 10.1126/science.1167106} {\bibfield  {journal} {\bibinfo
  {journal} {Science}\ }\textbf {\bibinfo {volume} {323}},\ \bibinfo {pages}
  {1329--1332} (\bibinfo {year} {2009})}\BibitemShut {NoStop}%
\bibitem [{\citenamefont {Kim}\ and\ \citenamefont
  {Khaliullin}(2017)}]{BJ2017}%
  \BibitemOpen
  \bibfield  {author} {\bibinfo {author} {\bibfnamefont {B.~J.}\ \bibnamefont
  {Kim}}\ and\ \bibinfo {author} {\bibfnamefont {Giniyat}\ \bibnamefont
  {Khaliullin}},\ }\bibfield  {title} {\enquote {\bibinfo {title} {Resonant
  inelastic x-ray scattering operators for ${t}_{2g}$ orbital systems},}\
  }\href {\doibase 10.1103/PhysRevB.96.085108} {\bibfield  {journal} {\bibinfo
  {journal} {Phys. Rev. B}\ }\textbf {\bibinfo {volume} {96}},\ \bibinfo
  {pages} {085108} (\bibinfo {year} {2017})}\BibitemShut {NoStop}%
\bibitem [{\citenamefont {Kim}\ \emph {et~al.}(2014)\citenamefont {Kim},
  \citenamefont {Daghofer}, \citenamefont {Said}, \citenamefont {Gog},
  \citenamefont {van~den Brink}, \citenamefont {Khaliullin},\ and\
  \citenamefont {Kim}}]{BJKim2014}%
  \BibitemOpen
  \bibfield  {author} {\bibinfo {author} {\bibfnamefont {Jungho}\ \bibnamefont
  {Kim}}, \bibinfo {author} {\bibfnamefont {M.}~\bibnamefont {Daghofer}},
  \bibinfo {author} {\bibfnamefont {A.~H.}\ \bibnamefont {Said}}, \bibinfo
  {author} {\bibfnamefont {T.}~\bibnamefont {Gog}}, \bibinfo {author}
  {\bibfnamefont {J.}~\bibnamefont {van~den Brink}}, \bibinfo {author}
  {\bibfnamefont {G.}~\bibnamefont {Khaliullin}}, \ and\ \bibinfo {author}
  {\bibfnamefont {B.~J.}\ \bibnamefont {Kim}},\ }\bibfield  {title} {\enquote
  {\bibinfo {title} {Excitonic quasiparticles in a spin--orbit mott
  insulator},}\ }\href {\doibase 10.1038/ncomms5453} {\bibfield  {journal}
  {\bibinfo  {journal} {Nature Communications}\ }\textbf {\bibinfo {volume}
  {5}},\ \bibinfo {pages} {4453} (\bibinfo {year} {2014})}\BibitemShut
  {NoStop}%
\bibitem [{\citenamefont {Revelli}\ \emph {et~al.}(2019)\citenamefont
  {Revelli}, \citenamefont {Moretti~Sala}, \citenamefont {Monaco},
  \citenamefont {Becker}, \citenamefont {Bohat{\'y}}, \citenamefont {Hermanns},
  \citenamefont {Koethe}, \citenamefont {Fr{\"o}hlich}, \citenamefont
  {Warzanowski}, \citenamefont {Lorenz}, \citenamefont {Streltsov},
  \citenamefont {van Loosdrecht}, \citenamefont {Khomskii}, \citenamefont
  {van~den Brink},\ and\ \citenamefont {Gr{\"u}ninger}}]{Revelli2019}%
  \BibitemOpen
  \bibfield  {author} {\bibinfo {author} {\bibfnamefont {A.}~\bibnamefont
  {Revelli}}, \bibinfo {author} {\bibfnamefont {M.}~\bibnamefont
  {Moretti~Sala}}, \bibinfo {author} {\bibfnamefont {G.}~\bibnamefont
  {Monaco}}, \bibinfo {author} {\bibfnamefont {P.}~\bibnamefont {Becker}},
  \bibinfo {author} {\bibfnamefont {L.}~\bibnamefont {Bohat{\'y}}}, \bibinfo
  {author} {\bibfnamefont {M.}~\bibnamefont {Hermanns}}, \bibinfo {author}
  {\bibfnamefont {T.~C.}\ \bibnamefont {Koethe}}, \bibinfo {author}
  {\bibfnamefont {T.}~\bibnamefont {Fr{\"o}hlich}}, \bibinfo {author}
  {\bibfnamefont {P.}~\bibnamefont {Warzanowski}}, \bibinfo {author}
  {\bibfnamefont {T.}~\bibnamefont {Lorenz}}, \bibinfo {author} {\bibfnamefont
  {S.~V.}\ \bibnamefont {Streltsov}}, \bibinfo {author} {\bibfnamefont
  {P.~H.~M.}\ \bibnamefont {van Loosdrecht}}, \bibinfo {author} {\bibfnamefont
  {D.~I.}\ \bibnamefont {Khomskii}}, \bibinfo {author} {\bibfnamefont
  {J.}~\bibnamefont {van~den Brink}}, \ and\ \bibinfo {author} {\bibfnamefont
  {M.}~\bibnamefont {Gr{\"u}ninger}},\ }\bibfield  {title} {\enquote {\bibinfo
  {title} {Resonant inelastic x-ray incarnation of young{\textquoteright}s
  double-slit experiment},}\ }\href {\doibase 10.1126/sciadv.aav4020}
  {\bibfield  {journal} {\bibinfo  {journal} {Science Advances}\ }\textbf
  {\bibinfo {volume} {5}} (\bibinfo {year} {2019}),\
  10.1126/sciadv.aav4020}\BibitemShut {NoStop}%
\bibitem [{\citenamefont {Savary}\ and\ \citenamefont
  {Senthil}(2015)}]{lucile_RIXS}%
  \BibitemOpen
  \bibfield  {author} {\bibinfo {author} {\bibfnamefont {Lucile}\ \bibnamefont
  {Savary}}\ and\ \bibinfo {author} {\bibfnamefont {T.}~\bibnamefont
  {Senthil}},\ }\bibfield  {title} {\enquote {\bibinfo {title} {Probing hidden
  orders with resonant inelastic x-ray scattering},}\ }\href@noop {} {\bibfield
   {journal} {\bibinfo  {journal} {arXiv:1506.04752}\ } (\bibinfo {year}
  {2015})}\BibitemShut {NoStop}%
\bibitem [{\citenamefont {Hal\'asz}\ \emph {et~al.}(2019)\citenamefont
  {Hal\'asz}, \citenamefont {Kourtis}, \citenamefont {Knolle},\ and\
  \citenamefont {Perkins}}]{rixs_method7}%
  \BibitemOpen
  \bibfield  {author} {\bibinfo {author} {\bibfnamefont {G\'abor~B.}\
  \bibnamefont {Hal\'asz}}, \bibinfo {author} {\bibfnamefont {Stefanos}\
  \bibnamefont {Kourtis}}, \bibinfo {author} {\bibfnamefont {Johannes}\
  \bibnamefont {Knolle}}, \ and\ \bibinfo {author} {\bibfnamefont {Natalia~B.}\
  \bibnamefont {Perkins}},\ }\bibfield  {title} {\enquote {\bibinfo {title}
  {Observing spin fractionalization in the kitaev spin liquid via temperature
  evolution of indirect resonant inelastic x-ray scattering},}\ }\href
  {\doibase 10.1103/PhysRevB.99.184417} {\bibfield  {journal} {\bibinfo
  {journal} {Phys. Rev. B}\ }\textbf {\bibinfo {volume} {99}},\ \bibinfo
  {pages} {184417} (\bibinfo {year} {2019})}\BibitemShut {NoStop}%
\bibitem [{\citenamefont {Marra}\ \emph {et~al.}(2016)\citenamefont {Marra},
  \citenamefont {van~den Brink},\ and\ \citenamefont {Sykora}}]{RIXS_SC}%
  \BibitemOpen
  \bibfield  {author} {\bibinfo {author} {\bibfnamefont {Pasquale}\
  \bibnamefont {Marra}}, \bibinfo {author} {\bibfnamefont {Jeroen}\
  \bibnamefont {van~den Brink}}, \ and\ \bibinfo {author} {\bibfnamefont
  {Steffen}\ \bibnamefont {Sykora}},\ }\bibfield  {title} {\enquote {\bibinfo
  {title} {Theoretical approach to resonant inelastic x-ray scattering in
  iron-based superconductors at the energy scale of the superconducting gap},}\
  }\href {\doibase 10.1038/srep25386} {\bibfield  {journal} {\bibinfo
  {journal} {Scientific Reports}\ }\textbf {\bibinfo {volume} {6}},\ \bibinfo
  {pages} {25386} (\bibinfo {year} {2016})}\BibitemShut {NoStop}%
\bibitem [{\citenamefont {Kourtis}(2016)}]{kourtis_2016}%
  \BibitemOpen
  \bibfield  {author} {\bibinfo {author} {\bibfnamefont {Stefanos}\
  \bibnamefont {Kourtis}},\ }\bibfield  {title} {\enquote {\bibinfo {title}
  {Bulk spectroscopic measurement of the topological charge of weyl nodes with
  resonant x rays},}\ }\href {\doibase 10.1103/PhysRevB.94.125132} {\bibfield
  {journal} {\bibinfo  {journal} {Phys. Rev. B}\ }\textbf {\bibinfo {volume}
  {94}},\ \bibinfo {pages} {125132} (\bibinfo {year} {2016})}\BibitemShut
  {NoStop}%
\bibitem [{\citenamefont {Hasan}\ and\ \citenamefont
  {Kane}(2010)}]{RevModPhys.82.3045}%
  \BibitemOpen
  \bibfield  {author} {\bibinfo {author} {\bibfnamefont {M.~Z.}\ \bibnamefont
  {Hasan}}\ and\ \bibinfo {author} {\bibfnamefont {C.~L.}\ \bibnamefont
  {Kane}},\ }\bibfield  {title} {\enquote {\bibinfo {title} {Colloquium:
  Topological insulators},}\ }\href {\doibase 10.1103/RevModPhys.82.3045}
  {\bibfield  {journal} {\bibinfo  {journal} {Rev. Mod. Phys.}\ }\textbf
  {\bibinfo {volume} {82}},\ \bibinfo {pages} {3045--3067} (\bibinfo {year}
  {2010})}\BibitemShut {NoStop}%
\bibitem [{\citenamefont {Qi}\ and\ \citenamefont {Zhang}(2011)}]{Qi_rmp}%
  \BibitemOpen
  \bibfield  {author} {\bibinfo {author} {\bibfnamefont {Xiao-Liang}\
  \bibnamefont {Qi}}\ and\ \bibinfo {author} {\bibfnamefont {Shou-Cheng}\
  \bibnamefont {Zhang}},\ }\bibfield  {title} {\enquote {\bibinfo {title}
  {Topological insulators and superconductors},}\ }\href {\doibase
  10.1103/RevModPhys.83.1057} {\bibfield  {journal} {\bibinfo  {journal} {Rev.
  Mod. Phys.}\ }\textbf {\bibinfo {volume} {83}},\ \bibinfo {pages}
  {1057--1110} (\bibinfo {year} {2011})}\BibitemShut {NoStop}%
\bibitem [{\citenamefont {Chiu}\ \emph {et~al.}(2016)\citenamefont {Chiu},
  \citenamefont {Teo}, \citenamefont {Schnyder},\ and\ \citenamefont
  {Ryu}}]{chiu_rmp}%
  \BibitemOpen
  \bibfield  {author} {\bibinfo {author} {\bibfnamefont {Ching-Kai}\
  \bibnamefont {Chiu}}, \bibinfo {author} {\bibfnamefont {Jeffrey C.~Y.}\
  \bibnamefont {Teo}}, \bibinfo {author} {\bibfnamefont {Andreas~P.}\
  \bibnamefont {Schnyder}}, \ and\ \bibinfo {author} {\bibfnamefont {Shinsei}\
  \bibnamefont {Ryu}},\ }\bibfield  {title} {\enquote {\bibinfo {title}
  {Classification of topological quantum matter with symmetries},}\ }\href
  {\doibase 10.1103/RevModPhys.88.035005} {\bibfield  {journal} {\bibinfo
  {journal} {Rev. Mod. Phys.}\ }\textbf {\bibinfo {volume} {88}},\ \bibinfo
  {pages} {035005} (\bibinfo {year} {2016})}\BibitemShut {NoStop}%
\bibitem [{\citenamefont {Chen}\ \emph {et~al.}(2009)\citenamefont {Chen},
  \citenamefont {Analytis}, \citenamefont {Chu}, \citenamefont {Liu},
  \citenamefont {Mo}, \citenamefont {Qi}, \citenamefont {Zhang}, \citenamefont
  {Lu}, \citenamefont {Dai}, \citenamefont {Fang}, \citenamefont {Zhang},
  \citenamefont {Fisher}, \citenamefont {Hussain},\ and\ \citenamefont
  {Shen}}]{Chen_ARPES2009}%
  \BibitemOpen
  \bibfield  {author} {\bibinfo {author} {\bibfnamefont {Y.~L.}\ \bibnamefont
  {Chen}}, \bibinfo {author} {\bibfnamefont {J.~G.}\ \bibnamefont {Analytis}},
  \bibinfo {author} {\bibfnamefont {J.-H.}\ \bibnamefont {Chu}}, \bibinfo
  {author} {\bibfnamefont {Z.~K.}\ \bibnamefont {Liu}}, \bibinfo {author}
  {\bibfnamefont {S.-K.}\ \bibnamefont {Mo}}, \bibinfo {author} {\bibfnamefont
  {X.~L.}\ \bibnamefont {Qi}}, \bibinfo {author} {\bibfnamefont {H.~J.}\
  \bibnamefont {Zhang}}, \bibinfo {author} {\bibfnamefont {D.~H.}\ \bibnamefont
  {Lu}}, \bibinfo {author} {\bibfnamefont {X.}~\bibnamefont {Dai}}, \bibinfo
  {author} {\bibfnamefont {Z.}~\bibnamefont {Fang}}, \bibinfo {author}
  {\bibfnamefont {S.~C.}\ \bibnamefont {Zhang}}, \bibinfo {author}
  {\bibfnamefont {I.~R.}\ \bibnamefont {Fisher}}, \bibinfo {author}
  {\bibfnamefont {Z.}~\bibnamefont {Hussain}}, \ and\ \bibinfo {author}
  {\bibfnamefont {Z.-X.}\ \bibnamefont {Shen}},\ }\bibfield  {title} {\enquote
  {\bibinfo {title} {Experimental realization of a three-dimensional
  topological insulator, bi<sub>2</sub>te<sub>3</sub>},}\ }\href {\doibase
  10.1126/science.1173034} {\bibfield  {journal} {\bibinfo  {journal}
  {Science}\ }\textbf {\bibinfo {volume} {325}},\ \bibinfo {pages} {178--181}
  (\bibinfo {year} {2009})}\BibitemShut {NoStop}%
\bibitem [{\citenamefont {Xia}\ \emph {et~al.}(2009)\citenamefont {Xia},
  \citenamefont {Qian}, \citenamefont {Hsieh}, \citenamefont {Wray},
  \citenamefont {Pal}, \citenamefont {Lin}, \citenamefont {Bansil},
  \citenamefont {Grauer}, \citenamefont {Hor}, \citenamefont {Cava},\ and\
  \citenamefont {Hasan}}]{Xia_ARPES}%
  \BibitemOpen
  \bibfield  {author} {\bibinfo {author} {\bibfnamefont {Y.}~\bibnamefont
  {Xia}}, \bibinfo {author} {\bibfnamefont {D.}~\bibnamefont {Qian}}, \bibinfo
  {author} {\bibfnamefont {D.}~\bibnamefont {Hsieh}}, \bibinfo {author}
  {\bibfnamefont {L.}~\bibnamefont {Wray}}, \bibinfo {author} {\bibfnamefont
  {A.}~\bibnamefont {Pal}}, \bibinfo {author} {\bibfnamefont {H.}~\bibnamefont
  {Lin}}, \bibinfo {author} {\bibfnamefont {A.}~\bibnamefont {Bansil}},
  \bibinfo {author} {\bibfnamefont {D.}~\bibnamefont {Grauer}}, \bibinfo
  {author} {\bibfnamefont {Y.~S.}\ \bibnamefont {Hor}}, \bibinfo {author}
  {\bibfnamefont {R.~J.}\ \bibnamefont {Cava}}, \ and\ \bibinfo {author}
  {\bibfnamefont {M.~Z.}\ \bibnamefont {Hasan}},\ }\bibfield  {title} {\enquote
  {\bibinfo {title} {Observation of a large-gap topological-insulator class
  with a single dirac cone on the surface},}\ }\href {\doibase
  10.1038/nphys1274} {\bibfield  {journal} {\bibinfo  {journal} {Nature
  Physics}\ }\textbf {\bibinfo {volume} {5}},\ \bibinfo {pages} {398--402}
  (\bibinfo {year} {2009})}\BibitemShut {NoStop}%
\bibitem [{\citenamefont {Hsieh}\ \emph {et~al.}(2009)\citenamefont {Hsieh},
  \citenamefont {Xia}, \citenamefont {Wray}, \citenamefont {Qian},
  \citenamefont {Pal}, \citenamefont {Dil}, \citenamefont {Osterwalder},
  \citenamefont {Meier}, \citenamefont {Bihlmayer}, \citenamefont {Kane},
  \citenamefont {Hor}, \citenamefont {Cava},\ and\ \citenamefont
  {Hasan}}]{ARPES_hsieh2009}%
  \BibitemOpen
  \bibfield  {author} {\bibinfo {author} {\bibfnamefont {D.}~\bibnamefont
  {Hsieh}}, \bibinfo {author} {\bibfnamefont {Y.}~\bibnamefont {Xia}}, \bibinfo
  {author} {\bibfnamefont {L.}~\bibnamefont {Wray}}, \bibinfo {author}
  {\bibfnamefont {D.}~\bibnamefont {Qian}}, \bibinfo {author} {\bibfnamefont
  {A.}~\bibnamefont {Pal}}, \bibinfo {author} {\bibfnamefont {J.~H.}\
  \bibnamefont {Dil}}, \bibinfo {author} {\bibfnamefont {J.}~\bibnamefont
  {Osterwalder}}, \bibinfo {author} {\bibfnamefont {F.}~\bibnamefont {Meier}},
  \bibinfo {author} {\bibfnamefont {G.}~\bibnamefont {Bihlmayer}}, \bibinfo
  {author} {\bibfnamefont {C.~L.}\ \bibnamefont {Kane}}, \bibinfo {author}
  {\bibfnamefont {Y.~S.}\ \bibnamefont {Hor}}, \bibinfo {author} {\bibfnamefont
  {R.~J.}\ \bibnamefont {Cava}}, \ and\ \bibinfo {author} {\bibfnamefont
  {M.~Z.}\ \bibnamefont {Hasan}},\ }\bibfield  {title} {\enquote {\bibinfo
  {title} {Observation of unconventional quantum spin textures in topological
  insulators},}\ }\href {\doibase 10.1126/science.1167733} {\bibfield
  {journal} {\bibinfo  {journal} {Science}\ }\textbf {\bibinfo {volume}
  {323}},\ \bibinfo {pages} {919--922} (\bibinfo {year} {2009})},\ \Eprint
  {http://arxiv.org/abs/https://www.science.org/doi/pdf/10.1126/science.1167733}
  {https://www.science.org/doi/pdf/10.1126/science.1167733} \BibitemShut
  {NoStop}%
\bibitem [{\citenamefont {Nishide}\ \emph {et~al.}(2010)\citenamefont
  {Nishide}, \citenamefont {Taskin}, \citenamefont {Takeichi}, \citenamefont
  {Okuda}, \citenamefont {Kakizaki}, \citenamefont {Hirahara}, \citenamefont
  {Nakatsuji}, \citenamefont {Komori}, \citenamefont {Ando},\ and\
  \citenamefont {Matsuda}}]{Nishide_ARPES2010}%
  \BibitemOpen
  \bibfield  {author} {\bibinfo {author} {\bibfnamefont {Akinori}\ \bibnamefont
  {Nishide}}, \bibinfo {author} {\bibfnamefont {Alexey~A.}\ \bibnamefont
  {Taskin}}, \bibinfo {author} {\bibfnamefont {Yasuo}\ \bibnamefont
  {Takeichi}}, \bibinfo {author} {\bibfnamefont {Taichi}\ \bibnamefont
  {Okuda}}, \bibinfo {author} {\bibfnamefont {Akito}\ \bibnamefont {Kakizaki}},
  \bibinfo {author} {\bibfnamefont {Toru}\ \bibnamefont {Hirahara}}, \bibinfo
  {author} {\bibfnamefont {Kan}\ \bibnamefont {Nakatsuji}}, \bibinfo {author}
  {\bibfnamefont {Fumio}\ \bibnamefont {Komori}}, \bibinfo {author}
  {\bibfnamefont {Yoichi}\ \bibnamefont {Ando}}, \ and\ \bibinfo {author}
  {\bibfnamefont {Iwao}\ \bibnamefont {Matsuda}},\ }\bibfield  {title}
  {\enquote {\bibinfo {title} {Direct mapping of the spin-filtered surface
  bands of a three-dimensional quantum spin hall insulator},}\ }\href {\doibase
  10.1103/PhysRevB.81.041309} {\bibfield  {journal} {\bibinfo  {journal} {Phys.
  Rev. B}\ }\textbf {\bibinfo {volume} {81}},\ \bibinfo {pages} {041309}
  (\bibinfo {year} {2010})}\BibitemShut {NoStop}%
\bibitem [{\citenamefont {Roushan}\ \emph {et~al.}(2009)\citenamefont
  {Roushan}, \citenamefont {Seo}, \citenamefont {Parker}, \citenamefont {Hor},
  \citenamefont {Hsieh}, \citenamefont {Qian}, \citenamefont {Richardella},
  \citenamefont {Hasan}, \citenamefont {Cava},\ and\ \citenamefont
  {Yazdani}}]{ARPES_STM_Roushan:2009}%
  \BibitemOpen
  \bibfield  {author} {\bibinfo {author} {\bibfnamefont {Pedram}\ \bibnamefont
  {Roushan}}, \bibinfo {author} {\bibfnamefont {Jungpil}\ \bibnamefont {Seo}},
  \bibinfo {author} {\bibfnamefont {Colin~V.}\ \bibnamefont {Parker}}, \bibinfo
  {author} {\bibfnamefont {Y.~S.}\ \bibnamefont {Hor}}, \bibinfo {author}
  {\bibfnamefont {D.}~\bibnamefont {Hsieh}}, \bibinfo {author} {\bibfnamefont
  {Dong}\ \bibnamefont {Qian}}, \bibinfo {author} {\bibfnamefont {Anthony}\
  \bibnamefont {Richardella}}, \bibinfo {author} {\bibfnamefont {M.~Z.}\
  \bibnamefont {Hasan}}, \bibinfo {author} {\bibfnamefont {R.~J.}\ \bibnamefont
  {Cava}}, \ and\ \bibinfo {author} {\bibfnamefont {Ali}\ \bibnamefont
  {Yazdani}},\ }\bibfield  {title} {\enquote {\bibinfo {title} {Topological
  surface states protected from backscattering by chiral spin texture},}\
  }\href {\doibase 10.1038/nature08308} {\bibfield  {journal} {\bibinfo
  {journal} {Nature}\ }\textbf {\bibinfo {volume} {460}},\ \bibinfo {pages}
  {1106--1109} (\bibinfo {year} {2009})}\BibitemShut {NoStop}%
\bibitem [{\citenamefont {Seo}\ \emph {et~al.}(2010)\citenamefont {Seo},
  \citenamefont {Roushan}, \citenamefont {Beidenkopf}, \citenamefont {Hor},
  \citenamefont {Cava},\ and\ \citenamefont {Yazdani}}]{STM_Seo2010}%
  \BibitemOpen
  \bibfield  {author} {\bibinfo {author} {\bibfnamefont {Jungpil}\ \bibnamefont
  {Seo}}, \bibinfo {author} {\bibfnamefont {Pedram}\ \bibnamefont {Roushan}},
  \bibinfo {author} {\bibfnamefont {Haim}\ \bibnamefont {Beidenkopf}}, \bibinfo
  {author} {\bibfnamefont {Y.~S.}\ \bibnamefont {Hor}}, \bibinfo {author}
  {\bibfnamefont {R.~J.}\ \bibnamefont {Cava}}, \ and\ \bibinfo {author}
  {\bibfnamefont {Ali}\ \bibnamefont {Yazdani}},\ }\bibfield  {title} {\enquote
  {\bibinfo {title} {Transmission of topological surface states through surface
  barriers},}\ }\href {\doibase 10.1038/nature09189} {\bibfield  {journal}
  {\bibinfo  {journal} {Nature}\ }\textbf {\bibinfo {volume} {466}},\ \bibinfo
  {pages} {343--346} (\bibinfo {year} {2010})}\BibitemShut {NoStop}%
\bibitem [{\citenamefont {Beidenkopf}\ \emph {et~al.}(2011)\citenamefont
  {Beidenkopf}, \citenamefont {Roushan}, \citenamefont {Seo}, \citenamefont
  {Gorman}, \citenamefont {Drozdov}, \citenamefont {Hor}, \citenamefont
  {Cava},\ and\ \citenamefont {Yazdani}}]{STM_Beidenkopf2011}%
  \BibitemOpen
  \bibfield  {author} {\bibinfo {author} {\bibfnamefont {Haim}\ \bibnamefont
  {Beidenkopf}}, \bibinfo {author} {\bibfnamefont {Pedram}\ \bibnamefont
  {Roushan}}, \bibinfo {author} {\bibfnamefont {Jungpil}\ \bibnamefont {Seo}},
  \bibinfo {author} {\bibfnamefont {Lindsay}\ \bibnamefont {Gorman}}, \bibinfo
  {author} {\bibfnamefont {Ilya}\ \bibnamefont {Drozdov}}, \bibinfo {author}
  {\bibfnamefont {Yew~San}\ \bibnamefont {Hor}}, \bibinfo {author}
  {\bibfnamefont {R.~J.}\ \bibnamefont {Cava}}, \ and\ \bibinfo {author}
  {\bibfnamefont {Ali}\ \bibnamefont {Yazdani}},\ }\bibfield  {title} {\enquote
  {\bibinfo {title} {Spatial fluctuations of helical dirac fermions on the
  surface of topological insulators},}\ }\href {\doibase 10.1038/nphys2108}
  {\bibfield  {journal} {\bibinfo  {journal} {Nature Physics}\ }\textbf
  {\bibinfo {volume} {7}},\ \bibinfo {pages} {939--943} (\bibinfo {year}
  {2011})}\BibitemShut {NoStop}%
\bibitem [{\citenamefont {Alpichshev}\ \emph {et~al.}(2010)\citenamefont
  {Alpichshev}, \citenamefont {Analytis}, \citenamefont {Chu}, \citenamefont
  {Fisher}, \citenamefont {Chen}, \citenamefont {Shen}, \citenamefont {Fang},\
  and\ \citenamefont {Kapitulnik}}]{STM_Alpichshev2010}%
  \BibitemOpen
  \bibfield  {author} {\bibinfo {author} {\bibfnamefont {Zhanybek}\
  \bibnamefont {Alpichshev}}, \bibinfo {author} {\bibfnamefont {J.~G.}\
  \bibnamefont {Analytis}}, \bibinfo {author} {\bibfnamefont {J.-H.}\
  \bibnamefont {Chu}}, \bibinfo {author} {\bibfnamefont {I.~R.}\ \bibnamefont
  {Fisher}}, \bibinfo {author} {\bibfnamefont {Y.~L.}\ \bibnamefont {Chen}},
  \bibinfo {author} {\bibfnamefont {Z.~X.}\ \bibnamefont {Shen}}, \bibinfo
  {author} {\bibfnamefont {A.}~\bibnamefont {Fang}}, \ and\ \bibinfo {author}
  {\bibfnamefont {A.}~\bibnamefont {Kapitulnik}},\ }\bibfield  {title}
  {\enquote {\bibinfo {title} {Stm imaging of electronic waves on the surface
  of ${\mathrm{bi}}_{2}{\mathrm{te}}_{3}$: Topologically protected surface
  states and hexagonal warping effects},}\ }\href {\doibase
  10.1103/PhysRevLett.104.016401} {\bibfield  {journal} {\bibinfo  {journal}
  {Phys. Rev. Lett.}\ }\textbf {\bibinfo {volume} {104}},\ \bibinfo {pages}
  {016401} (\bibinfo {year} {2010})}\BibitemShut {NoStop}%
\bibitem [{\citenamefont {Nadj-Perge}\ \emph {et~al.}(2014)\citenamefont
  {Nadj-Perge}, \citenamefont {Drozdov}, \citenamefont {Li}, \citenamefont
  {Chen}, \citenamefont {Jeon}, \citenamefont {Seo}, \citenamefont {MacDonald},
  \citenamefont {Bernevig},\ and\ \citenamefont {Yazdani}}]{STM_majorana}%
  \BibitemOpen
  \bibfield  {author} {\bibinfo {author} {\bibfnamefont {Stevan}\ \bibnamefont
  {Nadj-Perge}}, \bibinfo {author} {\bibfnamefont {Ilya~K.}\ \bibnamefont
  {Drozdov}}, \bibinfo {author} {\bibfnamefont {Jian}\ \bibnamefont {Li}},
  \bibinfo {author} {\bibfnamefont {Hua}\ \bibnamefont {Chen}}, \bibinfo
  {author} {\bibfnamefont {Sangjun}\ \bibnamefont {Jeon}}, \bibinfo {author}
  {\bibfnamefont {Jungpil}\ \bibnamefont {Seo}}, \bibinfo {author}
  {\bibfnamefont {Allan~H.}\ \bibnamefont {MacDonald}}, \bibinfo {author}
  {\bibfnamefont {B.~Andrei}\ \bibnamefont {Bernevig}}, \ and\ \bibinfo
  {author} {\bibfnamefont {Ali}\ \bibnamefont {Yazdani}},\ }\bibfield  {title}
  {\enquote {\bibinfo {title} {Observation of majorana fermions in
  ferromagnetic atomic chains on a superconductor},}\ }\href {\doibase
  10.1126/science.1259327} {\bibfield  {journal} {\bibinfo  {journal}
  {Science}\ }\textbf {\bibinfo {volume} {346}},\ \bibinfo {pages} {602--607}
  (\bibinfo {year} {2014})},\ \Eprint
  {http://arxiv.org/abs/https://www.science.org/doi/pdf/10.1126/science.1259327}
  {https://www.science.org/doi/pdf/10.1126/science.1259327} \BibitemShut
  {NoStop}%
\bibitem [{\citenamefont {Ruby}\ \emph {et~al.}(2015)\citenamefont {Ruby},
  \citenamefont {Pientka}, \citenamefont {Peng}, \citenamefont {von Oppen},
  \citenamefont {Heinrich},\ and\ \citenamefont {Franke}}]{STM_majroana2015}%
  \BibitemOpen
  \bibfield  {author} {\bibinfo {author} {\bibfnamefont {Michael}\ \bibnamefont
  {Ruby}}, \bibinfo {author} {\bibfnamefont {Falko}\ \bibnamefont {Pientka}},
  \bibinfo {author} {\bibfnamefont {Yang}\ \bibnamefont {Peng}}, \bibinfo
  {author} {\bibfnamefont {Felix}\ \bibnamefont {von Oppen}}, \bibinfo {author}
  {\bibfnamefont {Benjamin~W.}\ \bibnamefont {Heinrich}}, \ and\ \bibinfo
  {author} {\bibfnamefont {Katharina~J.}\ \bibnamefont {Franke}},\ }\bibfield
  {title} {\enquote {\bibinfo {title} {End states and subgap structure in
  proximity-coupled chains of magnetic adatoms},}\ }\href {\doibase
  10.1103/PhysRevLett.115.197204} {\bibfield  {journal} {\bibinfo  {journal}
  {Phys. Rev. Lett.}\ }\textbf {\bibinfo {volume} {115}},\ \bibinfo {pages}
  {197204} (\bibinfo {year} {2015})}\BibitemShut {NoStop}%
\bibitem [{\citenamefont {J{\"a}ck}\ \emph {et~al.}(2019)\citenamefont
  {J{\"a}ck}, \citenamefont {Xie}, \citenamefont {Li}, \citenamefont {Jeon},
  \citenamefont {Bernevig},\ and\ \citenamefont
  {Yazdani}}]{STM_majorana_edge2019}%
  \BibitemOpen
  \bibfield  {author} {\bibinfo {author} {\bibfnamefont {Berthold}\
  \bibnamefont {J{\"a}ck}}, \bibinfo {author} {\bibfnamefont {Yonglong}\
  \bibnamefont {Xie}}, \bibinfo {author} {\bibfnamefont {Jian}\ \bibnamefont
  {Li}}, \bibinfo {author} {\bibfnamefont {Sangjun}\ \bibnamefont {Jeon}},
  \bibinfo {author} {\bibfnamefont {B.~Andrei}\ \bibnamefont {Bernevig}}, \
  and\ \bibinfo {author} {\bibfnamefont {Ali}\ \bibnamefont {Yazdani}},\
  }\bibfield  {title} {\enquote {\bibinfo {title} {Observation of a majorana
  zero mode in a topologically protected edge channel},}\ }\href {\doibase
  10.1126/science.aax1444} {\bibfield  {journal} {\bibinfo  {journal}
  {Science}\ }\textbf {\bibinfo {volume} {364}},\ \bibinfo {pages} {1255--1259}
  (\bibinfo {year} {2019})},\ \Eprint
  {http://arxiv.org/abs/https://www.science.org/doi/pdf/10.1126/science.aax1444}
  {https://www.science.org/doi/pdf/10.1126/science.aax1444} \BibitemShut
  {NoStop}%
\bibitem [{\citenamefont {Benalcazar}\ \emph
  {et~al.}(2017{\natexlab{a}})\citenamefont {Benalcazar}, \citenamefont
  {Bernevig},\ and\ \citenamefont {Hughes}}]{Benalcazar2017}%
  \BibitemOpen
  \bibfield  {author} {\bibinfo {author} {\bibfnamefont {Wladimir~A.}\
  \bibnamefont {Benalcazar}}, \bibinfo {author} {\bibfnamefont {B.~Andrei}\
  \bibnamefont {Bernevig}}, \ and\ \bibinfo {author} {\bibfnamefont
  {Taylor~L.}\ \bibnamefont {Hughes}},\ }\bibfield  {title} {\enquote {\bibinfo
  {title} {Quantized electric multipole insulators},}\ }\href {\doibase
  10.1126/science.aah6442} {\bibfield  {journal} {\bibinfo  {journal}
  {Science}\ }\textbf {\bibinfo {volume} {357}},\ \bibinfo {pages} {61--66}
  (\bibinfo {year} {2017}{\natexlab{a}})}\BibitemShut {NoStop}%
\bibitem [{\citenamefont {Benalcazar}\ \emph
  {et~al.}(2017{\natexlab{b}})\citenamefont {Benalcazar}, \citenamefont
  {Bernevig},\ and\ \citenamefont {Hughes}}]{hughes2017}%
  \BibitemOpen
  \bibfield  {author} {\bibinfo {author} {\bibfnamefont {Wladimir~A.}\
  \bibnamefont {Benalcazar}}, \bibinfo {author} {\bibfnamefont {B.~Andrei}\
  \bibnamefont {Bernevig}}, \ and\ \bibinfo {author} {\bibfnamefont
  {Taylor~L.}\ \bibnamefont {Hughes}},\ }\bibfield  {title} {\enquote {\bibinfo
  {title} {Electric multipole moments, topological multipole moment pumping,
  and chiral hinge states in crystalline insulators},}\ }\href {\doibase
  10.1103/PhysRevB.96.245115} {\bibfield  {journal} {\bibinfo  {journal} {Phys.
  Rev. B}\ }\textbf {\bibinfo {volume} {96}},\ \bibinfo {pages} {245115}
  (\bibinfo {year} {2017}{\natexlab{b}})}\BibitemShut {NoStop}%
\bibitem [{\citenamefont {Fang}\ \emph {et~al.}(2012)\citenamefont {Fang},
  \citenamefont {Gilbert},\ and\ \citenamefont {Bernevig}}]{Fang_Chern_number}%
  \BibitemOpen
  \bibfield  {author} {\bibinfo {author} {\bibfnamefont {Chen}\ \bibnamefont
  {Fang}}, \bibinfo {author} {\bibfnamefont {Matthew~J.}\ \bibnamefont
  {Gilbert}}, \ and\ \bibinfo {author} {\bibfnamefont {B.~Andrei}\ \bibnamefont
  {Bernevig}},\ }\bibfield  {title} {\enquote {\bibinfo {title} {Bulk
  topological invariants in noninteracting point group symmetric insulators},}\
  }\href {\doibase 10.1103/PhysRevB.86.115112} {\bibfield  {journal} {\bibinfo
  {journal} {Phys. Rev. B}\ }\textbf {\bibinfo {volume} {86}},\ \bibinfo
  {pages} {115112} (\bibinfo {year} {2012})}\BibitemShut {NoStop}%
\bibitem [{\citenamefont {Fu}\ and\ \citenamefont {Kane}(2007)}]{inv_TI}%
  \BibitemOpen
  \bibfield  {author} {\bibinfo {author} {\bibfnamefont {Liang}\ \bibnamefont
  {Fu}}\ and\ \bibinfo {author} {\bibfnamefont {C.~L.}\ \bibnamefont {Kane}},\
  }\bibfield  {title} {\enquote {\bibinfo {title} {Topological insulators with
  inversion symmetry},}\ }\href {\doibase 10.1103/PhysRevB.76.045302}
  {\bibfield  {journal} {\bibinfo  {journal} {Phys. Rev. B}\ }\textbf {\bibinfo
  {volume} {76}},\ \bibinfo {pages} {045302} (\bibinfo {year}
  {2007})}\BibitemShut {NoStop}%
\bibitem [{\citenamefont {Schindler}\ \emph {et~al.}(2018)\citenamefont
  {Schindler}, \citenamefont {Cook}, \citenamefont {Vergniory}, \citenamefont
  {Wang}, \citenamefont {Parkin}, \citenamefont {Bernevig},\ and\ \citenamefont
  {Neupert}}]{hinge_HOTI}%
  \BibitemOpen
  \bibfield  {author} {\bibinfo {author} {\bibfnamefont {Frank}\ \bibnamefont
  {Schindler}}, \bibinfo {author} {\bibfnamefont {Ashley~M.}\ \bibnamefont
  {Cook}}, \bibinfo {author} {\bibfnamefont {Maia~G.}\ \bibnamefont
  {Vergniory}}, \bibinfo {author} {\bibfnamefont {Zhijun}\ \bibnamefont
  {Wang}}, \bibinfo {author} {\bibfnamefont {Stuart S.~P.}\ \bibnamefont
  {Parkin}}, \bibinfo {author} {\bibfnamefont {B.~Andrei}\ \bibnamefont
  {Bernevig}}, \ and\ \bibinfo {author} {\bibfnamefont {Titus}\ \bibnamefont
  {Neupert}},\ }\bibfield  {title} {\enquote {\bibinfo {title} {Higher-order
  topological insulators},}\ }\href {\doibase 10.1126/sciadv.aat0346}
  {\bibfield  {journal} {\bibinfo  {journal} {Science Advances}\ }\textbf
  {\bibinfo {volume} {4}},\ \bibinfo {pages} {eaat0346} (\bibinfo {year}
  {2018})},\ \Eprint
  {http://arxiv.org/abs/https://www.science.org/doi/pdf/10.1126/sciadv.aat0346}
  {https://www.science.org/doi/pdf/10.1126/sciadv.aat0346} \BibitemShut
  {NoStop}%
\bibitem [{\citenamefont {Su}\ \emph {et~al.}(1979)\citenamefont {Su},
  \citenamefont {Schrieffer},\ and\ \citenamefont
  {Heeger}}]{PhysRevLett.42.1698}%
  \BibitemOpen
  \bibfield  {author} {\bibinfo {author} {\bibfnamefont {W.~P.}\ \bibnamefont
  {Su}}, \bibinfo {author} {\bibfnamefont {J.~R.}\ \bibnamefont {Schrieffer}},
  \ and\ \bibinfo {author} {\bibfnamefont {A.~J.}\ \bibnamefont {Heeger}},\
  }\bibfield  {title} {\enquote {\bibinfo {title} {Solitons in
  polyacetylene},}\ }\href {\doibase 10.1103/PhysRevLett.42.1698} {\bibfield
  {journal} {\bibinfo  {journal} {Phys. Rev. Lett.}\ }\textbf {\bibinfo
  {volume} {42}},\ \bibinfo {pages} {1698--1701} (\bibinfo {year}
  {1979})}\BibitemShut {NoStop}%
\bibitem [{\citenamefont {Jin}\ and\ \citenamefont {Liu}(2020)}]{KHJIN2020}%
  \BibitemOpen
  \bibfield  {author} {\bibinfo {author} {\bibfnamefont {Kyung-Hwan}\
  \bibnamefont {Jin}}\ and\ \bibinfo {author} {\bibfnamefont {Feng}\
  \bibnamefont {Liu}},\ }\bibfield  {title} {\enquote {\bibinfo {title} {1d
  topological phases in transition-metal monochalcogenide nanowires},}\ }\href
  {\doibase 10.1039/D0NR03529G} {\bibfield  {journal} {\bibinfo  {journal}
  {Nanoscale}\ }\textbf {\bibinfo {volume} {12}},\ \bibinfo {pages}
  {14661--14667} (\bibinfo {year} {2020})}\BibitemShut {NoStop}%
\bibitem [{\citenamefont {Potel}\ \emph {et~al.}(1980)\citenamefont {Potel},
  \citenamefont {Chevrel}, \citenamefont {Sergent}, \citenamefont {Armici},
  \citenamefont {Decroux},\ and\ \citenamefont {Fischer}}]{SSH_mat_1}%
  \BibitemOpen
  \bibfield  {author} {\bibinfo {author} {\bibfnamefont {M.}~\bibnamefont
  {Potel}}, \bibinfo {author} {\bibfnamefont {R.}~\bibnamefont {Chevrel}},
  \bibinfo {author} {\bibfnamefont {M.}~\bibnamefont {Sergent}}, \bibinfo
  {author} {\bibfnamefont {J.C.}\ \bibnamefont {Armici}}, \bibinfo {author}
  {\bibfnamefont {M.}~\bibnamefont {Decroux}}, \ and\ \bibinfo {author}
  {\bibfnamefont {{\O}.}~\bibnamefont {Fischer}},\ }\bibfield  {title}
  {\enquote {\bibinfo {title} {New pseudo-one-dimensional metals: M2mo6se6 (m =
  na, in, k, ti), m2mo6s6 (m = k, rb, cs), m2mo6te6 (m = in, ti)},}\ }\href
  {\doibase https://doi.org/10.1016/0022-4596(80)90505-8} {\bibfield  {journal}
  {\bibinfo  {journal} {Journal of Solid State Chemistry}\ }\textbf {\bibinfo
  {volume} {35}},\ \bibinfo {pages} {286--290} (\bibinfo {year}
  {1980})}\BibitemShut {NoStop}%
\bibitem [{\citenamefont {Tarascon}\ \emph {et~al.}(1984)\citenamefont
  {Tarascon}, \citenamefont {Hull},\ and\ \citenamefont {DiSalvo}}]{SSH_mat_2}%
  \BibitemOpen
  \bibfield  {author} {\bibinfo {author} {\bibfnamefont {J.M.}\ \bibnamefont
  {Tarascon}}, \bibinfo {author} {\bibfnamefont {G.W.}\ \bibnamefont {Hull}}, \
  and\ \bibinfo {author} {\bibfnamefont {F.J.}\ \bibnamefont {DiSalvo}},\
  }\bibfield  {title} {\enquote {\bibinfo {title} {A facile synthesis of pseudo
  one-monodimensional ternary molybdenum chalcogenides m2mo6x6 (x = se,te; m =
  li,na..cs)},}\ }\href {\doibase https://doi.org/10.1016/0025-5408(84)90054-0}
  {\bibfield  {journal} {\bibinfo  {journal} {Materials Research Bulletin}\
  }\textbf {\bibinfo {volume} {19}},\ \bibinfo {pages} {915--924} (\bibinfo
  {year} {1984})}\BibitemShut {NoStop}%
\bibitem [{\citenamefont {Hor}\ \emph {et~al.}(1985)\citenamefont {Hor},
  \citenamefont {Fan}, \citenamefont {Chou}, \citenamefont {Meng},
  \citenamefont {Chu}, \citenamefont {Tarascon},\ and\ \citenamefont
  {Wu}}]{SSH_mat_3}%
  \BibitemOpen
  \bibfield  {author} {\bibinfo {author} {\bibfnamefont {P.H}\ \bibnamefont
  {Hor}}, \bibinfo {author} {\bibfnamefont {W.C}\ \bibnamefont {Fan}}, \bibinfo
  {author} {\bibfnamefont {L.S}\ \bibnamefont {Chou}}, \bibinfo {author}
  {\bibfnamefont {R.L}\ \bibnamefont {Meng}}, \bibinfo {author} {\bibfnamefont
  {C.W}\ \bibnamefont {Chu}}, \bibinfo {author} {\bibfnamefont {J.M}\
  \bibnamefont {Tarascon}}, \ and\ \bibinfo {author} {\bibfnamefont {M.K}\
  \bibnamefont {Wu}},\ }\bibfield  {title} {\enquote {\bibinfo {title} {Study
  of the metal-semiconductor transition in rb2mo6se6, rb2mo6te6 and cs2mo6te6
  under pressures},}\ }\href {\doibase
  https://doi.org/10.1016/0038-1098(85)90722-7} {\bibfield  {journal} {\bibinfo
   {journal} {Solid State Communications}\ }\textbf {\bibinfo {volume} {55}},\
  \bibinfo {pages} {231--235} (\bibinfo {year} {1985})}\BibitemShut {NoStop}%
\bibitem [{\citenamefont {Hsieh}\ \emph {et~al.}(2008)\citenamefont {Hsieh},
  \citenamefont {Qian}, \citenamefont {Wray}, \citenamefont {Xia},
  \citenamefont {Hor}, \citenamefont {Cava},\ and\ \citenamefont
  {Hasan}}]{BiSb2009}%
  \BibitemOpen
  \bibfield  {author} {\bibinfo {author} {\bibfnamefont {D.}~\bibnamefont
  {Hsieh}}, \bibinfo {author} {\bibfnamefont {D.}~\bibnamefont {Qian}},
  \bibinfo {author} {\bibfnamefont {L.}~\bibnamefont {Wray}}, \bibinfo {author}
  {\bibfnamefont {Y.}~\bibnamefont {Xia}}, \bibinfo {author} {\bibfnamefont
  {Y.~S.}\ \bibnamefont {Hor}}, \bibinfo {author} {\bibfnamefont {R.~J.}\
  \bibnamefont {Cava}}, \ and\ \bibinfo {author} {\bibfnamefont {M.~Z.}\
  \bibnamefont {Hasan}},\ }\bibfield  {title} {\enquote {\bibinfo {title} {A
  topological dirac insulator in a quantum spin hall phase},}\ }\href {\doibase
  10.1038/nature06843} {\bibfield  {journal} {\bibinfo  {journal} {Nature}\
  }\textbf {\bibinfo {volume} {452}},\ \bibinfo {pages} {970--974} (\bibinfo
  {year} {2008})}\BibitemShut {NoStop}%
\bibitem [{\citenamefont {Taskin}\ and\ \citenamefont
  {Ando}(2009)}]{BiSb2009a}%
  \BibitemOpen
  \bibfield  {author} {\bibinfo {author} {\bibfnamefont {A.~A.}\ \bibnamefont
  {Taskin}}\ and\ \bibinfo {author} {\bibfnamefont {Yoichi}\ \bibnamefont
  {Ando}},\ }\bibfield  {title} {\enquote {\bibinfo {title} {Quantum
  oscillations in a topological insulator
  ${\text{bi}}_{1\ensuremath{-}x}{\text{sb}}_{x}$},}\ }\href {\doibase
  10.1103/PhysRevB.80.085303} {\bibfield  {journal} {\bibinfo  {journal} {Phys.
  Rev. B}\ }\textbf {\bibinfo {volume} {80}},\ \bibinfo {pages} {085303}
  (\bibinfo {year} {2009})}\BibitemShut {NoStop}%
\bibitem [{\citenamefont {Wieder}\ \emph {et~al.}(2020)\citenamefont {Wieder},
  \citenamefont {Wang}, \citenamefont {Cano}, \citenamefont {Dai},
  \citenamefont {Schoop}, \citenamefont {Bradlyn},\ and\ \citenamefont
  {Bernevig}}]{C4_TI_material}%
  \BibitemOpen
  \bibfield  {author} {\bibinfo {author} {\bibfnamefont {Benjamin~J.}\
  \bibnamefont {Wieder}}, \bibinfo {author} {\bibfnamefont {Zhijun}\
  \bibnamefont {Wang}}, \bibinfo {author} {\bibfnamefont {Jennifer}\
  \bibnamefont {Cano}}, \bibinfo {author} {\bibfnamefont {Xi}~\bibnamefont
  {Dai}}, \bibinfo {author} {\bibfnamefont {Leslie~M.}\ \bibnamefont {Schoop}},
  \bibinfo {author} {\bibfnamefont {Barry}\ \bibnamefont {Bradlyn}}, \ and\
  \bibinfo {author} {\bibfnamefont {B.~Andrei}\ \bibnamefont {Bernevig}},\
  }\bibfield  {title} {\enquote {\bibinfo {title} {Strong and fragile
  topological dirac semimetals with higher-order fermi arcs},}\ }\href
  {\doibase 10.1038/s41467-020-14443-5} {\bibfield  {journal} {\bibinfo
  {journal} {Nature Communications}\ }\textbf {\bibinfo {volume} {11}},\
  \bibinfo {pages} {627} (\bibinfo {year} {2020})}\BibitemShut {NoStop}%
\bibitem [{\citenamefont {Ament}\ \emph
  {et~al.}(2011{\natexlab{b}})\citenamefont {Ament}, \citenamefont
  {Khaliullin},\ and\ \citenamefont {van~den Brink}}]{rixs_method6}%
  \BibitemOpen
  \bibfield  {author} {\bibinfo {author} {\bibfnamefont {Luuk J.~P.}\
  \bibnamefont {Ament}}, \bibinfo {author} {\bibfnamefont {Giniyat}\
  \bibnamefont {Khaliullin}}, \ and\ \bibinfo {author} {\bibfnamefont {Jeroen}\
  \bibnamefont {van~den Brink}},\ }\bibfield  {title} {\enquote {\bibinfo
  {title} {Theory of resonant inelastic x-ray scattering in iridium oxide
  compounds: Probing spin-orbit-entangled ground states and excitations},}\
  }\href {\doibase 10.1103/PhysRevB.84.020403} {\bibfield  {journal} {\bibinfo
  {journal} {Phys. Rev. B}\ }\textbf {\bibinfo {volume} {84}},\ \bibinfo
  {pages} {020403} (\bibinfo {year} {2011}{\natexlab{b}})}\BibitemShut
  {NoStop}%
\bibitem [{\citenamefont {Ament}\ \emph {et~al.}(2007)\citenamefont {Ament},
  \citenamefont {Forte},\ and\ \citenamefont {van~den Brink}}]{rixs_method4}%
  \BibitemOpen
  \bibfield  {author} {\bibinfo {author} {\bibfnamefont {Luuk J.~P.}\
  \bibnamefont {Ament}}, \bibinfo {author} {\bibfnamefont {Filomena}\
  \bibnamefont {Forte}}, \ and\ \bibinfo {author} {\bibfnamefont {Jeroen}\
  \bibnamefont {van~den Brink}},\ }\bibfield  {title} {\enquote {\bibinfo
  {title} {Ultrashort lifetime expansion for indirect resonant inelastic x-ray
  scattering},}\ }\href {\doibase 10.1103/PhysRevB.75.115118} {\bibfield
  {journal} {\bibinfo  {journal} {Phys. Rev. B}\ }\textbf {\bibinfo {volume}
  {75}},\ \bibinfo {pages} {115118} (\bibinfo {year} {2007})}\BibitemShut
  {NoStop}%
\bibitem [{\citenamefont {Haverkort}(2010)}]{haverkort_RIXS_NSF}%
  \BibitemOpen
  \bibfield  {author} {\bibinfo {author} {\bibfnamefont {M.~W.}\ \bibnamefont
  {Haverkort}},\ }\bibfield  {title} {\enquote {\bibinfo {title} {Theory of
  resonant inelastic x-ray scattering by collective magnetic excitations},}\
  }\href {\doibase 10.1103/PhysRevLett.105.167404} {\bibfield  {journal}
  {\bibinfo  {journal} {Phys. Rev. Lett.}\ }\textbf {\bibinfo {volume} {105}},\
  \bibinfo {pages} {167404} (\bibinfo {year} {2010})}\BibitemShut {NoStop}%
\bibitem [{\citenamefont {Benjamin}\ \emph {et~al.}(2014)\citenamefont
  {Benjamin}, \citenamefont {Klich},\ and\ \citenamefont
  {Demler}}]{non_SF_RIXS}%
  \BibitemOpen
  \bibfield  {author} {\bibinfo {author} {\bibfnamefont {David}\ \bibnamefont
  {Benjamin}}, \bibinfo {author} {\bibfnamefont {Israel}\ \bibnamefont
  {Klich}}, \ and\ \bibinfo {author} {\bibfnamefont {Eugene}\ \bibnamefont
  {Demler}},\ }\bibfield  {title} {\enquote {\bibinfo {title} {Single-band
  model of resonant inelastic x-ray scattering by quasiparticles in
  high-${T}_{c}$ cuprate superconductors},}\ }\href {\doibase
  10.1103/PhysRevLett.112.247002} {\bibfield  {journal} {\bibinfo  {journal}
  {Phys. Rev. Lett.}\ }\textbf {\bibinfo {volume} {112}},\ \bibinfo {pages}
  {247002} (\bibinfo {year} {2014})}\BibitemShut {NoStop}%
\bibitem [{SI()}]{SI}%
  \BibitemOpen
  \href@noop {} {}\bibinfo {note} {Supplementary material will be uploaded by
  authors.}\BibitemShut {Stop}%
\bibitem [{\citenamefont {King-Smith}\ and\ \citenamefont
  {Vanderbilt}(1993)}]{PhysRevB.47.1651}%
  \BibitemOpen
  \bibfield  {author} {\bibinfo {author} {\bibfnamefont {R.~D.}\ \bibnamefont
  {King-Smith}}\ and\ \bibinfo {author} {\bibfnamefont {David}\ \bibnamefont
  {Vanderbilt}},\ }\bibfield  {title} {\enquote {\bibinfo {title} {Theory of
  polarization of crystalline solids},}\ }\href {\doibase
  10.1103/PhysRevB.47.1651} {\bibfield  {journal} {\bibinfo  {journal} {Phys.
  Rev. B}\ }\textbf {\bibinfo {volume} {47}},\ \bibinfo {pages} {1651--1654}
  (\bibinfo {year} {1993})}\BibitemShut {NoStop}%
\bibitem [{\citenamefont {Resta}(1998)}]{PhysRevLett.80.1800}%
  \BibitemOpen
  \bibfield  {author} {\bibinfo {author} {\bibfnamefont {Raffaele}\
  \bibnamefont {Resta}},\ }\bibfield  {title} {\enquote {\bibinfo {title}
  {Quantum-mechanical position operator in extended systems},}\ }\href
  {\doibase 10.1103/PhysRevLett.80.1800} {\bibfield  {journal} {\bibinfo
  {journal} {Phys. Rev. Lett.}\ }\textbf {\bibinfo {volume} {80}},\ \bibinfo
  {pages} {1800--1803} (\bibinfo {year} {1998})}\BibitemShut {NoStop}%
\bibitem [{\citenamefont {Fu}\ \emph {et~al.}(2007)\citenamefont {Fu},
  \citenamefont {Kane},\ and\ \citenamefont {Mele}}]{Fu-Kane-Mele2007}%
  \BibitemOpen
  \bibfield  {author} {\bibinfo {author} {\bibfnamefont {Liang}\ \bibnamefont
  {Fu}}, \bibinfo {author} {\bibfnamefont {C.~L.}\ \bibnamefont {Kane}}, \ and\
  \bibinfo {author} {\bibfnamefont {E.~J.}\ \bibnamefont {Mele}},\ }\bibfield
  {title} {\enquote {\bibinfo {title} {Topological insulators in three
  dimensions},}\ }\href {\doibase 10.1103/PhysRevLett.98.106803} {\bibfield
  {journal} {\bibinfo  {journal} {Phys. Rev. Lett.}\ }\textbf {\bibinfo
  {volume} {98}},\ \bibinfo {pages} {106803} (\bibinfo {year}
  {2007})}\BibitemShut {NoStop}%
\bibitem [{\citenamefont {Benalcazar}\ \emph {et~al.}(2019)\citenamefont
  {Benalcazar}, \citenamefont {Li},\ and\ \citenamefont {Hughes}}]{hughes2019}%
  \BibitemOpen
  \bibfield  {author} {\bibinfo {author} {\bibfnamefont {Wladimir~A.}\
  \bibnamefont {Benalcazar}}, \bibinfo {author} {\bibfnamefont {Tianhe}\
  \bibnamefont {Li}}, \ and\ \bibinfo {author} {\bibfnamefont {Taylor~L.}\
  \bibnamefont {Hughes}},\ }\bibfield  {title} {\enquote {\bibinfo {title}
  {Quantization of fractional corner charge in ${C}_{n}$-symmetric higher-order
  topological crystalline insulators},}\ }\href {\doibase
  10.1103/PhysRevB.99.245151} {\bibfield  {journal} {\bibinfo  {journal} {Phys.
  Rev. B}\ }\textbf {\bibinfo {volume} {99}},\ \bibinfo {pages} {245151}
  (\bibinfo {year} {2019})}\BibitemShut {NoStop}%
\bibitem [{\citenamefont {Bradlyn}\ \emph {et~al.}(2017)\citenamefont
  {Bradlyn}, \citenamefont {Elcoro}, \citenamefont {Cano}, \citenamefont
  {Vergniory}, \citenamefont {Wang}, \citenamefont {Felser}, \citenamefont
  {Aroyo},\ and\ \citenamefont {Bernevig}}]{bradlyn2017topological}%
  \BibitemOpen
  \bibfield  {author} {\bibinfo {author} {\bibfnamefont {Barry}\ \bibnamefont
  {Bradlyn}}, \bibinfo {author} {\bibfnamefont {L}~\bibnamefont {Elcoro}},
  \bibinfo {author} {\bibfnamefont {Jennifer}\ \bibnamefont {Cano}}, \bibinfo
  {author} {\bibfnamefont {MG}~\bibnamefont {Vergniory}}, \bibinfo {author}
  {\bibfnamefont {Zhijun}\ \bibnamefont {Wang}}, \bibinfo {author}
  {\bibfnamefont {C}~\bibnamefont {Felser}}, \bibinfo {author} {\bibfnamefont
  {MI}~\bibnamefont {Aroyo}}, \ and\ \bibinfo {author} {\bibfnamefont
  {B~Andrei}\ \bibnamefont {Bernevig}},\ }\bibfield  {title} {\enquote
  {\bibinfo {title} {Topological quantum chemistry},}\ }\href@noop {}
  {\bibfield  {journal} {\bibinfo  {journal} {Nature}\ }\textbf {\bibinfo
  {volume} {547}},\ \bibinfo {pages} {298--305} (\bibinfo {year}
  {2017})}\BibitemShut {NoStop}%
\bibitem [{\citenamefont {Kitaev}(2009)}]{Kitaev_periodic}%
  \BibitemOpen
  \bibfield  {author} {\bibinfo {author} {\bibfnamefont {Alexei}\ \bibnamefont
  {Kitaev}},\ }\bibfield  {title} {\enquote {\bibinfo {title} {Periodic table
  for topological insulators and superconductors},}\ }\bibfield  {booktitle}
  {\emph {\bibinfo {booktitle} {AIP Conference Proceedings}},\ }\href {\doibase
  10.1063/1.3149495} {\bibfield  {journal} {\bibinfo  {journal} {AIP Conference
  Proceedings}\ }\textbf {\bibinfo {volume} {1134}},\ \bibinfo {pages} {22--30}
  (\bibinfo {year} {2009})}\BibitemShut {NoStop}%
\bibitem [{\citenamefont {Kruthoff}\ \emph {et~al.}(2017)\citenamefont
  {Kruthoff}, \citenamefont {de~Boer}, \citenamefont {van Wezel}, \citenamefont
  {Kane},\ and\ \citenamefont {Slager}}]{PhysRevX.7.041069}%
  \BibitemOpen
  \bibfield  {author} {\bibinfo {author} {\bibfnamefont {Jorrit}\ \bibnamefont
  {Kruthoff}}, \bibinfo {author} {\bibfnamefont {Jan}\ \bibnamefont {de~Boer}},
  \bibinfo {author} {\bibfnamefont {Jasper}\ \bibnamefont {van Wezel}},
  \bibinfo {author} {\bibfnamefont {Charles~L.}\ \bibnamefont {Kane}}, \ and\
  \bibinfo {author} {\bibfnamefont {Robert-Jan}\ \bibnamefont {Slager}},\
  }\bibfield  {title} {\enquote {\bibinfo {title} {Topological classification
  of crystalline insulators through band structure combinatorics},}\ }\href
  {\doibase 10.1103/PhysRevX.7.041069} {\bibfield  {journal} {\bibinfo
  {journal} {Phys. Rev. X}\ }\textbf {\bibinfo {volume} {7}},\ \bibinfo {pages}
  {041069} (\bibinfo {year} {2017})}\BibitemShut {NoStop}%
\bibitem [{\citenamefont {Wang}\ \emph {et~al.}(2019)\citenamefont {Wang},
  \citenamefont {Wieder}, \citenamefont {Li}, \citenamefont {Yan},\ and\
  \citenamefont {Bernevig}}]{bernevig_2019}%
  \BibitemOpen
  \bibfield  {author} {\bibinfo {author} {\bibfnamefont {Zhijun}\ \bibnamefont
  {Wang}}, \bibinfo {author} {\bibfnamefont {Benjamin~J.}\ \bibnamefont
  {Wieder}}, \bibinfo {author} {\bibfnamefont {Jian}\ \bibnamefont {Li}},
  \bibinfo {author} {\bibfnamefont {Binghai}\ \bibnamefont {Yan}}, \ and\
  \bibinfo {author} {\bibfnamefont {B.~Andrei}\ \bibnamefont {Bernevig}},\
  }\bibfield  {title} {\enquote {\bibinfo {title} {Higher-order topology,
  monopole nodal lines, and the origin of large fermi arcs in transition metal
  dichalcogenides $x{\mathrm{te}}_{2}$ ($x=\mathrm{Mo},\mathrm{W}$)},}\ }\href
  {\doibase 10.1103/PhysRevLett.123.186401} {\bibfield  {journal} {\bibinfo
  {journal} {Phys. Rev. Lett.}\ }\textbf {\bibinfo {volume} {123}},\ \bibinfo
  {pages} {186401} (\bibinfo {year} {2019})}\BibitemShut {NoStop}%
\bibitem [{\citenamefont {Choi}\ \emph {et~al.}(2020)\citenamefont {Choi},
  \citenamefont {Xie}, \citenamefont {Chen}, \citenamefont {Park},
  \citenamefont {Song}, \citenamefont {Yoon}, \citenamefont {Kim},
  \citenamefont {Taniguchi}, \citenamefont {Watanabe}, \citenamefont {Kim},
  \citenamefont {Fong}, \citenamefont {Ali}, \citenamefont {Law},\ and\
  \citenamefont {Lee}}]{Choi:2020wa}%
  \BibitemOpen
  \bibfield  {author} {\bibinfo {author} {\bibfnamefont {Yong-Bin}\
  \bibnamefont {Choi}}, \bibinfo {author} {\bibfnamefont {Yingming}\
  \bibnamefont {Xie}}, \bibinfo {author} {\bibfnamefont {Chui-Zhen}\
  \bibnamefont {Chen}}, \bibinfo {author} {\bibfnamefont {Jinho}\ \bibnamefont
  {Park}}, \bibinfo {author} {\bibfnamefont {Su-Beom}\ \bibnamefont {Song}},
  \bibinfo {author} {\bibfnamefont {Jiho}\ \bibnamefont {Yoon}}, \bibinfo
  {author} {\bibfnamefont {B.~J.}\ \bibnamefont {Kim}}, \bibinfo {author}
  {\bibfnamefont {Takashi}\ \bibnamefont {Taniguchi}}, \bibinfo {author}
  {\bibfnamefont {Kenji}\ \bibnamefont {Watanabe}}, \bibinfo {author}
  {\bibfnamefont {Jonghwan}\ \bibnamefont {Kim}}, \bibinfo {author}
  {\bibfnamefont {Kin~Chung}\ \bibnamefont {Fong}}, \bibinfo {author}
  {\bibfnamefont {Mazhar~N.}\ \bibnamefont {Ali}}, \bibinfo {author}
  {\bibfnamefont {Kam~Tuen}\ \bibnamefont {Law}}, \ and\ \bibinfo {author}
  {\bibfnamefont {Gil-Ho}\ \bibnamefont {Lee}},\ }\bibfield  {title} {\enquote
  {\bibinfo {title} {Evidence of higher-order topology in multilayer wte2 from
  josephson coupling through anisotropic hinge states},}\ }\href {\doibase
  10.1038/s41563-020-0721-9} {\bibfield  {journal} {\bibinfo  {journal} {Nature
  Materials}\ }\textbf {\bibinfo {volume} {19}},\ \bibinfo {pages} {974--979}
  (\bibinfo {year} {2020})}\BibitemShut {NoStop}%
\end{thebibliography}%


\begin{thebibliography}{0}%
\makeatletter
\providecommand \@ifxundefined [1]{%
 \@ifx{#1\undefined}
}%
\providecommand \@ifnum [1]{%
 \ifnum #1\expandafter \@firstoftwo
 \else \expandafter \@secondoftwo
 \fi
}%
\providecommand \@ifx [1]{%
 \ifx #1\expandafter \@firstoftwo
 \else \expandafter \@secondoftwo
 \fi
}%
\providecommand \natexlab [1]{#1}%
\providecommand \enquote  [1]{``#1''}%
\providecommand \bibnamefont  [1]{#1}%
\providecommand \bibfnamefont [1]{#1}%
\providecommand \citenamefont [1]{#1}%
\providecommand \href@noop [0]{\@secondoftwo}%
\providecommand \href [0]{\begingroup \@sanitize@url \@href}%
\providecommand \@href[1]{\@@startlink{#1}\@@href}%
\providecommand \@@href[1]{\endgroup#1\@@endlink}%
\providecommand \@sanitize@url [0]{\catcode `\\12\catcode `\$12\catcode
  `\&12\catcode `\#12\catcode `\^12\catcode `\_12\catcode `\%12\relax}%
\providecommand \@@startlink[1]{}%
\providecommand \@@endlink[0]{}%
\providecommand \url  [0]{\begingroup\@sanitize@url \@url }%
\providecommand \@url [1]{\endgroup\@href {#1}{\urlprefix }}%
\providecommand \urlprefix  [0]{URL }%
\providecommand \Eprint [0]{\href }%
\providecommand \doibase [0]{http://dx.doi.org/}%
\providecommand \selectlanguage [0]{\@gobble}%
\providecommand \bibinfo  [0]{\@secondoftwo}%
\providecommand \bibfield  [0]{\@secondoftwo}%
\providecommand \translation [1]{[#1]}%
\providecommand \BibitemOpen [0]{}%
\providecommand \bibitemStop [0]{}%
\providecommand \bibitemNoStop [0]{.\EOS\space}%
\providecommand \EOS [0]{\spacefactor3000\relax}%
\providecommand \BibitemShut  [1]{\csname bibitem#1\endcsname}%
\let\auto@bib@innerbib\@empty
\end{thebibliography}%

\end{document}


\title{Supplemental Information for ``Metrology of band topology via resonant inelastic x-ray scattering"}

\author{Sangjin Lee}
\thanks{Electronic Address: sangjin.lee@apctp.org}
\affiliation{Asia Pacific Center for Theoretical Physics, Pohang 37673, Republic of Korea}
\affiliation{Center for Artificial Low Dimensional Electronic Systems, Institute for Basic Science (IBS), Pohang 37673, Republic of Korea}

\author{Kyung-Hwan Jin}
\affiliation{Center for Artificial Low Dimensional Electronic Systems, Institute for Basic Science (IBS), Pohang 37673, Republic of Korea}

\author{Byungmin Kang}
\affiliation{School of Physics, Korea Institute for Advanced Study, Seoul 02455, Korea}

\author{B. J. Kim}
\thanks{Electronic Address: bjkim6@postech.ac.kr }
\affiliation{Department of Physics, Pohang University of Science and Technology (POSTECH), Pohang 37673, Republic of Korea}
\affiliation{Center for Artificial Low Dimensional Electronic Systems, Institute for Basic Science (IBS), Pohang 37673, Republic of Korea}

\author{Gil Young Cho}
\thanks{Electronic Address: gilyoungcho@postech.ac.kr}
\affiliation{Asia Pacific Center for Theoretical Physics, Pohang 37673, Republic of Korea}
\affiliation{Department of Physics, Pohang University of Science and Technology (POSTECH), Pohang 37673, Republic of Korea}
\affiliation{Center for Artificial Low Dimensional Electronic Systems, Institute for Basic Science (IBS), Pohang 37673, Republic of Korea}

\date{\today}
\maketitle

\tableofcontents

\maketitle
\setcounter{table}{0}  
  \renewcommand{\thetable}{S\arabic{table}} 
  \setcounter{figure}{0} 
  \renewcommand{\thefigure}{S\arabic{figure}}
  \setcounter{equation}{0} 
  \renewcommand{\theequation}{S\arabic{equation}}

\section{RIXS Intensity}
Here we present a brief review of the theoretical formulation of the RIXS spectral intensity\cite{rmp}. 
Our starting point is the RIXS sepctral intensity formula with momentum transfer $\bf{q}$ and energy transfer $\Delta\omega$.  
\begin{align}
\mathcal{I} (\Delta \bm{q}, \Delta \omega) = \sum_{f} \left|\mathcal{A}_{fi}\left( {\bf{k}}_f- {\bf{k}}_i= \Delta {\bf{q}}   , \omega_f- \omega_i = \Delta \omega    , {\bf{\epsilon}}_i , {\bf{\epsilon}}_f  \right)   \right|^2 \delta(E_f - E_i - \Delta \omega ) ,
\end{align} 
which sums over allowed states $f$. Here $\mathcal{A}_{fi}$ is a RIXS quantum amplitude which is a function of incident (outgoing) photon momentum ${\bf{k}}_{i(f)}$ with energy $\omega_{i(f)}$ and polarization ${\bf{\epsilon}}_{i(f)}$. Next, $E_i$ and $E_f$ are the energy of many-electron states before and after the scattering process. 

\subsection{RIXS quantum  amplitude}
\begin{figure}[h!]
	\includegraphics[width=.5\textwidth]{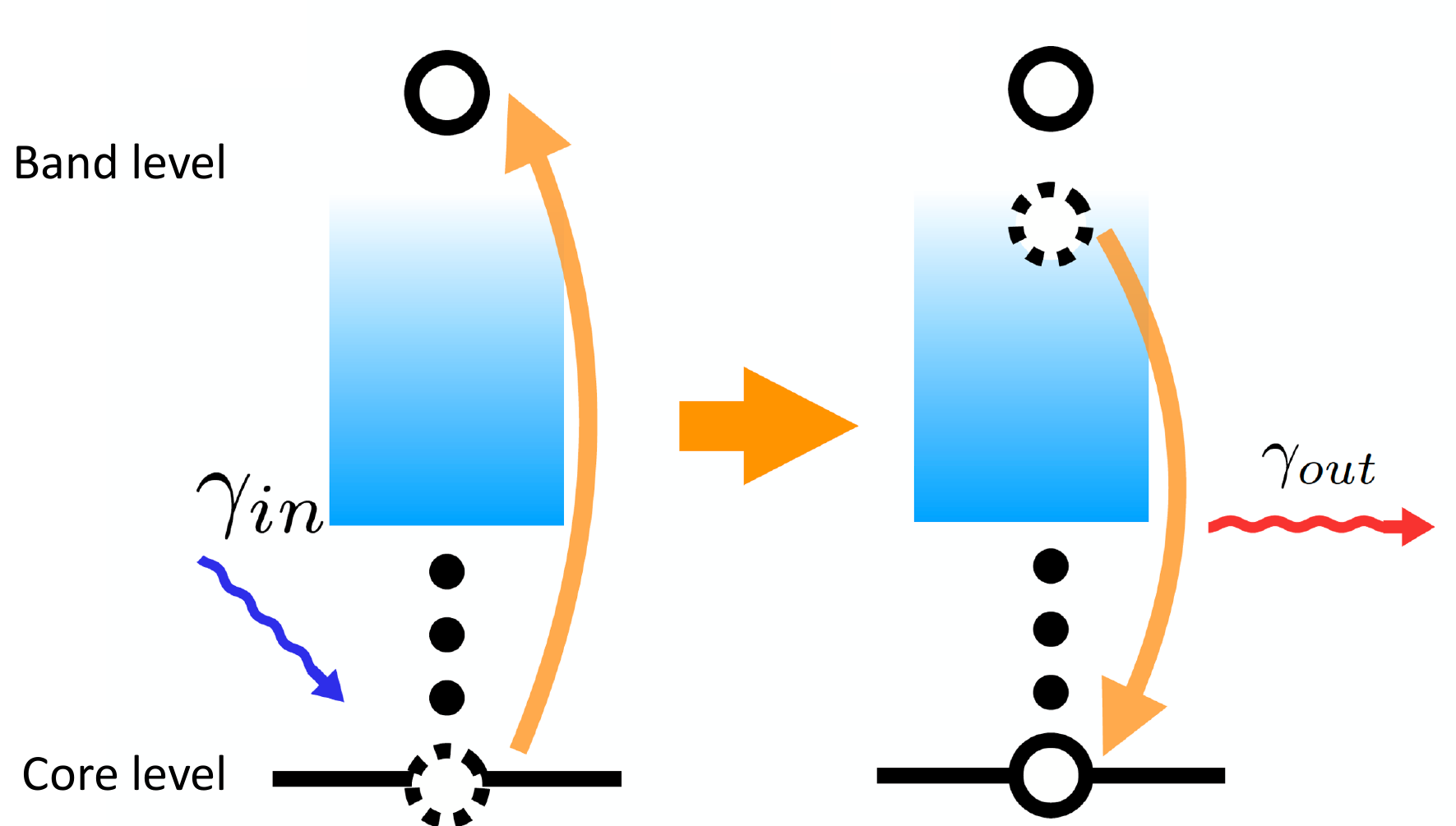}
		\caption{RIXS process. An incident photon transfers the core electron to an empty state, which leaves a core hole and an electron excitation. This configuration decays rather quickly and the core hole is filled back by another electron. This leaves an electron-hole pair excitation.}
	\label{fig:rixsprocess}
\end{figure}

Formally, the quantum amplitude $\mathcal{A}_{fi}$ is determined by internal processes (See [Fig.\ref{fig:rixsprocess}]).
\begin{align}
\mathcal{A}_{fi} = \langle	 f |\hat{D}({\bf k}_f,{\bf \epsilon}_f)\; \hat{G}_{\text{core-hole}}(\omega) \;\hat{D}^\dagger({\bf k}_i, {\bf \epsilon}_i)   |g \rangle, \label{A}
\end{align}
where $\hat{{D}}$ is a transition operator that transits a core electron to an empty state, i.e. a state in the conduction band. Here $\hat{G}_{\text{core-hole}}(\omega)$ is the green's function of the whole system with a core hole and the excited electron. It is instructive to perform the spectral decomposition of $\hat{G}_{\text{core-hole}}(\omega)$: 
\begin{align}
\hat{G}_{\text{core-hole}} (\omega)&= \sum_{n}  \frac{ |n \rangle \langle n|}{\omega- E_{n} +i \Gamma_n},
\end{align}
where $E_{n}$ is the energy of an intermediate state with a core hole and an excited electron, and $\Gamma_n$ is an inverse of the lifetime of the intermediate state. 
We assume that the core-hole lifetime is very short, e.g. it is roughly $O(1)$ fs in 4d/5d materials \cite{Mo,lifetime} which makes the excitation remain at the same site during the scattering. Near the resonant condition, we can effectively take all the $\Gamma_n$ as a single number $\Gamma$ i.e., $\Gamma_n\approx \Gamma \gg |\omega-E_n |$ which is known as the fast collision approximation \cite{rmp,FCA1,FCA2,FCA3} 
\begin{align}
\hat{G}_{\text{core-hole}} (\omega)&\simeq \sum_{n}  \frac{ |n \rangle \langle n|}{\omega- E_{n} +i \Gamma}.
\label{G}
\end{align}
Here, $\hat{G}_{\text{core-hole}}(\omega)$ as written above can incorporate non-trivial dynamics of the intermediate states, for example the excitonic effect (interaction between the core hole and excited electron). \cite{demler,FCA2,excitonic3,excitonic4}

Next we identify the dipole transition operator $\hat{{D}}$. 
In general, we have 
\begin{align}
\hat{D}({\bf{k}}, \hat\epsilon) &= \sum_{{\bf{R}},\ell,\ell',\sigma,\sigma'}   e^{i {\bf {k} \cdot \bf{R}}}  \langle c; \ell, \sigma |\hat{\epsilon}\cdot\vec{r} |d;\ell', \sigma' \rangle   c_{{\bf{R}},\ell,\sigma} d^\dagger_{{\bf{R}},\ell',\sigma'} +\text{h.c}, \label{Dipole}
\end{align}
which transits a core electron $c_{{\bf{R}}, \ell ,\sigma}$  to another valence state ($ d_{{\bf{R}},\ell',\sigma'}$), where $\ell$ represents the orbital of the relevant electronic states and $\sigma$ represents the electronic spins. Here, $d_{\bf{R},{\ell,\sigma}}$ are the fermions which forms the basis for the tight-binding models whose topology we wish to investigate in the main text. Note that the information encoded in this operator $\hat{D}({\bf{k}}, \hat\epsilon)$ is atomic, which is in principle independent to the detailed values of the hopping integrals of the low-energy tight-binding models.  

Plugging Eq.\eqref{G} and Eq.\eqref{Dipole} into Eq.\eqref{A}, we obtain
\begin{align}
\mathcal{A}_{fi} = \sum_{\bm{R} } \sum_{n}\sum_{(\ell, \sigma), (\ell', \sigma')}  g_n(\omega)  M_{\hat{\epsilon}_{i},\hat{\epsilon}_{f},n;(\ell' \sigma'), (\ell, \sigma)} \langle f | d_{\bm{R},\ell',\sigma'}d^\dagger_{\bm{R},\ell,\sigma}  e^{i\bm{q}\cdot \bm{R}} |g\rangle \label{RIXS-Supp}
\end{align}
where $g^{-1}_n(\omega) = \omega - E_n + i\Gamma$ and $\bm{q} = \bm{k}_f - \bm{k}_i$. Depending on the matrix elements of $M_{\hat{\epsilon}_{i},\hat{\epsilon}_{f},n,\sigma,\sigma'}$, one can access either the non-spin-flip RIXS (which allows only the transition between the electronic states of $\sigma = \sigma'$) or the spin-flip RIXS (which allows the transitions between the states $\sigma \neq \sigma'$). 

When the polarizations of the photons are carefully chosen, or when the RIXS transition itself does not allow the spin flip (e.g. Bi $2s \to 6p$ transition), we can access the non-spin-flip RIXS \cite{demler}, i.e. $M_{\hat{\epsilon}_{i},\hat{\epsilon}_{f},n;(\ell', \sigma'),(\ell, \sigma)} = M_{\hat{\epsilon}_{i},\hat{\epsilon}_{f},n; \ell}  \delta_{\sigma, \sigma'}$. Thus we find  
\begin{align}
\mathcal{A}_{fi} = \sum_{\bm{R} } \sum_{n}\sum_{\sigma}  g_n(\omega)  M_{\hat{\epsilon}_{i},\hat{\epsilon}_{f},n} \cdot \langle f | d_{\bm{R},\sigma}d^\dagger_{\bm{R},\sigma}  e^{i\bm{q}\cdot \bm{R}} |g\rangle = \mathcal{C}(\omega)  \sum_{\sigma} \sum_{\bm{R} }\langle f | d_{\bm{R},\sigma}d^\dagger_{\bm{R},\sigma}  e^{i\bm{q}\cdot \bm{R}} |g\rangle \label{RIXS-Supp-Final}
\end{align}
In the above, we have suppressed notating $\ell$ because we have assumed that there is one particular $\ell$, e.g. $5d_{xz}$, which is relevant both in a particular RIXS edge (that we would like to utilize) and selected polarizations of the photons. 
Here, let us emphasize that $d_{\bm{R},\ell, \sigma}$ has nothing to do with the Wannier states of certain bands. This is obvious because $d_{\bm{R},\ell, \sigma}$, which is atomic and localized in real space, is clearly not an eigenstate of generic Hamiltonian. Thus it is not the real-space representation of certain band. Nevertheless, to have a desired transition, it is certainly needed to require $d_{\bm{R}, \ell, \sigma}$ to have a finite overlap with the target states included in the transition.
Here, $\mathcal{C}(\omega) = \sum_{n} g_n(\omega)  M_{\hat{\epsilon}_{i},\hat{\epsilon}_{f},n}$ which depends on the details such as $g_{n}(\omega)$ and also the polarizations of photons $(\hat{\epsilon}_{i},\hat{\epsilon}_{f})$. The RHS of Eq.\eqref{RIXS-Supp-Final} is the equation Eq.(2) in our main text. In the following subsection \ref{Polarization-RIXS}, we will work out $M_{\hat{\epsilon}_{i},\hat{\epsilon}_{f},n;(\ell' \sigma'), (\ell, \sigma)}$ for the transition $2p \to 5d$, which are relevant for the W-based SSH materials A${}_2$W${}_6$S${}_6$ (See \ref{SSH material}). There, we will also show that in principle, the spin-flip and non-spin-flip RIXS can be selected by appropriately choosing the polarizations of incident and scattered photons.



\subsection{Example: 2p $\to$ 5$d_{xz}$ transition}\label{Polarization-RIXS}

As a concrete example, we will work out the details of the RIXS transition $2p \to 5d_{xz}$, which may be relevant for A${}_2$W${}_6$S${}_6$ \ref{SSH material} with 5$d$ element W. We will also discuss the effect of the polarizations of the photons. 

Let us start with the core-hole state. In the 5$d$ elements, the core-hole spin-orbit coupling can be strong. Because of this spin-orbit coupling, the core-hole states are splitted into the two different angular-momentum sectors $J=\frac{3}{2}, \frac{1}{2}$, which are energetically separated. We will focus on the $J=\frac{3}{2}$ states below, although the discussion can be easily generalized to the case with $J=\frac{1}{2}$.

Within the dipole approximation, we will need the following matrix element to calculate the RIXS amplitude:  
\begin{align}
\langle 5d_{xz} |\hat{\epsilon} \cdot \vec{r} | 2p_{i=x,y,z},\sigma=\uparrow,\downarrow \rangle,
\end{align}
where $\hat\epsilon$ is the polarization of the photon involved in the transition. We introduce the following polarization basis for the convenience  
\begin{align*}
&|m=0 \rangle = |2p_z\rangle , \quad |m=\pm1 \rangle  = \frac{1}{\sqrt{2}} \left(| 2p_x \rangle \pm i |2p_y\rangle \right),\\
&\hat{r}_{\pm1} =  \frac{1}{\sqrt{2}} \left(  \hat{x} \pm i \hat{y} \right), \quad \hat r_0  = \hat{z}.
\end{align*}
During the RIXS process, the total angular momentum is conserved even in the strong spin-orbit coupled systems. Hence, we use the total angular momentum basis  
\begin{align*}
&\Big|J=\frac{3}{2},m_J=\frac{3}{2}\Big\rangle = |m=1,\uparrow\rangle ,\\
&\Big|J=\frac{3}{2},m_J=\frac{1}{2}\Big\rangle = \sqrt{\frac{1}{3}} |m=1,\downarrow\rangle  +\sqrt{\frac{2}{3}} |m=0,\uparrow\rangle,\\
&\Big|J=\frac{3}{2},m_J=-\frac{1}{2}\Big\rangle = \sqrt{\frac{2}{3}} |m=0,\downarrow\rangle  +\sqrt{\frac{1}{3}} |m=-1,\uparrow\rangle,\\
&\Big|J=\frac{3}{2},m_J=-\frac{3}{2}\Big\rangle = |m=-1,\downarrow\rangle ,\\
&\Big|J=\frac{1}{2},m_J=\frac{1}{2}\Big\rangle = - \sqrt{\frac{1}{3}} |m=0,\uparrow\rangle  +\sqrt{\frac{2}{3}} |m=1,\downarrow\rangle,\\
&\Big|J=\frac{1}{2},m_J=-\frac{1}{2}\Big\rangle =  \sqrt{\frac{1}{3}} |m=0,\downarrow\rangle  -\sqrt{\frac{2}{3}} |m=-1,\uparrow\rangle. 
\end{align*}
Using this basis, we will compute relevant matrix elements associated with $2p\rightarrow 5d_{xz}$, which are 
\begin{align}
\langle 5d_{xz} |\hat{\epsilon} \cdot \vec{r} | m=\ell\rangle = M_\ell \hat{\epsilon}\cdot\hat{r}_\ell
\end{align}
where
\begin{align} 
M_{\pm1} =M_{xz} = \frac{1}{\sqrt{2}}  \langle5d_{xz}| r_x | 2p_z \rangle , \quad M_{0}=M_{zx} = \langle5d_{xz}| r_z | 2p_x \rangle.
\end{align}
Plugging these, we find the dipole transition operator 
\begin{align}
\hat{D}(\bm{q}) &= \sum_{\bm{R},\ell ,\sigma} e^{-i \bm{q}\cdot \bm{R}} M_\ell \hat{\epsilon}\cdot \hat{r}_\ell d^\dagger_{\bm{R},\sigma}  p_{\bm{R},\ell,\sigma},\nonumber\\
&= M_{xz}\sum_{\bm{R}} e^{-i \bm{q}\cdot \bm{R}} \hat{\epsilon}\cdot \left(  \hat{r}_1 \left( d^\dagger_{\bm{R},\uparrow} p_{\bm{R},m_J = \frac{3}{2}}  + \sqrt{\frac{1}{3}}     d^\dagger_{\bm{R},\downarrow} p_{i,m_J =\frac{1}{2}}  \right)      + \hat{r}_{-1}  \left(  d^\dagger_{\bm{R},\downarrow} p_{\bm{R},m_J = -\frac{3}{2}} +\sqrt{\frac{1}{3}} d^\dagger_{\bm{R},\uparrow} p_{\bm{R},m_J = -\frac{1}{2}}      \right)   \right)\nonumber\\
&\quad+\sqrt{\frac{2}{3}} M_{zx}\sum_{i} e^{-i \bm{q}\cdot \bm{R}} \hat{\epsilon}\cdot  \hat{r}_0  \left( d^\dagger_{\bm{R},\uparrow} p_{\bm{R},m_J = \frac{1}{2}}  +    d^\dagger_{\bm{R},\downarrow} p_{\bm{R},m_J =-\frac{1}{2}}  \right).
\end{align}

Inserting this to the RIXS amplitude (whose energy transfer $\omega$ is tuned to induce the transition between $2p$ and $5d_{xy}$), we find 
\begin{align}
{\mathcal{A}}_{fi}(\bm{q},\omega) 
&=\sum_{\bm{R}} \sum_{\alpha,\beta=\{ \uparrow,\downarrow\} }   g(\omega) \chi_{\alpha\beta}\langle f   |  d_{\bm{R},\alpha}  d^\dagger_{\bm{R},\beta} e^{- i  \bm{q}\cdot \bm{R} } | g \rangle 
\end{align}
where $\chi_{\alpha \beta}$ for $2p \rightarrow 5d_{xz}$ is given by  
\begin{align}
&\chi_{\uparrow\uparrow}= \chi^*_{\downarrow\downarrow} = |M_{xz}|^2\left( (\hat{\epsilon}_f \cdot \hat{r}_1 )^* (\hat{\epsilon}_i \cdot \hat{r}_1 )  + \frac{1}{3} (\hat{\epsilon}_f \cdot \hat{r}_{-1} )^*(\hat{\epsilon}_i \cdot \hat{r}_{-1} )  \right) + \frac{2}{3} |M_{zx}|^2  (\hat{\epsilon}_f \cdot \hat{r}_0 )^*(\hat{\epsilon}_i \cdot \hat{r}_0 ) ,\nonumber\\
&\chi_{\uparrow\downarrow}= \chi^*_{\downarrow\uparrow}= \frac{\sqrt{2}}{3} \left( M_{xz}M_{zx}^* (\hat{\epsilon}_f \cdot \hat{r}_0 )^*(\hat{\epsilon}_i \cdot \hat{r}_1) +M^*_{xz}M_{zx}  (\hat{\epsilon}_f \cdot \hat{r}_1 )^*(\hat{\epsilon}_i \cdot \hat{r}_0) \right). 
\label{matrixelement2}
\end{align}
This $\chi_{\alpha,\beta}$ can be actually further recast into 
\begin{align}
\chi_{\alpha\beta } = m_{NSF}\delta_{\alpha \beta}  \hat{\epsilon}_{f}^* \cdot \hat{\epsilon_{i}} +m_{SF}\vec{\sigma}_{\alpha \beta} \cdot\left( \hat{\epsilon}_{f}^* \times \hat{\epsilon_{i}}\right).
\label{spin-RIXS}
\end{align}
where $m_{NSF}$ and $m_{SF}$ are the corresponding coefficients (which can be written out explicitly in terms of $M_{xz}$ and $M_{zx}$).\cite{lucile} Hence, when $\hat{\epsilon}_{f}^* \times \hat{\epsilon}_{i} = 0$, only the non-spin-flip process will be allowed. With this, we finally find 
\begin{align}
{\mathcal{A}}_{fi}(\bm{q},\omega) 
&=\sum_{\bm{R}} \sum_{\alpha,\beta=\{ \uparrow,\downarrow\} }  g(\omega)\chi_{\alpha\beta} \langle f   |  d_{\bm{R},\alpha}  d^\dagger_{\bm{R},\beta} e^{- i  \bm{q}\cdot \bm{R} } | g \rangle = \mathcal{C}(\omega) \sum_{\bm{R}} \sum_{\sigma }   \langle f   |  d_{\bm{R},\sigma}  d^\dagger_{\bm{R},\sigma} e^{- i  \bm{q}\cdot \bm{R} } | g \rangle,
\end{align}
where we have identified $g(\omega) \chi_{\alpha\beta } = g(\omega)m_{NSF}\delta_{\alpha \beta} $ as $\mathcal{C}(\omega)$.

%

\section{1D Su-Schrieffer-Heeger(SSH) model}
\subsection{Review of SSH model}
\label{section:SSH}
Here, we present a brief review of the SSH model \cite{ssh}.  
We will temporarily suppress the spin index because the model has the spin-rotational symmetry. The SSH Hamiltonian is given as  
\begin{align}
 H_{{SSH}}=  -\sum_{x \in \mathbb{Z}} \Big( t_1 c_{x,1}^\dagger c_{x,2}  + t_2 c_{x,2}^\dagger c_{x+1,1}  + \text{h.c} \Big), 
\end{align}
where $t_1(t_2)$ is an intra(inter) site hopping parameter. $c^\dagger_{x,\alpha}$ creates electrons at $x_\alpha=x+d_{\alpha}$ with $d_{\alpha=1,2}=(-1)^\alpha  d$. Hence, the unitcell size is normalized to be ``$1$". Within a unitcell, there are two sublattices labeled by $\alpha$, whose positions are $\mathbb{Z} \pm d$. The real-space position of the sublattices is not often discussed in detail because it does not affect the symmetry and topology of the model. However, the RIXS intensity depends on the distance between the sublattices and so we will need to keep track of this below. 

Much of the important physics of the model can be well illustrated in momentum space. Hence, we proceed to the momentum space by performing the Fourier transformation 
\begin{align}
c_{x,\alpha} = \sum_k c_{k,\alpha}  e^{i k x}. 
\end{align} 
With this in hand, the Hamiltonian can be written as   
\begin{align}
H_{SSH} = \sum_{k} \bm{c}^{\dagger}_k h_{SSH}(k) \bm{c}_k,
\end{align} 
where 
\begin{align}
h_{SSH}(k) = -
\begin{pmatrix}
0&   t_1+ t_2 e^{-i k }\\
 t_1+ t_2 e^{i k}&0\\
\end{pmatrix}, \label{ssh}
\end{align}
and $\bm{c}^{T}_k = (c_{k,1}, c_{k,2})$. 
Next, we diagoanalize the Hamiltonian, $\gamma_{\eta,s}(k) = U_{\eta \alpha} c_{k,\alpha,s}$ such that 
\begin{align}
{U}^\dagger = 
\begin{bmatrix}
\psi_{c}(k)| \psi_{v}(k)
\end{bmatrix}, 
\label{U}
\end{align}
where $\psi_{c,v}(k)$ are the Bloch functions. With this, we obtain the standard results 
\begin{align}
H_{SSH}&= \sum_{k ,\eta =c,v } \varepsilon_\eta (k) \gamma^\dagger_{\eta }(k) \gamma_{\eta }(k), \quad\quad 
\varepsilon_{\eta =c,v} (k)= \pm \sqrt{ t_1^2+t_2^2 +2 t_1 t_2 \cos k}.
\end{align}

\subsection{Symmetry}
There can be two symmetries in the model: the reflection symmetry and the chiral symmety. Among the two, we will use only the reflection symmetry to derive the RIXS intensity later. That is, although we discuss the chiral symmetry for completeness, our results apply to the systems without the chiral symmetry.  

Let us start with the reflection symmetry, which is represented as $\hat{\mathcal{R}} \doteq \sigma^1$.  
\begin{align}
\mathcal{R}: \bm{c}_k \to \hat{\mathcal{R}}\cdot \bm{c}_{-k} = \sigma^1 \cdot \bm{c}_{-k}.
\end{align}
We will enforce the reflection symmetry, e.g. $\hat{\mathcal{R}} \cdot H_{SSH}\cdot \hat{ \mathcal{R}}^{-1} = H_{SSH}$.


Secondly, the chiral symmetry is given by $\hat{\Pi} \doteq \sigma^3$, which is the unitary matrix part of the symmetry operation on the fermion fields 
\begin{align}
\Pi: \bm{c}_k \to \hat{\Pi}\cdot \bm{c}_{k} = \sigma^3 \cdot \bm{c}_{k}.
\end{align}
This is in fact one of the symmetries of the simplest SSH chain model as written above, i.e., $\hat{\Pi} \cdot H_{SSH} \cdot \hat{\Pi}^{-1} = -H_{SSH}$. However, we will always assume that \textit{the chiral symmetry is explicitly broken by some perturbations}. 


\subsection{Details of the Proof for Eq.(6) of the main text}\label{Proof-SSH}
Here we will present some details for our proof of Eq.(6) in our main text. 
Our starting point is the non-spin-flip RIXS quantum amplitude,
\begin{align}
\mathcal{A}_{f}(q)  =& \mathcal{C}(\omega) \sum_{x\in \mathbb{Z},\alpha, s=\{ \uparrow,\downarrow\} } \langle f |c_{x,\alpha, s} c^\dagger_{x,\alpha ,s} e^{i q x_\alpha} | GS\rangle,
\label{amplitude}
\end{align}
where $\alpha$ is the sublattice index, and $s$ is the spin. On this, we plug $c_{x,\alpha,s=\uparrow,\downarrow} = \sum_{k,\eta} U_{\alpha \eta}^\dagger \gamma_{\eta,s}(k) e^{i k x}$ of Eq.\eqref{U} to find 
\begin{align}
\mathcal{A}_{f}(q)=& \mathcal{C}(\omega) \sum_{x \in \mathbb{Z},\alpha} \sum_{k,k', \mu,\nu,s} \langle f|  U_{\alpha \mu}^{\dagger}(k) U_{\nu \alpha}(k') \gamma_{\mu,s}(k)  \gamma_{\nu,s}^\dagger(k') e^{-i (k'-k+q) x} e^{-i q d_\alpha}  |g\rangle , \\
=& \mathcal{C}(\omega) \sum_{k,s} \langle f|  (U(k)  \hat{ \mathcal{M}}_ {k+q,k}  U^\dagger(k+q))_{\nu\mu}\gamma_{\mu,s}(k+q)  \gamma^\dagger_{\nu,s}(k) |g \rangle ,
\label{a}
\end{align}
where we have introduced a diagonal matrix $\hat{ \mathcal{M}}_ {k+q,k}$ whose $\alpha$-th component is $e^{-i q d_{\alpha=1,2}}$, i.e.,
\begin{align}
\hat{ \mathcal{M}}_ {k+q,k}=
\begin{pmatrix}
e^{i q d}&0\\
0& e^{-i q d}
\end{pmatrix} \label{M}
\end{align}
This matrix depends solely on the real space distance between the two sublattices, but not on the Hamiltonians. Before proceeding further, we briefly comment on the gauge dependence of $\hat{\mathcal{M}}_{k+q, k}$. $\hat{\mathcal{M}}_{k+q, k}$ depends on our gauge choice of the Bloch functions. For example, it is not invariant under the U(1) phase rotation of the Bloch functions. This is because it is not the quantum amplitude $\mathcal{A}_{f}$ which is measure in experiment. It is the spectral intensity $\mathcal{I} \sim |\mathcal{A}_f|^2$, which is measured in experiment. 



%
From this, the RIXS intensity is 
\begin{align}
\mathcal{I}(q,\omega) =2|\mathcal{C}(\omega)|^2 \sum'_{f} |\psi_{c}^\dagger(k+q) \hat{\mathcal{M}}_{k+q,q} \psi_{v}(k)|^2. 
\end{align}
Here ${ \displaystyle\sum'}$ represents a sum over the final states that satisfy the energy-momentum conservation and the factor of 2 comes from the spin degeneracy. 

\begin{figure}[t!]
	\includegraphics[width=.4\textwidth]{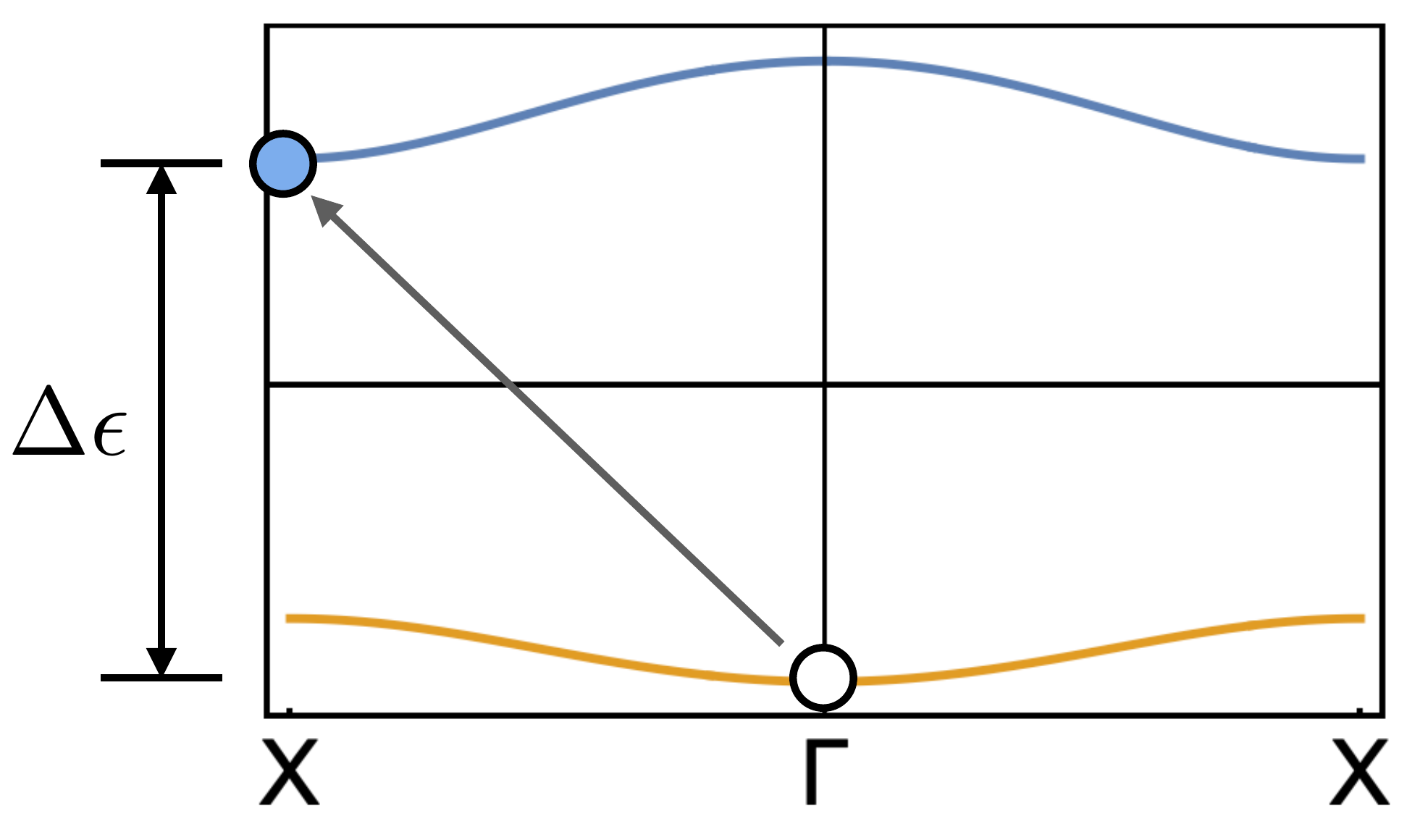}
		\caption{RIXS channel of SSH model, which we utilize to diagnose the topology.}
	\label{fig:ssh}
\end{figure}

If we further specify the energy and momentum transfer of the RIXS as $\Delta\epsilon =\epsilon_c (X) - \epsilon_v (\Gamma)$ and $q_n=(2n+1)\pi$ ($n\in\mathbb{Z}$), the allow transition is fixed to be $\Gamma \rightarrow X$, i.e. from the state at $\Gamma$ of the valence band to the state at $X$ of the conduction band. See [Fig. \ref{fig:ssh}]. Hence, the RIXS amplitude is 
\begin{align}
\mathcal{I} (q_n,\Delta \epsilon)
=2|\mathcal{C}(\Delta \epsilon)|^2 |\psi_{c}^\dagger (X)    \hat{\mathcal{M}}_{q_n,0}                 \psi_{v} (\Gamma)|^2,
\end{align}
where $\hat{\mathcal{M}}_{q_n,0} $ is obtained from Eq.\eqref{M}.
\begin{align}
 \hat{\mathcal{M}}_{k+q_n, \pi} = 
\begin{pmatrix}
e^{i q_n d}&0\\
0& e^{-i q_n d}
\end{pmatrix}= \cos q_n d + i \sin q_n d  \cdot \hat{\Pi}. 
\label{ssh-form}
\end{align}
where 
$\hat{\Pi} \cdot \hat{ \mathcal{M}}_ {q_n} \cdot \hat{\Pi} = \hat{ \mathcal{M}}_ {q_n}$ where $\hat{\Pi} = \sigma^3$ is the chiral operator. Note again that although the chiral symmetry operator appears in the discussion, we do not assume that the Hamiltonians respect the chiral symmetry.



The corresponding RIXS amplitude (with a fixed spin state) is 
\begin{align}
\mathcal{A}_f (q_n,\Delta \epsilon)
&=\mathcal{C}(\Delta \epsilon)  \psi_{c}^\dagger (X)    \hat{\mathcal{M}}_{q_n,0}                 \psi_{v} (\Gamma) \nonumber\\ 
&=\mathcal{C}(\Delta \epsilon) \Big( \psi_{c}^\dagger (X) \psi_{v} (\Gamma) \cos q_n d + i \psi_{c}^\dagger (X)  \hat{\Pi} \psi_{v} (\Gamma) \sin q_n d  \Big), \nonumber 
\end{align}
and its intensity 
\begin{align}
\mathcal{I}_f (q_n,\Delta \epsilon)
=2 |\mathcal{C}(\Delta \epsilon)|^2 \Big| \psi_{c}^\dagger (X) \psi_{v} (\Gamma) \cos q_n d + i \psi_{c}^\dagger (X)  \hat{\Pi} \psi_{v} (\Gamma) \sin q_n d  \Big|^2, \nonumber 
\end{align}
where the factor $2$ is obtained by counting the two spin states, i.e., $s=\uparrow, \downarrow$. 

Our goal is to show the following two: 
\begin{align}
|\psi_{c}^\dagger (X) \psi_{v} (\Gamma)|^2 = 1, ~ |\psi_{c}^\dagger (X)   \hat{\Pi}  \psi_{v} (\Gamma)|^2=0 \text{ for } \mathcal{P}=1/2 ~ \Rightarrow ~ \mathcal{I}_f (q_n,\Delta \epsilon) = 2|\mathcal{C}(\Delta \epsilon)|^2  \cos^2 q_n d \text{ for } \mathcal{P}=1/2, \label{P=1/2}
\end{align} 
and 
\begin{align}
|\psi_{c}^\dagger (X) \psi_{v} (\Gamma)|^2 = 0, ~ |\psi_{c}^\dagger (X)   \hat{\Pi}  \psi_{v} (\Gamma)|^2=1 \text{ for } \mathcal{P}=0 ~ \Rightarrow ~ \mathcal{I}_f (q_n,\Delta \epsilon) = 2|\mathcal{C}(\Delta \epsilon)|^2  \sin^2 q_n d \text{ for } \mathcal{P}=0. \label{P=0}
\end{align} 
If we show the LHS of Eqs.\eqref{P=1/2}, \eqref{P=0}, then the RHS trivially follows and also Eq.(6) of the main text. 

We start with proving $\psi^\dagger_{\eta = c/v} (k_{*}) \hat{\Pi} \psi_{\eta}(k_{*})=0$ for $k_{*}= X, \Gamma$. This will be useful throughout this supplemental material. This can be shown as
\begin{align*}
\psi^\dagger_{\eta }(k_{*}) \hat{\Pi}\psi_{\eta}(k_{*}) &= \psi^\dagger_{\eta }(k_{*}) \hat{\Pi} \hat{\mathcal{R} }\hat{ \mathcal{R} } \psi_{\eta}(k_{*}),\\
&= \mathcal{R}_{\eta}(k_{*}) \psi^\dagger_{\eta }k_{*}) \hat{\Pi} \hat{\mathcal{R} } \psi_{\eta}(k_{*}) \nonumber,\\
&= -\mathcal{R}_{\eta}(k_{*}) \psi^\dagger_{\eta }(k_{*})  \hat{\mathcal{R} }\hat{\Pi}  \psi_{\eta}(k_{*})\nonumber,\\
&= -\mathcal{R}^2_{\eta}(k_{*}) \psi^\dagger_{\eta }(k_{*}) \hat{\Pi}  \psi_{\eta}(k_{*})\nonumber ,\\
&=-\psi^\dagger_{\eta }(k_{*}) \hat{\Pi}  \psi_{\eta}(k_{*}).
\end{align*}
In the third line, we used $\{ \hat{\Pi},\hat{\mathcal{R}} \}=0$. This implies $\psi_{c/v}(k_{*}) \perp \hat{\Pi} \psi_{c/v}(k_{*})$. Combining this with the fact that $\{\psi_{c}(k_{*}), \psi_v(k_{*})\}$ forms an orthonormal basis for the two dimensional complex plane, we conclude 
 \begin{align}
\psi_{c/v}(k_{*}) \perp \hat{\Pi} \psi_{c/v}(k_{*}) \Rightarrow \psi_{v/c}(k_{*}) \parallel \hat{\Pi} \psi_{c/v}(k_{*}).
\label{ssh-chiral}
\end{align}
That is, $\hat{\Pi} \psi_{c/v}(k_{*})$ is identical to $\psi_{v/c}(k_{*})$ up to a phase. This immeidately implies $\mathcal{R}_{c}(k_*) = - \mathcal{R}_{v}(k_*)$, which we used in our main text. This follows from $\psi_{c}(k_{*}) =  c \hat{\Pi} \psi_{v}(k_{*})$ (with a complex phase factor $|c| =1$) and $\{ \hat{\Pi},\hat{\mathcal{R}} \}=0$.

Another useful fact is that if $|\Psi^{\dagger}\psi_v(k_*)|^2 =1$ for a normalized vector $\Psi$, then $|\Psi^{\dagger}\Pi \psi_v(k_*)|^2 =1$. This is because $\{\psi_{v}(k_{*}), \Pi \psi_v(k_{*})\}$ is an orthonormal basis for the two-dimensional complex vector space. Similary, if $|\Psi^{\dagger}\psi_v(k_*)|^2 =0$ for a normalized vector $\Psi$, then $|\Psi^{\dagger}\Pi \psi_v(k_*)|^2 =1$. 

This immediately implies that for $\mathcal{P}=0$, $|\psi_{c}^\dagger (X)   \hat{\Pi}  \psi_{v} (\Gamma)|^2 =1$ because we have proved $|\psi_{c}^\dagger (X) \psi_{v} (\Gamma)|^2 = 0$ in our main text. This completes our proof for the $\mathcal{P}=0$ case Eq.\eqref{P=0}. 

On the other hand, for $\mathcal{P}=1/2$, we need to prove the LHS of Eq.\eqref{P=1/2}. For this, we first prove $\psi_{c}^\dagger (X)   \hat{\Pi}  \psi_{v} (\Gamma) =0$  
\begin{align}
\psi_{c}^\dagger (X)   \hat{\Pi}  \psi_{v} (\Gamma)  &= \psi_{c}^\dagger (X)   \hat{\Pi}  \mathcal{R}^{-1} \mathcal{R}\psi_{v} (\Gamma)\nonumber\\ 
& = - \psi_{c}^\dagger (X)     \hat{\mathcal{R}}^{-1} \hat{\Pi} \hat{\mathcal{R}}\psi_{v} (\Gamma) \nonumber\\ 
& = - \mathcal{R}_v(\Gamma) \mathcal{R}_c(X)  \psi_{c}^\dagger (X)   \hat{\Pi}  \psi_{v} (\Gamma)  \nonumber\\ 
&= \mathcal{R}_v(\Gamma) \mathcal{R}_v (X)  \psi_{c}^\dagger (X)   \hat{\Pi}  \psi_{v} (\Gamma) \nonumber\\ 
&= e^{2\pi i \mathcal{P}}\psi_{c}^\dagger (X)   \hat{\Pi}  \psi_{v} (\Gamma), 
\end{align}
where we have used sequentially $\{ \hat{\mathcal{R}}, \hat{\Pi} \} = 0$, $\mathcal{R}_{c}(k_*) = - \mathcal{R}_{v}(k_*)$, and $ \mathcal{R}_{v}(\Gamma)  \mathcal{R}_{v}(X) = \exp(2\pi i \mathcal{P})$. Now, we see that for $\mathcal{P}=1/2$, $\psi_{c}^\dagger (X)   \hat{\Pi}  \psi_{v} (\Gamma) =0$, which immediately implies $|\psi_{c}^\dagger (X) \psi_{v} (\Gamma)|^2 = 1$. This completes our proof for the $\mathcal{P}=1/2$ case Eq.\eqref{P=1/2}.

Finally, we can summarize the RIXS intensity formula Eqs.\eqref{P=0}, \eqref{P=1/2} into 
\begin{equation}
 \mathcal{I}\Big(q_n, \Delta \epsilon \Big) = 2|\mathcal{C}(\Delta \epsilon)|^2\sin^2 \left(q_n d +\mathcal{P}\pi \right)=
    \begin{cases}
      2|\mathcal{C}(\Delta \epsilon)|^2\cos^2 \Big((2n+1)\pi d\Big) & \text{for $\mathcal{P}= 1/2$ mod 1}, \\
      2|\mathcal{C}(\Delta \epsilon)|^2\sin^2 \Big((2n+1)\pi d\Big) & \text{for $\mathcal{P}= 0$ mod 1}. 
    \end{cases}       
\end{equation}


\begin{figure}[t!]
	\includegraphics[width=.6\textwidth]{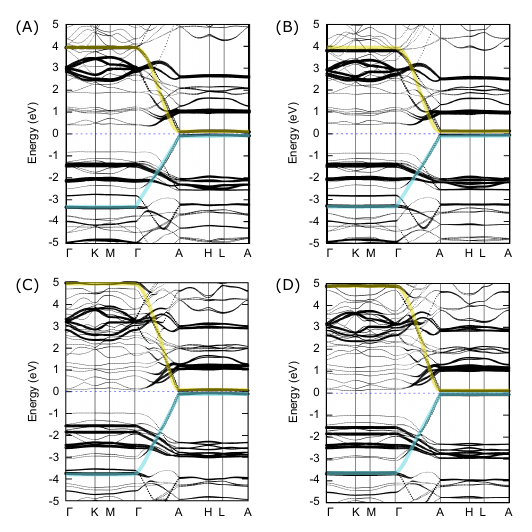}
		\caption{ DFT calculated band structure of (A) Rb$_{2}$Mo$_{6}$S$_{6}$, (B) Cs$_2$Mo$_6$S$_6$, (C) Rb$_2$W$_6$S$_6$ and (D) Cs$_2$W$_6$S$_6$. The size of the black circle represents the contribution of M $d_{xz}$, $d_{yz}$ orbitals. The valence and conduction SSH bands are highlighted by blue and yellow colors, respectively.  } 
	\label{fig:dft_bands}
\end{figure}

\subsection{Numerical Details for [Fig.1(C)] in the main text}
Let us provide some numerical details for the plot [Fig.1(C)] in the main text. 

\paragraph{Simulation of the RIXS intensity:} The RIXS intensity data (the filled circles) are generated by adding the two contributions. 
\begin{align}
\mathcal{I}(q_n, \omega) = \mathcal{I}_0 (q_n, \omega) + \delta \mathcal{I}_{\text{random}}, 
\end{align}
where $\mathcal{I}_0 (q_n, \omega)$ is the RIXS intensity obtained by applying Eq.(3) of the main text to the SSH model, and $\delta \mathcal{I}_{\text{random}}$ is the random noise. $\delta \mathcal{I}_{\text{random}}$ is drawn from the uniform distribution $[-0.2, 0.2] \times |\mathcal{C}(\Delta \epsilon)|^2$ (which is rather large). Hence, $\mathcal{I}_0 (q_n, \omega)$ is the idealistic, perfect RIXS signal on the clean SSH chain. For the topological case $\mathcal{P}=1/2$ (red circles), the parameters of the SSH Hamiltonian were $t_1 = 0.1$ and $t_2=1$. For the trivial case $\mathcal{P}=0$, the parameters were $t_1 =1$ and $t_2 = 0.1$ (so that $\Delta \epsilon$ is the same for the both cases). We set $d=0.24$ for both the cases. 

\paragraph{Fitting procedure:} We have attempted to fit the data with $\mathcal{I} (q_n)  = A \sin^2 (q_n B+P) + C$. 

For the ideal, noiseless signal of trivial case (without the noise $\delta \mathcal{I}_{\text{random}}$), we expect to find 
\begin{align}
A=2, B=0.24, C=0, P=0, 
\label{ssh-ideal-triv}
\end{align} 
which is the dotted green line in [Fig.1(C)] of the main text. As the result of the fitting (the solid green line), we found 
\begin{align}
A=1.98, B=0.24, C=0.02, P=0.00,
\end{align} 
which are pretty close to the ideal values (\ref{ssh-ideal-triv}). More importantly, we succesfully diagnose the band topology $\mathcal{P} = P/\pi =0$.

For the ideal, noiseless signal of trivial case (without the noise $\delta \mathcal{I}_{\text{random}}$), we expect to find 
\begin{align}
A=2, B=0.24, C=0, P=1.58, 
\label{ssh-ideal-top}
\end{align} 
which is the dotted red line in [Fig.1(C)] of the main text. As the result of the fitting (the red solid line), we found 
\begin{align}
A=1.99, B=0.23, C=-0.04, P=1.58, 
\end{align} 
which are pretty close to the ideal values (\ref{ssh-ideal-top}). More importantly, we succesfully diagnose the band topology $\mathcal{P} = P/\pi =1/2$.

 \subsection{Four-Band Model: Polarization per each filled band}
In this section, we consider a four-band, non-chiral symmetric model and explicitly demonstrate the applicability of our protocol, which allows us to read off the band topology per each filled band. For the clarity and brevity of our discussion, we will temporarily omit the spin indices.  

\begin{figure}[h!]
	\includegraphics[width=.8\textwidth]{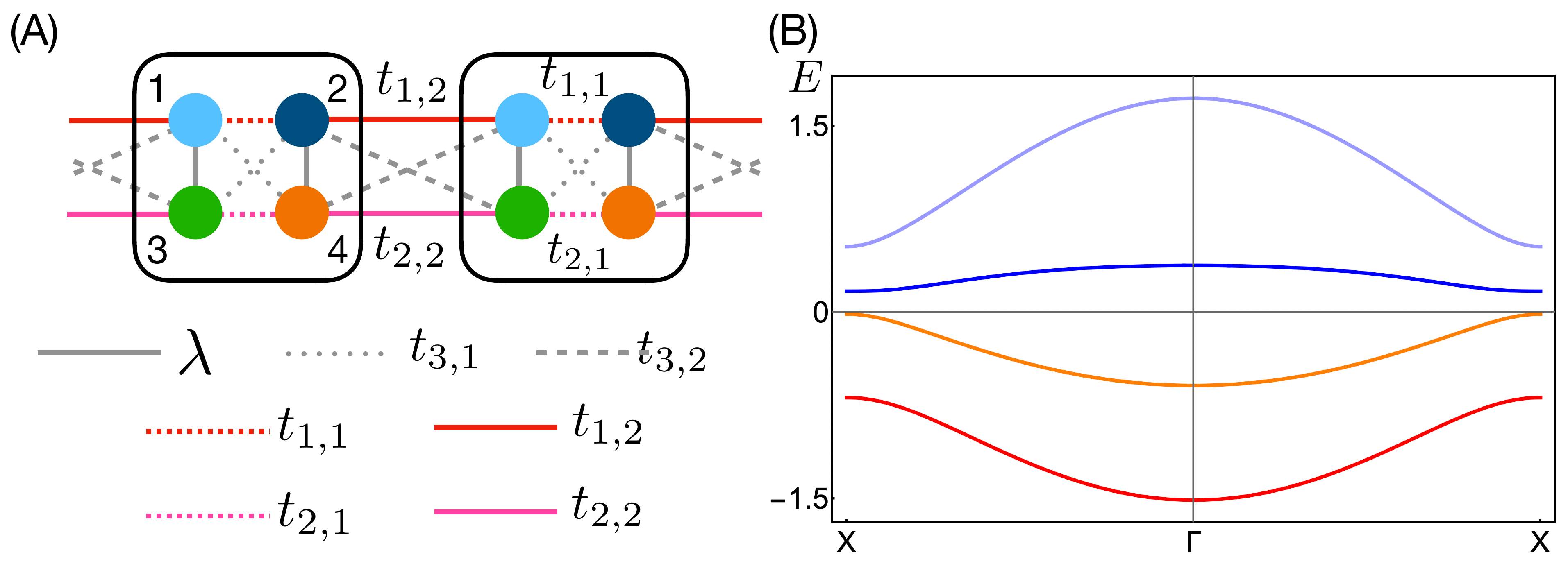}
		\caption{(A) The hopping pattern of four-band ladder model Eq.\eqref{Ladder}. Red(pink) dotted line represents the intra-unitcell hopping of the upper(lower) chain, $t_{1,1}(t_{2,1})$. Red(pink) solid line, on the other hand, represents the inter-unitcell hopping of the upper(lower) chain, $t_{1,2}(t_{2,2})$. Hopping between the two chains ($t_{3,1},t_{3,2}, \lambda)$ are represented by the gray dotted, solid and dashed lines. (B) Generic band structure of the four-band ladder model Eq.\eqref{Ladder}. In the plot, we used the arbitrarily-chosen values of the parameters, e.g. $(t_{1,1},t_{1,2},t_{2,1},t_{2,2}, t_{3,1},t_{3,2},\lambda) = (1,0.5,0.4,0.2,0.1,0.2,0.2)$. Note that the particle-hole symmetry in the spectrum is absent.}
	\label{fig:ladder_SSH}
\end{figure}

The model is basing on the two-leg ladder generalization of the SSH chain.  
\begin{align}
 H_{{ladder}}&=  H_{1,SSH} + H_{2,SSH} + H_{1,2},\label{Ladder}\\
 H_{1,SSH}&=-\sum_{x \in \mathbb{Z}} \Big( t_{1,1} c_{x,1}^\dagger c_{x,2}  + t_{1,2} c_{x,2}^\dagger c_{x+1,1}  + \text{h.c} \Big), \nonumber\\
 H_{2,SSH}&=-\sum_{x \in \mathbb{Z}} \Big( t_{2,1} c_{x,3}^\dagger c_{x,4}  + t_{2,2} c_{x,4}^\dagger c_{x+1,3}  + \text{h.c} \Big), \nonumber \\
H_{1,2}&=-\sum_{x \in \mathbb{Z},i=1,2} \Big( \lambda c_{x,i}^\dagger c_{x,i+2} + t_{3,1} c^\dagger_{x,i} c_{x,5-i} +t_{3,2} c^\dagger_{x,i} c_{x+1,5-i}  + \text{h.c} \Big), \nonumber
\end{align}
where $t_{a,1}(t_{a,2})$ is an intra(inter)-unitcell hopping parameter of the $a$-th SSH chain. On the other hand, $(t_{3,1},t_{3,2},\lambda)$ are the interchain hopping parameters, see [Fig.\ref{fig:ladder_SSH} (A)] for the pictorial representation of the model. Here, $c_{x,\alpha}^\dagger$ creates an electron, whose real-space position in $x$ is given by $x_{\alpha}  = x   + d_{\alpha}$ with $d_{\alpha = 1,2,3,4} = (-1)^\alpha d$. 
By performing Fourier transformation, 
\begin{align}
c_{x,\alpha} = \sum_{k} c_{k,\alpha} e^{ik x}, \nonumber
\end{align}
the ladder Hamiltonian is written as
\begin{align}
H_{ladder-SSH} =\sum_{k} \bm{c}^\dagger_{k} h_{ladder}(k) \bm{c}_{k}, \nonumber
\end{align}
where
\begin{align}
h_{a={1,2},SSH}(k)&=- \frac{1}{2} \left(\tau_0 -(-1)^a \tau_3\right) \big(  \left( t_{a,1} +t_{a,2} \cos k \right)\sigma_1  +t_{a,2} \sin k  \sigma_2      \big)\\
h_{1,2}(k)&=   \lambda \tau_1\sigma_0 + \left(t_{3,1} + t_{3,2} \cos k \right) \tau_1\sigma_1 +  t_{3,2} \sin k \tau_1\sigma_2\\
%
%
h_{ladder-SSH}(k)&=h_{1,SSH}(k) + h_{2,SSH}(k)+h_{1,2}(k)\\
&=-\begin{pmatrix}
0&   t_{1,1}+ t_{1,2} e^{-i k }&\lambda &t_{3,1} + t_{3,2} e^{-ik}  \\
 t_{1,1}+ t_{1,2} e^{i k}&0&t_{3,1} + t_{3,2} e^{-ik} &\lambda \\
\lambda &t_{3,1} + t_{3,2} e^{ik} & 0&   t_{2,1}+ t_{2,2} e^{-i k } \\
t_{3,1} + t_{3,2} e^{ik} &\lambda &   t_{2,1}+ t_{2,2} e^{i k}&0\\
\end{pmatrix}, \nonumber
\end{align}
with $\bm{c}^T_k = (c_{k,1},c_{k,2},c_{k,3},c_{k,4})$. Here, $\tau_{a}$ is the Pauli matrix acting on the chain index and $\sigma_{i}$ acts on the sublattice index of each SSH chains. The $h_{ladder}(k)$ can be diagonalized via $\gamma_{\eta}(k) = U_{\eta\alpha} c_{k,\alpha}$ where
\begin{align}
U^\dagger = \begin{bmatrix}
\psi_{c,1}(k)|\psi_{c,2}(k)|\psi_{v,1}(k)|\psi_{v,2}(k)
\end{bmatrix}. \nonumber
\end{align}
Then, $\psi_{\eta=c,v,\alpha=1,2}(k)$ corresponds to the Bloch state of $h_{ladder}$. With this, we find the diagnonalized Hamiltonian
\begin{align}
H_{ladder} =\sum_{k,\eta=1,2,3,4} \epsilon_{\eta}(k) \gamma^\dagger_{\eta} (k)\gamma_\eta(k).  \nonumber
\end{align}
The band structure with generic parameters can be found in [Fig.\ref{fig:ladder_SSH} (B)]. As obvious from the spectrum, the chiral symmetry is explicitly broken and there is no particle-hole symmetry. More explicitly, we can show that the chiral symmetry $\hat{\Pi} _{chiral}\doteq \tau_3 \sigma_3$ is explicitly broken, because $\{\hat{\Pi}_{chiral} , H_{ladder} \} \neq0$ when $t_{3,1}$ and $t_{3,2}$ are finite. We will assume that the two lowest bands are filled. 
 
Here, we will show how the band topology for each filled band can be read off from the momentum oscillations of the RIXS intensity. The quantum amplitude with the momentum transfer $q$ in this model can be obtained by 
\begin{align*}
\mathcal{A}_f(q, \omega ) = \mathcal{C}(\omega) \sum_{x\in \mathbb{Z},\alpha} \langle f | c_{x,\alpha}c^\dagger_{x,\alpha} e^{i q x_\alpha} |g\rangle .
\end{align*}
By substituting $c_{x,\alpha} = \sum_{k } U^\dagger_{\alpha\eta}\gamma_{\eta} (k) e^{i k x} $, we obtain the amplitude in terms of the Bloch states, 
\begin{align}
\mathcal{A}_f(q, \omega)= \mathcal{C}(\omega) \sum_{k} \langle f | \left( U(k) \hat{\mathcal{M}}_{k+q,k} U^\dagger(k+q)  \right)_{\nu\mu} \gamma_\mu (k+q) \gamma^\dagger_{\nu}(k) |g\rangle ,
\label{ladder}
\end{align}
where $\hat{\mathcal{M}}_{k+q,k}= \cos \left( q d\right)  1_{4\times4} + i \sin \left( q d\right)\hat{\Pi}_{sub} $ is generated from the spatial separation of the sublattices along $x$-direction of the model. Here, $\hat{\Pi}_{sub}\doteq \tau_0 \sigma_3$. Note that $\hat{\Pi}_{sub}$ anticommutes with the reflection symmetry operator $\hat{\mathcal{R}}\doteq\tau_0 \sigma_1$ and also $\hat{\Pi}_{sub}$ is distinct from the chiral symmetry $\hat{\Pi} _{chiral}\doteq \tau_3 \sigma_3$. 

\begin{figure}[h!]
	\includegraphics[width=.7\textwidth]{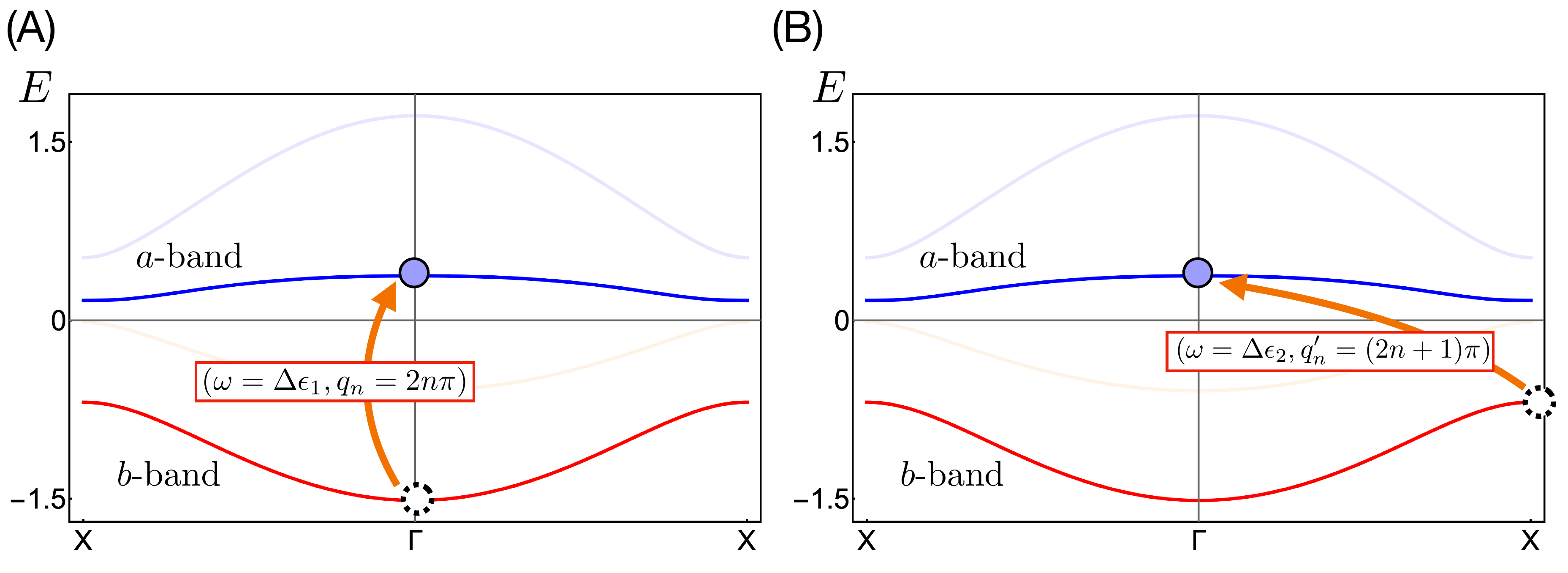}
		\caption{Our protocol to read off the topological index of $b$-band from the RIXS intensities. There are two transitions for us to determine the band topology of the b-band. (A) the RIXS transition with the energy transfer $\omega_1=\Delta\epsilon_1 = \epsilon_{a}(\Gamma) - \epsilon_{b}(\Gamma)$ and momentum transfer $q_n = 2 n \pi$ $(n\in\mathbb{Z})$ (A), which connects $\Gamma$ point of the $b$-band and $\Gamma$ point of the $a$-band. (B) Another transition with energy transfer $\omega_2=\Delta \epsilon_2 = \epsilon_{a}(\Gamma) - \epsilon_{b}(X)$, and momentum transfer $q'_n = (2 n+1) \pi$ (B), which connects $X$ point of the $b$-band and $\Gamma$ point of the $a$-band. }
	\label{fig:ladder_SSH_transition}
\end{figure}

Implementing the Bloch functions explicitly, we find 
\begin{align}
\mathcal{A}_f(q, \omega_{fi})= \mathcal{C}(\omega_{fi}) \Big(\psi^{\dagger}_{\text{final}}(k+q)\psi_{\text{initial}}(k) \cdot \cos \left( q d\right) + i \psi^{\dagger}_{\text{final}}(k+q)\hat{\Pi}_{sub} \psi_{\text{initial}}(k) \cdot \sin \left( q d\right)  \Big),
\label{Ladder-RIXS}
\end{align}
in which $\psi_{\text{final}/\text{initial}}(p)$ is the Bloch function of the final and initial bands at the momentum $p$ after/before the scattering, and $\omega_{f,i}$ is the energy transfer required to connect the two states. For example, we can consider the scattering processes like in [SFig. \ref{fig:ladder_SSH_transition}]. Below, we will be interested in the cases, where $k$ and $k+q$ are both the reflection symmetric. I.e., $k$ will be either $\Gamma$ or $X$, and $q = 2\pi n + \pi$ or $2\pi n$. These are exactly the processes in [SFig. \ref{fig:ladder_SSH_transition}].

Given this expression, it is straightforward to show that if the RIXS intensity shows $\cos^2(q d)$-like oscillation, the product of reflection eigenvalues of the initial and final states that participate in the RIXS intensity is 1. That is,   
\begin{align}
I(q) \propto \cos^2(q d) \quad \rightarrow \quad \mathcal{R}(k) \mathcal{R}(k+q) =1. 
\end{align}
Similarly, 
\begin{align}
I(q) \propto \sin^2(q d) \quad \rightarrow \quad \mathcal{R}(k) \mathcal{R}(k+q) =-1, 
\end{align}
because $\{ \hat{\mathcal{R}} ,\hat{\Pi}_{sub}  \}=0$. (Remind that $k$ and $k+q$ are both the reflection symmetric.)

The proof for these is the following. For example, for the RIXS intensity with the $\cos^2(q d)$-like oscillations, we see that from Eq.\eqref{Ladder-RIXS}, $(|\psi^\dagger_{\text{final}}(k+q)\psi_{\text{initial}}(k)|,|\psi^\dagger_{\text{final}}(k+q)\hat{\Pi}_{sub}  \psi_{\text{initial}}(k)| ) =(1,0)$. Then,
\begin{align*}
\psi^\dagger_{\text{final}}(k+q)\psi_{\text{initial}}(k) &= \psi^\dagger_{\text{final}}(k+q)\hat{\mathcal{R}}^\dagger \hat{\mathcal{R}} \psi_{\text{initial}}(k) \\
&=\mathcal{R}_{\text{final}}(k+q)\mathcal{R}_{\text{initial}}(k)\psi^\dagger_{\text{final}}(k+q)\psi_{\text{initial}}(k) \neq 0. 
\end{align*}
In the above, we have used the fact that $k$ and $k+q$ are both the reflection symmetric, i.e. $X$ or $\Gamma$. Hence, $\psi(k+q)$ and $\psi(k)$ are the eigenvectors of $\hat{\mathcal{R}}$ with the eigenvalues $\mathcal{R}_{\text{final}}(k+q)$ and $\mathcal{R}_{\text{initial}}(k)$. From the above expression, we note that $\mathcal{R}_{\text{final}}(k+q)\mathcal{R}_{\text{initial}}(k)$ is either $+1$ or $-1$. If it were $-1$, then $\psi^\dagger_{\text{final}}(k+q)\psi_{\text{initial}}(k)$ would have been zero. Hence, $\mathcal{R}_{\text{final}}(k+q)\mathcal{R}_{\text{initial}}(k)$ should be $+1$. Similarly, one can prove the case of $I(q) \propto \sin^2(q d)$.

\subsubsection*{Polarization per each filled band}
With the above, we can read off the polarization of any filled band from the RIXS intensity via the two types of the momentum transfers, $q_n = 2n \pi $ and $q'_n = (2n+1)  \pi $ $(n\in \mathbb{Z})$ [SFig. \ref{fig:ladder_SSH_transition}]. For example, we will consider the b-band in [SFig. \ref{fig:ladder_SSH_transition}], while keeping the $\Gamma$ point at the a-band [SFig. \ref{fig:ladder_SSH_transition}] as the reference. Essentially, these two transitions will allow us to determine 
\begin{align}
\mathcal{R}_a(\Gamma)\mathcal{R}_b (\Gamma)  
\end{align}
from the energy-momentum transfer in (A) of [SFig. \ref{fig:ladder_SSH_transition}], and 
\begin{align}
\mathcal{R}_a(\Gamma)\mathcal{R}_b (X)  
\end{align}
from the energy-momentum transfer in (B) of [SFig. \ref{fig:ladder_SSH_transition}], from their momentum oscillations. With these, we can determine the polarization of the b-band via 
\begin{align}
\exp(2\pi i \mathcal{P}_b ) = \mathcal{R}_b (\Gamma) \mathcal{R}_b (X)  = \mathcal{R}_a(\Gamma)\mathcal{R}_b (\Gamma)   \cdot \mathcal{R}_a(\Gamma)\mathcal{R}_b (X),   
\end{align}
because $\mathcal{R}_a(\Gamma)^2 =1$.  

\begin{figure}[h!]
	\includegraphics[width=.7\textwidth]{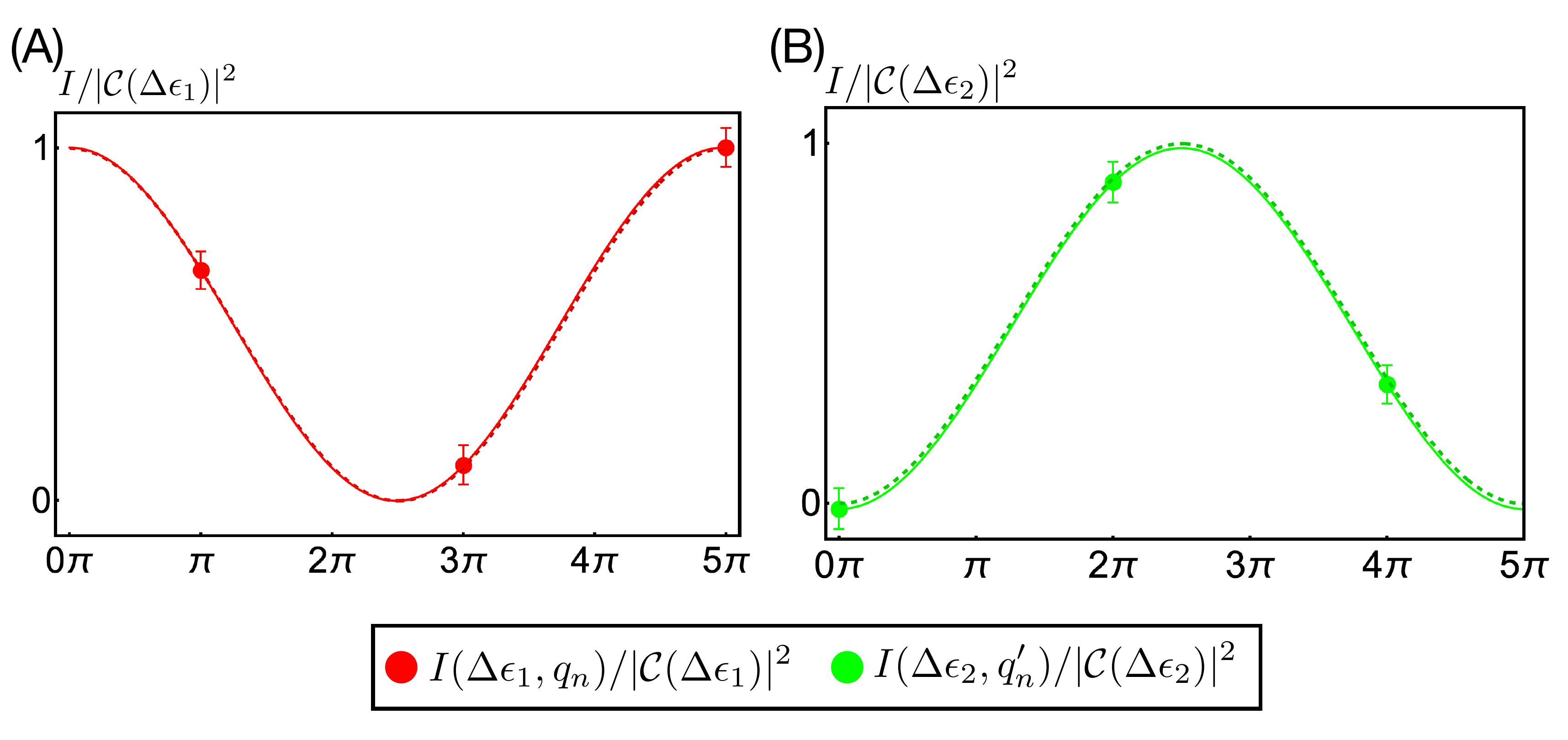}	
		\caption{RIXS intensity oscillation with respect to momentum transfer.  Simulated RIXS intensity with noises as a function of $q_n=2n\pi $ (A) and $q'_n=(2n+1)\pi$ (B) $(n\in \mathbb{Z})$. The simulated random noises taken from uniform distribution $[-0.1,0.1]\times |\mathcal{C}(\Delta \epsilon_{1,2})|^2$.  Dots with error bars show simulated RIXS data with momentum transfer and we try to fit with simulated data (solid lines) where ideal behaivors of RIXS intensity are dashed lines. Since functional forms contain information of $\mathcal{R}_{a}(\Gamma)\mathcal{R}_{b}(\Gamma)$ and $\mathcal{R}_{a}(\Gamma)\mathcal{R}_{b}(X)$, collecting a product of reflection eigenvalues determines polarization of $b$-band.  }
	\label{fig:ladder_SSH_oscillation}
\end{figure}

As a concrete illustration, we simulate a set of the RIXS intensities with the white noises taken from the uniform distribution, $[-0.1,0.1]\times |\mathcal{C}(\Delta \epsilon)|^2$ [SFig. \ref{fig:ladder_SSH_oscillation}]. As in the previous section, we can obtain RIXS intensity oscillations as a function of momentum transfer. For these, we try to fit the simulated RIXS intensity data into $A \sin^2(B q_n )+C$, which gives
\begin{align}
\mathcal{I}(q_n=2n\pi,\omega_1=\Delta \epsilon_1 )/|\mathcal{C}(\Delta \epsilon_1)|^2 = 1.00 \sin^2 (0.20 q_n ) +0.00,
\end{align}
for $q_n= 2n \pi $ case with $\Delta\epsilon_1= \epsilon_{a}(\Gamma) - \epsilon_{b}(\Gamma)$. This determines 
\begin{align}
\mathcal{R}_a(\Gamma)\mathcal{R}_b (\Gamma) = -1.  
\end{align}
On the other hand, for $q'_n = (2n+1)\pi$ case, we try to fit simulated RIXS intensity data into $A \cos^2(B q'_n )+C$, which gives
\begin{align}
\mathcal{I}(q'_n=(2n+1)\pi ,\omega_2 =\Delta \epsilon_2 ) /|\mathcal{C}(\Delta \epsilon_2)|^2  = 1.01 \cos^2 (0.20 q'_n ) +0.00
\end{align}
where $\Delta \epsilon_2 = \epsilon_{a}(\Gamma) - \epsilon_{b}(X)$[SFig. \ref{fig:ladder_SSH_oscillation} (B)]. This determines 
\begin{align}
\mathcal{R}_a(\Gamma)\mathcal{R}_b (X) = +1.  
\end{align}
Hence, we find 
\begin{align}
\exp(2\pi i \mathcal{P}_b ) = \mathcal{R}_b (\Gamma) \mathcal{R}_b (X)  = \mathcal{R}_a(\Gamma)\mathcal{R}_b (\Gamma)   \cdot \mathcal{R}_a(\Gamma)\mathcal{R}_b (X) = -1,    
\end{align}
which is consistent with the fact that the b-band carries the polarization of $1/2$ mod $1$. 

Note that the above process can be repeated for the other filled band, and hence we can determine the band topology per each filled band. Next, here we used $\Gamma$-point of the a-band as the reference, as an illustration. However, one can obviously choose other high-symmetric points in other bands as the reference. Finally, it is straighforward to generalize the above to other band topology and read off the band-wise topology, e.g. 3D topological band insulators, which we explicitly demonstrate in the section \ref{3dtbi-supp}. 


\subsection{Material candidates of SSH model and DFT calculation}\label{SSH material}
We conduct DFT calculation for searching realistic materials that host SSH model. Specifically, a family of materials with the formula A$_2$M$_6$X$_6$, where A = (Na, K, Rb, Cs, In, Tl), M = (Mo, W) and X = (S, Se, Te) can be a suitable material candidate. These compounds were experimentally synthesized via the reaction of the corresponding alkali metal tetrathiomolybdates with stoichiometric amounts of M and X. They share the same crystalline structure with the space group $P$6$_3$/$m$ (No. 176). The structure features close-packed one-dimensional M$_6$X$_6$ wires, consisting of face-sharing M$_6$ octahedra surrounded by X atoms. The M$_6$X$_6$ wires are oriented along the $z$ direction with screw rotation symmetry, arranged in a trigonal lattice in the $x$-$y$ plane. The A atoms are intercalated in the large holes between the M$_6$X$_6$ wires. The A$_2$M$_6$X$_6$ structure contains the two-fold screw rotation symmetry $S_{2z}$ and the time reversal symmetry $T$.

[Fig. \ref{fig:dft_bands}] shows the calculated band structure of selected molybdenum and tungsten monochalcogenide compounds with dimerization whose unit cell size is $4.3\si{\angstrom}\sim 4.5\si{\angstrom}$\cite{ssh_material}. Without dimerization, one observes a nodal surface located at the $k_z$=$\pi$ plane, protected by the antiunitary symmetry $TS_{2z}$. The nodal surface is very flat due to the weak inter-wire hopping, and it sits almost exactly at the Fermi level. The spin-orbit coupling (for the valence electrons) is also relatively small, which we estimate to be $\mathcal{O}($1-$10)$meV. Such a band structure would be ideal for realizing the SSH model and employing the RIXS experiment. 

All the bands near the Fermi level are contributed from the $d$ orbitals of the transition-metal (Mo or W) atoms. Especially, the flat nodal surface states mainly originate from $d_{xz}$ and $d_{yz}$ orbitals of Mo or W. When we allow the dimerization, there are two different M-M bonding distances along the chain axis. Due to the Peierls instability, the dimer formation of lower symmetry is found to be more energetically stable. This dimerization results in a structural deformation $d/a_0 \approx 0.24$. The dimerization also induces a gap at the nodal surface with the gap size of 0.16$\sim$0.2 eV. Depending on the unit cell construction, there are two different topologically distinct phases. One can consider the L-edges of transition metal Mo (W), i.e., $2p\to 4d$ $(5d)$, the transition metal monochalcogenide compounds A$_2$M$_6$X$_6$ can be used to test our theory.

\subsection{A few minor comments} 
Here we leave a few minor comments on our RIXS intensity formula, which are interesting to note.

\subsubsection{quantization of ``normalized" RIXS intensity}
Let us define the normalized RIXS intensity 
\begin{align}
\mathcal{J} (q, \omega) = \frac{\mathcal{I}(q, \omega)}{2|\mathcal{C}(\omega)|^2}, 
\end{align}
which essentially factors out the energy dependence from the usual RIXS intensity.  

Using this, we can summarize the result of our RIXS intensity formula for the transition [Fig.\ref{fig:ssh}]: 
\begin{align}
 \mathcal{J} (q_n, \Delta \epsilon) = \sin^{2}(q_n d + \mathcal{P}\pi), 
\end{align}
which is entirely fixed by the topology $\mathcal{P}$, but not by the details of the Hamiltonian. For instance, even if one changes the detailed values of the tight-binding parameters, $\mathcal{J} (q_n, \Delta \epsilon)$ remains the same and forms the plateau.

\begin{figure}[h!]
	\includegraphics[width=.6\textwidth]{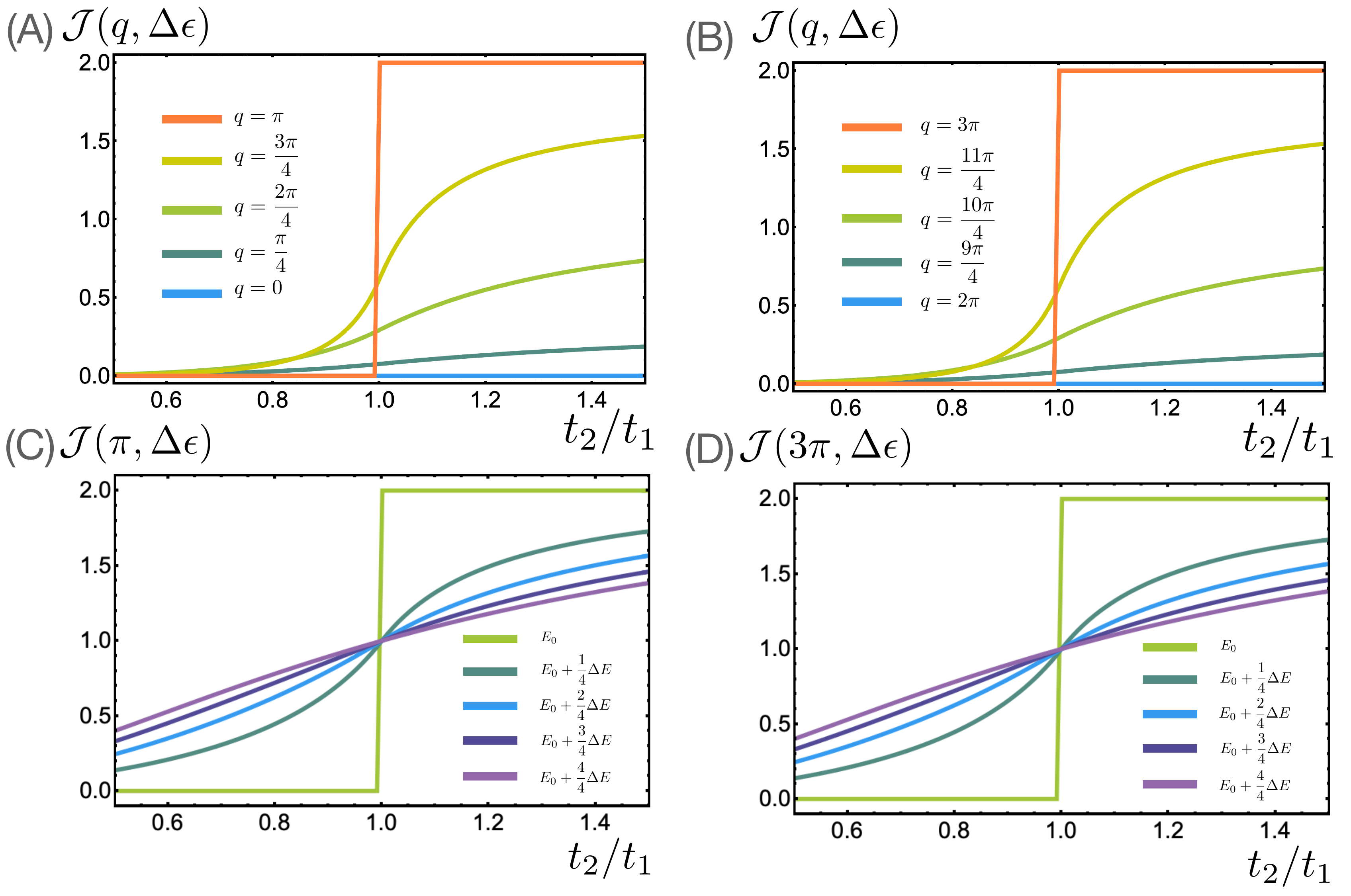}
		\caption{Non-topological RIXS intensity of the SSH model with different energy and momentum transfer. 
		(A-B) RIXS intensity with the fixed energy transfer $\Delta\epsilon = \epsilon_c(X)- \epsilon_v(\Gamma)$. Here we have changed the momentum transfers from $q=0$ to $q=\pi$ (A), from $q=2\pi$ to $q=3\pi$ (B). 
		(C-D) RIXS intensity with the fixed momentum transfer $q=\pi$ (C), $q=3\pi$ (D), while we tune the energy transfer from $\Delta \epsilon = E_0=\epsilon_c(X)-\epsilon_v(\Gamma)$ to $\Delta\epsilon= E_0 +\Delta E$. The increment in each step is $\frac{1}{4}\Delta E=\frac{1}{4} \Big( \text{Max}\big(\epsilon_c(k+\pi)-\epsilon_v(k)\big)-\text{Min}\big(\epsilon_c(k+\pi)-\epsilon_v(k)\big)\Big)$.} 
	\label{fig:SSH_cut}
\end{figure}

See the orange lines in [Fig.\ref{fig:SSH_cut} (A), (B)] for the normalized RIXS intensity at $q_n = \pi$ (A) and $q_n = 3\pi$ (B). Here we took $d=0$. We made this plot by numerically calculating the normalized RIXS intensity as we tune the ratio between the hopping parameters $t_2/t_1$. For $t_2/t_1 <1$, the ground state is trivial, i.e. $\mathcal{P}=0$. For  $t_2/t_1 >1$, the ground state is topological, i.e. $\mathcal{P}=1/2$. Hence, this explains the steps at $t_2/t_1 =1$. 

Such ``quantization" of this normalized RIXS intensity is unique to the particular energy and momentum transfers as depicted in [Fig.\ref{fig:ssh}]. One can easily guess this from the main text, where the reflection symmetry and its relation to the polarization $\mathcal{P}$ have been essential in deriving our RIXS intensity formula Eq.(6) of the main text.

To check this, we numerically calculated the normalized RIXS intensity of different energy and momentum transfers while varying the ratio $t_2/t_1$. We set $d=0$ for convenience (here the atomic unitcell size is $a_0 =1$). We have performed the two following numerical experiments. First, we fix the energy transfer of the RIXS intensity at $\Delta\epsilon=\epsilon_c(X)-\epsilon_v(\Gamma)$ and vary the momentum transfer from $q=0$ to $\pi$ [Fig. \ref{fig:SSH_cut} (A)], $q=2\pi$ to  $3\pi$ [Fig. \ref{fig:SSH_cut} (B)]. One can clearly note that except the orange lines (which are precisely the channels that we studied in our manuscript), all the others smoothly evolve and thus depend on the details of the band structures. Secondly, we change the energy transfer from $\Delta\epsilon=E_0= \epsilon_c(X)-\epsilon_v(\Gamma)$ to $\Delta\epsilon= \text{Max}\left(\epsilon_c(k+\pi)-\epsilon_v(k)\right)$ by $\frac{1}{4}\Delta E=\frac{1}{4} \Big( \text{Max}\big(\epsilon_c(k+\pi)-\epsilon_v(k)\big)-\text{Min}\big(\epsilon_c(k+\pi)-\epsilon_v(k)\big)\Big)$ while we fixed the momentum transfer $q=\pi$ [Fig. \ref{fig:SSH_cut} (C)], $q=3\pi$ [Fig. \ref{fig:SSH_cut} (D)]. Again, except the green lines (which are the channels that we studied in our manuscript), every others smoothly evolve and depend on the detailed band structures. These clearly demonstrate that the RIXS intensity at all the other energy and momentum (except the channel in [Fig.\ref{fig:ssh}]) are non-topological and depend on the details of the band structure.

\subsubsection{effect of redefining the unitcell}
\begin{figure}[h!]
	\includegraphics[width=.7\textwidth]{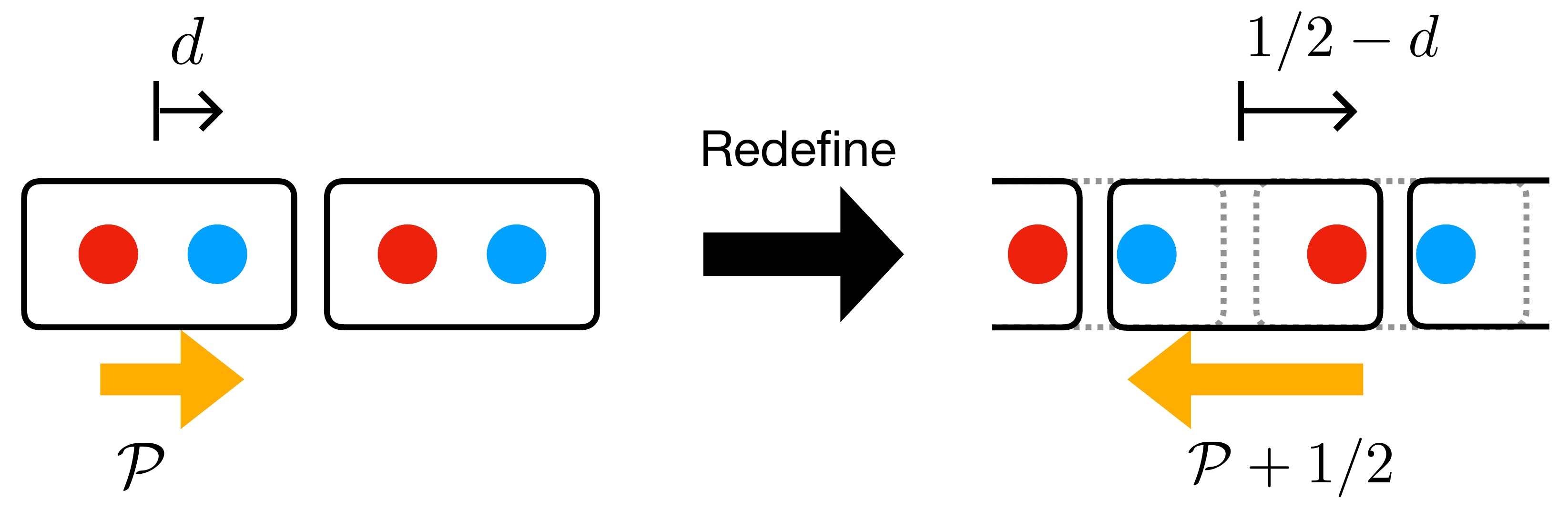}
		\caption{ Unitcell dependence of Polarization. If we redefine the unitcell from the left panel to the right one by shifting the center of the unitcell by the half lattice constant $a_0 =1$, $\mathcal{P}$ should also be shifted by the half. Such covariant transformation gives the exactly-same RIXS intensity. }
	\label{fig:unicell}
\end{figure}

It is well known that in the SSH chain, the two distinct band topology $\mathcal{P}=0$ and $\mathcal{P}=1/2$ can be exchanged by simply redefining the unitcell. That is, if we redefine unit cell by shifting $d\to d-1/2$, then polarizaiton $\mathcal{P}$ (mod 1) should also be shifted $\mathcal{P} \to \mathcal{P} -1/2 $. Motivated from this, one can ask if our RIXS intensity formula is consistent with this. In fact, it is so because we can rewrite it as 
\begin{align}
\mathcal{I} (q_n, \Delta \epsilon) = 2|\mathcal{C}(\Delta\epsilon)|^2 \sin^2 \left(q_n d +\mathcal{P}\pi \right) =   2|\mathcal{C}(\Delta\epsilon)|^2 \sin^2 \left(q_n \left(d-\frac{1}{2}\right)  +\left(\mathcal{P}-\frac{1}{2} \right)\pi \right). 
\end{align}
Hence, the intensity remains the same under the redefinition of unit cells as it should be (because the RIXS intensity is physical, gauge-invariant). See [Fig.\ref{fig:unicell}] for an illustration. 


\subsubsection{effect of spin-flip RIXS on the SSH materials}
Here we will show that our conclusion remains the same even when the spin-flip RIXS is included. For example, when one cannot control the polarization of photons, then the spin-flip RIXS may occur simultaneously with the non-spin-flip RIXS.  We will use Eq. \ref{spin-RIXS}\cite{lucile} to calculate the intensity for the spin-flip RIXS. For example, one possible form of the spin-flip RIXS is   
\begin{align}
\mathcal{I}_{\text{spin-flip}} (q,\omega) =|\mathcal{C}_{\text{spin-flip}}(\omega)|^2 \sum'_{f} |\psi_{c,s}^\dagger(k+q) \left( \hat{\mathcal{M}}_{k+q,q}\otimes \sigma_0 \right)\left(Id_{2\times 2}\otimes \sigma^\pm_{ss'}\right) \psi_{v,s'}(k)|^2. 
\end{align}
Here, we note that $ \hat{\mathcal{M}}_{k+q,q}$ is independent of the spin and that $\sigma^\pm_{s,s'}$ signifies the spin-flip process. Because of the spin-flip process, $\psi_{c,s}(k+q) $ and $\psi_{v,s'}(k)$ with $s\neq s'$ appear in the expression. However, the form of the Bloch function is independent of the spin state, this spin-flip RIXS intensity is the same as the non-spin-flip RIXS up to a multiplicative constant. That is,
\begin{align}
\mathcal{I}_{\text{spin-flip}} (q, \omega) =  \text{const.} \times \mathcal{I}_{\text{non-spin-flip}} (q, \omega). 
\end{align}
Hence, even with the spin-flip process, we can still use the same formula, i.e. Eq.(6) in our main text, to diagnose the band topology of the SSH materials.

\section{2D quadrupole insulator }
\subsection{Review of Quadrupole Insulator Model}
In this supplemental material, we will review some basics of the $C_4$-symmetric quadrupole insulator(QI). 
In this subsection, we will temporarily suppress the spin index, which we will recover later when we discuss the RIXS intensity. 

Our starting point is the QI Hamiltonian, \cite{multipole} 
\begin{align}
H_{QI} &= - \sum_{\bm{r}\in \mathbb{Z}^2} t_1 \left(c_{\bm{r}, 1}^\dagger c_{\bm{r}, 3} +c_{\bm{r},2}^\dagger c_{\bm{r}, 4} +c_{\bm{r}, 1}^\dagger c_{\bm{r}, 4} -c_{\bm{r}, 2}^\dagger c_{\bm{r}, 3} \right) + \text{h.c} \nonumber\\
&\quad - \sum_{\bm{r}\in \mathbb{Z}^2} t_2 \left(c_{\bm{r}, 1}^\dagger c_{\bm{r}+\hat{x}, 3} +c_{\bm{r}+\hat{x},2}^\dagger c_{\bm{r}, 4} +c_{\bm{r}, 1}^\dagger c_{\bm{r}+\hat{y}, 4} -c_{\bm{r}+\hat{y}, 2}^\dagger c_{\bm{r}, 3} \right) + \text{h.c},
\end{align}
where $c_{\bm{r},\alpha}^\dagger$ creates an electron at ${\bf{r_\alpha}}= \bm{r} + \bm{d}_{\alpha}$ with
\begin{align}
{\bf{r_1}} = {\bf{r}}+ (d,d),\quad {\bf{r_2}} = {\bf{r}}+ (-d,-d),\quad {\bf{r_3}} = {\bf{r}}+ (-d,d),\quad {\bf{r_4}} = {\bf{r}}+ (d,-d).
\end{align}
and $t_1(t_2)$ is intra(inter)site hopping parameter. We move to the momentum space by performing Fourier transformation, $c_{\bm{r},\alpha} = \sum_{k,\alpha} c_{\bm{k}, \alpha} e^{i \bm{k}\cdot \bm{r}}$ 
\begin{align}
H_{QI} = \sum_{{\bf k}} {\bf {c}^\dagger_{{\bf k}}} h_{QI}({\bf k}) {\bf {c}_{{\bf k}}} ,
\end{align}
with
\begin{align}
h_{QI}(\bm{k}) =
\begin{pmatrix}
0&0&  \left(t_1+t_2 e^{i k_x}\right)&  \left(t_1+t_2 e^{i k_y}\right) \\
0&0& - \left(t_1+t_2 e^{-i k_y}\right)& \left(t_1+t_2 e^{-i k_x}\right) \\
\left(t_1+t_2 e^{-i k_x}\right)&  -\left(t_1+t_2 e^{i k_y}\right)&0&0\\
\left(t_1+t_2 e^{-i k_y}\right) & \left(t_1+t_2 e^{i k_x}\right)&0&0
\end{pmatrix}.
\label{hQI}
\end{align}
and ${{\bf{c}}^T_{\bf k}} =  (c_{{\bf{k}},1},c_{{\bf{k}},2},c_{{\bf{k}},3},c_{{\bf{k}},4})$. 

By diagonalizing the Hamiltonian, we finally find 
\begin{align}
H_{QI}  = \sum_{k, \eta=c,v,\xi=1,2} \varepsilon_{\xi,\eta} (k) \gamma^\dagger_{\xi,\eta} (k) \gamma_{\xi,\eta} (k), \quad \varepsilon_{\xi=1,2,\eta=c,v} (k) = \pm \sqrt{ 2t_1^2+ 2 t_2^2 + 2t_1 t_2  ( \cos k_x + \cos k_y)}.
\end{align}

\subsection{Symmetry}
There are two symmetries of the insulator, which we will often use: $C_4$ rotation symmetry and chiral symmetry. (In the QI, we assume that there are perturbations to the Hamiltonian, which break the chiral symmetry explicitly. Hence, although we discuss the chiral symmetry operator here, we do not impose it to the Hamiltonian.) 

First, the $C_4$ rotation symmetry is given by the matrix $\hat{\mathcal{C}}_4$
\begin{align}
\hat{\mathcal{C}}_4 \doteq 
\begin{pmatrix}
0&\sigma_0 \\
-i \sigma_2&0
\end{pmatrix}.
\label{C4}
\end{align}
When the symmetry is applied to the fermion field, we find:  
\begin{align}
C_4: \bm{c}_k \to \hat{\mathcal{C}}_4 \cdot \bm{c}_{r_4 [\bm{k}]} = 
\begin{pmatrix}
0&\sigma_0 \\
-i \sigma_2&0
\end{pmatrix}
 \cdot \bm{c}_{r_4 [\bm{k}]},
\end{align}
where $r_{4} : (k_x,k_y) \rightarrow (-k_y ,k_x)$. 

Secondly, the chiral symmetry is given by the matrix 
\begin{align}
\hat{\Pi} \doteq \sigma_3 \otimes\sigma_0 =
\begin{pmatrix}
1&0&0&0\\
0&1&0&0\\
0&0&-1&0\\
0&0&0&-1\\
\end{pmatrix}.
\end{align}
When the chiral symmetry is applied to the fermion field, we find: 
\begin{align}
\Pi: \bm{c}_k \to \hat{\Pi}\cdot \bm{c}_{k} = \left( \sigma_3 \otimes\sigma_0\right) \cdot \bm{c}_{k}.
\end{align}
 
\subsection{Topology} 
We present the $C_4$ rotation eigenvalues which are useful to calculate the RIXS intensity formula. See [Table \ref{quadruphase}]. The quadrupole moment, $\mathcal{Q}_{xy}$ of the insulator, satisfies the relation Eq. \eqref{quadrupole} \cite{multipole}:
\begin{align}
\exp\left({2\pi i \mathcal{Q}_{xy}}\right)  = r_{4,\eta}^\xi(\Gamma)^{*} r_{4,\eta}^\xi(M)=r_{4,\eta}^\xi(\Gamma)^{*} r_{4,\eta}^\xi(M).
\label{quadrupole}
\end{align}
Here $(\xi = \pm)$ is the index for the degenerate states at the high-symmetry points $\Gamma$ and $M$, which can be resolved by the $C_4$ eigenvalues. 

\begin{table}[t!]
 \begin{tabular}{c|| c |c } 
Bands& $\Gamma\left({\bf{q}}=\left(0,0\right)\right)$ & $ M\left({\bf{q}}=\left(\pi,\pi\right)\right)$  \\ 
 \hline
$(v,-)$ &   $ \text{sign}(t_1+t_2)e^{-i \frac{3}{4}\pi}$ & $ \text{sign}(t_1+t_2)e^{2\pi i \mathcal{Q}_{xy}}  e^{-i \frac{3}{4}\pi}$         \\
\hline
$(v,+)$ &  $ \text{sign}(t_1+t_2)e^{+i \frac{3}{4}\pi}$ & $ \text{sign}(t_1+t_2)e^{2\pi i \mathcal{Q}_{xy}}  e^{+i \frac{3}{4}\pi}$         \\
\hline
$(c,-)$ & $- \text{sign}(t_1+t_2)e^{-i \frac{3}{4}\pi}$ &$- \text{sign}(t_1+t_2)e^{2\pi i \mathcal{Q}_{xy}}  e^{-i \frac{3}{4}\pi}$         \\
\hline
$(c,+)$ &  $- \text{sign}(t_1+t_2)e^{+i \frac{3}{4}\pi}$ & $- \text{sign}(t_1+t_2)e^{2\pi i \mathcal{Q}_{xy}}  e^{+i \frac{3}{4}\pi}$         \\
\hline
\end{tabular}
\caption{$C_4$ eigenvalues at symmetric points. We denote a $C_4$ eigenvalue of a band($\eta=c,v$). Here, $(\xi=\pm)$ represents the index for the degenerate states at $\bm{k}^*=  \Gamma, M$ as $r_{4,\eta=c,v}^{\xi=\pm} (\bm{k}^*)$. When $|t_2|>|t_1|$, $\mathcal{Q}_{xy}=1/2$ mod 1 and otherwise $0$ mod 1.}
\label{quadruphase}
\end{table}
 
\subsection{RIXS intensity of QI model}  
We present the calculation of the $\mathcal{I}(q,\omega)$ of the QI model. This can be viewed as the proof for the formula presented in the main text. Here we will explicitly keep track of the spin. Our starting point is the Bloch state $c_{{\bf r},\alpha,s=\uparrow,\downarrow} = \sum_{{\bf k},\eta} U_{\alpha \eta}^\dagger \gamma_{\eta,s}({\bf{k}}) e^{i {\bf k\cdot r}_\alpha }$, which we insert into the formula for the non-spin-flip RIXS quantum amplitude, 
\begin{align}
\mathcal{A}_{fi}(\bm{q},\omega)=& \mathcal{C}(\omega)\sum_{{\bf{r}} \in \mathbb{Z}^2,\alpha,s=\{\uparrow,\downarrow \} }  \langle f| c_{{\bf{r}},\alpha,s} c_{{\bf{r}},\alpha,s}^\dagger e^{-i {    \bf{q}\cdot {\bf{r}}}_\alpha}|g\rangle,  \nonumber\\
=& \mathcal{C}(\omega)\sum_{{\bf{r}} \in \mathbb{Z}^2 ,\alpha} \sum_{ {\bf k,k'}, \mu,\nu,s}  \langle f| U_{\alpha \mu}^{\dagger}({\bf k}) U_{\nu \alpha}({\bf k'} ) \gamma_{\mu,s}({\bf k})  \gamma_{\nu,s}^\dagger({\bf k}') e^{-i \bf{(k'-k+q)}\cdot {\bf{r}}} e^{-i \bf{q \cdot d_\alpha}} |g\rangle,  \nonumber\\
=& \mathcal{C}(\omega)\sum_{{\bf k},s} \langle f|(U({\bf k})  \hat{ \mathcal{M}}_ {{\bf k+q,k}}    U^\dagger({\bf k+q} ))_{\nu\mu}\gamma_{\mu,s}({\bf k+q})  \gamma^\dagger_{\nu,s}({\bf k})  |g\rangle , \label{a}
\end{align}
where 
\begin{align}
\hat{\mathcal{M}}_{{\bf{k+q},\bf{k}}}=
\begin{pmatrix}
e^{i (q_x+q_y) d}&0&0&0\\
0&e^{-i (q_x+q_y) d}&0&0\\
0&0&e^{i (-q_x+q_y) d}&0\\
0&0&0&e^{i (q_x-q_y) d}
\end{pmatrix}.
\end{align}
The corresponding RIXS intensity of the QI model is 
\begin{align}
\mathcal{I} ({\bf{q}},\omega) =  2 |\mathcal{C}(\omega)|^2 \sum'_{f} |\psi_{c}^\dagger ({\bf{k+q}}) \hat{\mathcal{M}}_{{\bf k+q,q}}  \psi_{v}({\bf{k}})|^2 ,
\end{align}
where factor 2 came from the double degeneracy of the spin.
Here, we consider ${\bf{q}}_n =  (2n+1)(\pi,\pi)$, $n\in\mathbb{Z}$, and $\omega = \Delta \epsilon = \epsilon_{c}(M) - \epsilon_v (\Gamma)$. 
This fixes the RIXS channel as $\Gamma \rightarrow M$ 
\begin{align}
\mathcal{I} ({\bf{q}}_n,\Delta \epsilon)&= 2|\mathcal{C}(\Delta \epsilon)|^2  \sum_{ \eta,\eta'= \pm} | \psi^\dagger_{c,\eta}(M) \hat{\mathcal{M}} _{{\bf 0} + {\bf{q}_n} ,{\bf 0}} \psi_{v,\eta'} (\Gamma)|^2.
\end{align}
To proceed, we note that $\psi_{\xi=c/v,\eta}(\bm{k}^* = \Gamma,M)$ is an eigenstate of $C_4$ rotation with the eigenvalue $r^\xi_{4,\eta}(\bm{k}^*)$. This in fact fixes the wavefunction up to a complex phase factor:  
\begin{align}
\psi^T_{\xi,\eta }(\bm{k}^* ) = \frac{1}{2} \left(1,-r^\xi_{4,\eta}(\bm{k}^*)^2,r^\xi_{4,\eta}(\bm{k}^*) ,-r^\xi_{4,\eta}(\bm{k}^*)^3  \right), 
\end{align}
where $r^\xi_{4,\eta}(\bm{k}^*)$ is a $C_4$ eigenvalue of the wavefunction at $\bm{k}^*=\Gamma,M$. See [Table. \ref{quadruphase}], for example. We insert this into $\mathcal{I}({\bf{q}}_n,\Delta \epsilon)$ to finally find 
\begin{align}
 \mathcal{I}({\bf{q}}_n,\Delta \epsilon) &= |\mathcal{C}(\Delta \epsilon)|^2 \times \Big\{~ \Big| \left( \cos 2 (2n+1)\pi d -  e^{2\pi i \mathcal{Q}_{xy}}\right)\Big|^2 +   \sin^2(2(2n+1)\pi d) ~\Big\}\\
&=  4|\mathcal{C}(\Delta \epsilon)|^2 \sin^2\Big( \big(  \left(2n+1\right )d +   \mathcal{Q}_{xy} \big) \pi \Big)=
\begin{cases}
4|\mathcal{C}(\Delta \epsilon)|^2\sin^2 (2n+1)\pi d & \text{for }|t_1|>|t_2| \;(\mathcal{Q}_{xy}=0), \\    
4|\mathcal{C}(\Delta \epsilon)|^2\cos^2 (2n+1)\pi d& \text{for } |t_1|<|t_2|\;(\mathcal{Q}_{xy}=\frac{1}{2}),
\end{cases}.
\label{QI-RIXS}
\end{align}
We expect our theory can be tested in XY (X=Ge, Y=S,Se)  with Ge K-edge $(1s\to4p)$\cite{C41}. 
\begin{figure}[h!]
	\includegraphics[width=.7\textwidth]{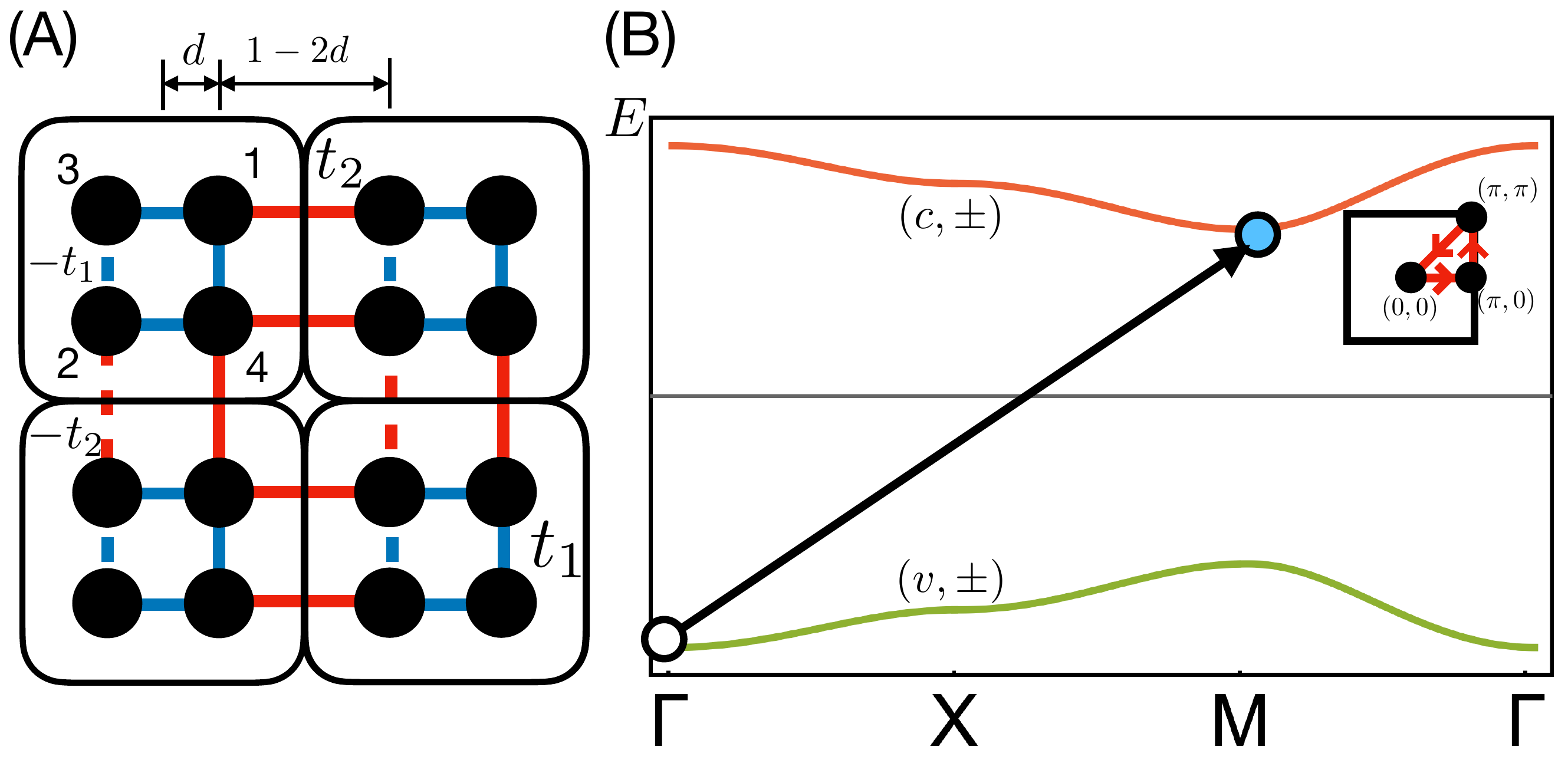}
	\caption{(A) Hopping pattern of QI model. $t_2(t_1)$ is the intersite(intrasite) hopping parameter and dotted lines represents different sign of hopping parameter (so that there are $\pi$ flux for all the plaquettes). We present intersite(intrasite) hopping as a red(blue) line. When $t_2>t_1 $, $H_{\text{QI}}$ hosts a topologically nontrivial phase. Otherwise, it is trivial. (B) Along the high-symmetry points, we plot the band structure of $H_{\text{QI}}$. Each bands are two-fold degenerate (degeneracy index $\pm$). At the half filling, $H_{\text{QI}}$ is an insulator. We consider the RIXS transition from $\Gamma\to M$. An empty circle represents the hole in the valence band and the filled circles presents the created electron.}
	\label{fig:QI_band}
\end{figure}

\subsection{Numerical simulations of RIXS intensity for QI model}
\begin{figure}[t!]
	\includegraphics[width=.6\textwidth]{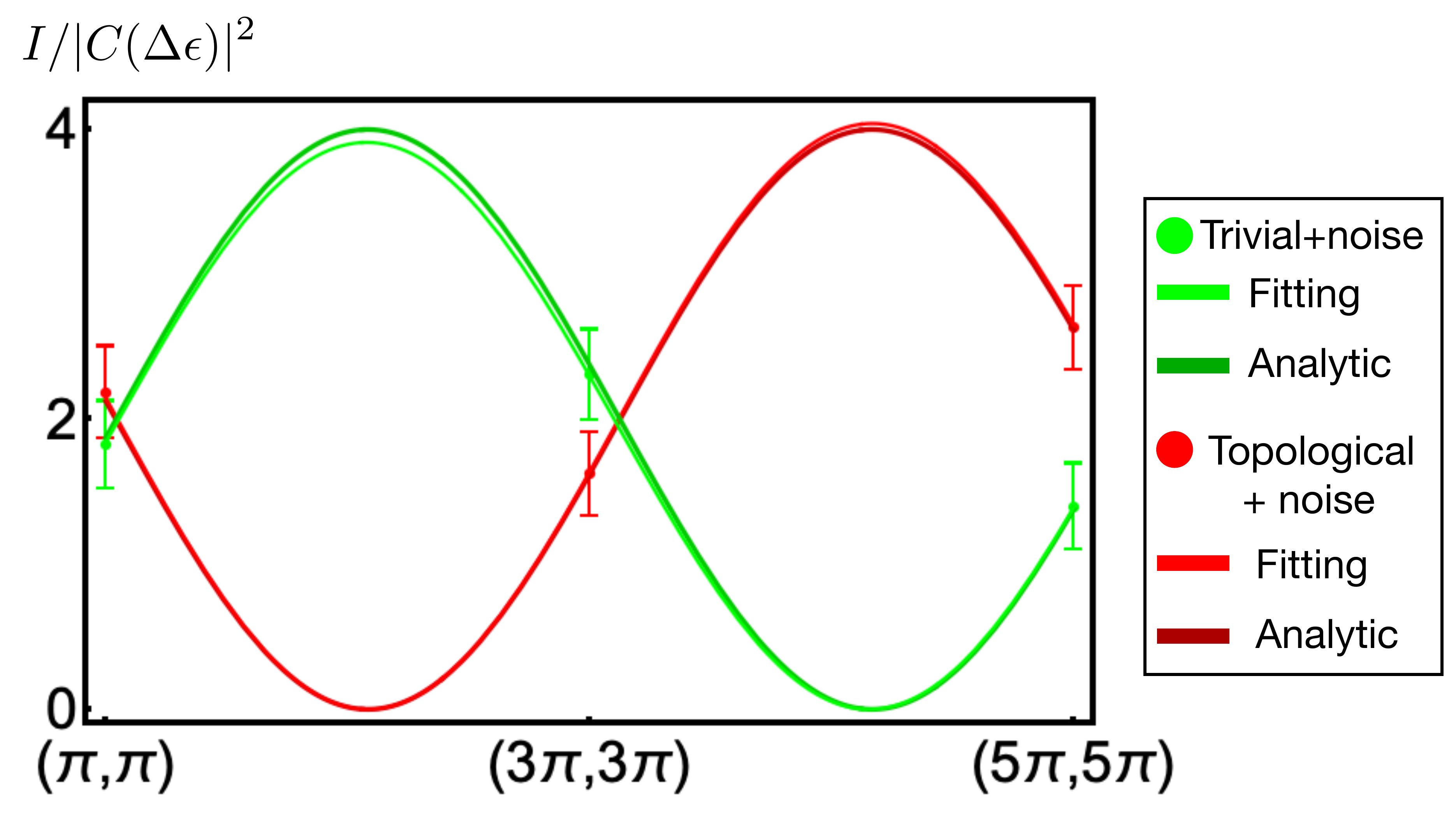}
		\caption{{ RIXS intensity of QI model and Fitting. }} 
	\label{fig:QI_momentum}
\end{figure}

\paragraph{Simulation of the RIXS intensity:} The RIXS intensity data with momentum transfer $\bm{q}_n=(2n+1)(\pi,\pi)$ are generated by adding the two contributions. See the simulated data at [Fig.\ref{fig:QI_momentum}]. 
\begin{align}
\mathcal{I}(\bm{q}_n, \omega) = \mathcal{I}_0 (\bm{q}_n, \omega) + \delta \mathcal{I}_{\text{random}}, 
\end{align}
where $\mathcal{I}_0 (\bm{q}_n, \omega)$ is the RIXS intensity obtained by applying Eq.\ref{QI-RIXS}, and $\delta \mathcal{I}_{\text{random}}$ is the random noise. $\delta \mathcal{I}_{\text{random}}$ is drawn from the uniform distribution $[-0.5, 0.5] \times |\mathcal{C}(\Delta \epsilon)|^2$. Hence, $\mathcal{I}_0 (\bm{q}_n, \omega)$ is the analytic, perfect RIXS signal on the clean QI model (this is obtained by applying Eq.(3) of the main text to the QI model). For the topological case $\mathcal{Q}_{xy}=1/2$ (red circles), the parameters of the QI model were $t_1 = 0.1$ and $t_2=1$. For the trivial case $\mathcal{Q}_{xy}=0$, the parameters were $t_1 =1$ and $t_2 = 0.1$ (so that $\Delta \epsilon$ is the same for the both cases). We set $d=0.24$ for both the cases.\cite{GeSe} The error bars are due to the noises $\delta \mathcal{I}_{\text{random}}$ that we included. 

\paragraph{Fitting procedure:} We have attempted to fit the data with $\mathcal{I} (\bm{q}_n)/|C(\Delta\epsilon)|^2= A \sin^2\big((2n+1)\pi B +Q  \big)+C$, [Cf. Eq.\ref{QI-RIXS}]. Note that the topological index $\mathcal{Q}_{xy}$ is given as $Q/\pi$. See the result of the fitting in [Fig.\ref{fig:QI_momentum}]. 

For the ideal, noiseless signal of the trivial case (without the noise $\delta \mathcal{I}_{\text{random}}$), we expect to find 
\begin{align}
A=4, B=0.24, C=0, Q=0
\label{QI-ideal-triv}
\end{align} 
which is the darker green line in [Fig.\ref{fig:QI_momentum}]. As the result of the fitting (the light green line in [Fig.\ref{fig:QI_momentum}]), we found 
\begin{align}
A=3.89, B=0.23, C=0.04, Q=0.00,
\end{align} 
which are pretty close to the ideal values (\ref{QI-ideal-triv}). In particular, the fitting gives us the correct quadrupole moment $\mathcal{Q}_{xy}=Q/\pi =0$. 

For the ideal, noiseless signal of the topological case (without the noise $\delta \mathcal{I}_{\text{random}}$), we expect to find 
\begin{align}
A=4, B=0.24, C=0,Q=1.58
\label{QI-ideal-top}
\end{align} 
which is the darker red line in [Fig.\ref{fig:QI_momentum}]. 
As the result of the fitting (the light red line in [Fig.\ref{fig:QI_momentum}]), we found 
\begin{align}
A=3.99, B=0.24, C=0.03, Q=1.56,
\end{align} 
which are pretty close to the ideal values (\ref{QI-ideal-top}). In particular, the fitting results in the almost correct quadrupole moment $\mathcal{Q}_{xy}=Q/\pi \approx 0.49$. If one remembers that the quadrupole moment should be quantized, one can confidently infer $\mathcal{Q}_{xy}=1/2$ from the fitting. 
 
\section{2D C$_4$ symmetric higher-order topological insulator }
\subsection{Review of 2D C$_4$ symmetric higher-order topological insulator}
\begin{figure}[h!]
	\includegraphics[width=.7\textwidth]{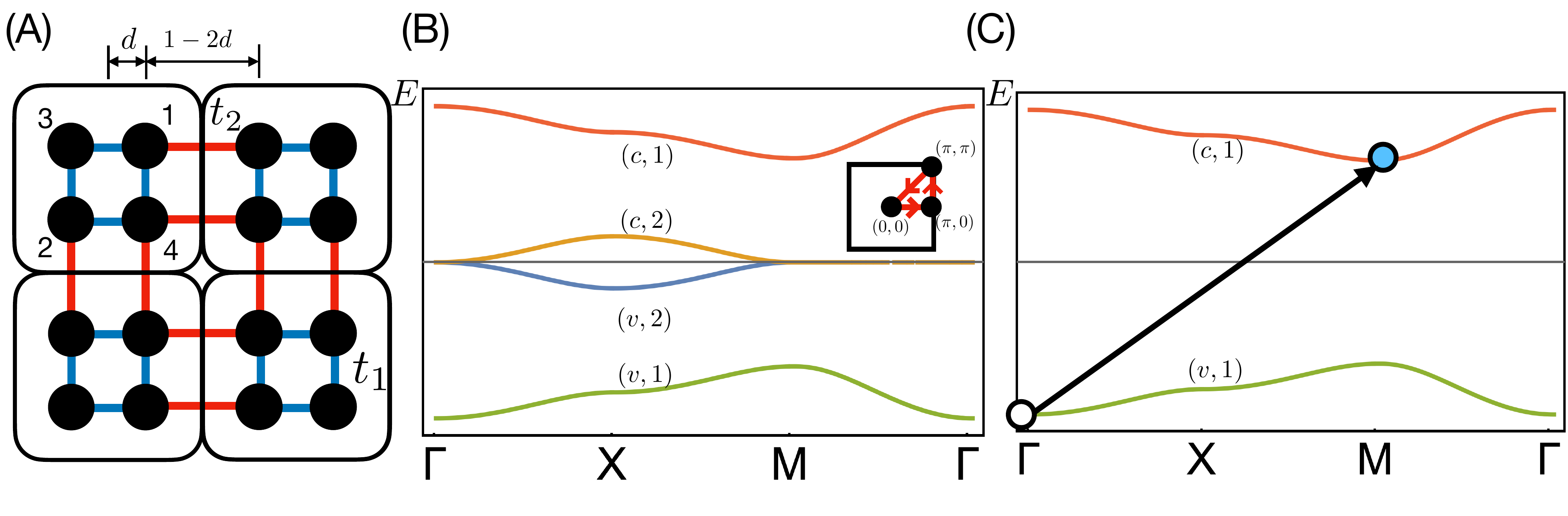}
	\caption{(A) Hopping pattern of $C_4$ symmetric insulator model $H^{(4)}_{1b}$. $t_2(t_1)$ is the intersite(intrasite) hopping parameter. We present intersite(intrasite) hopping as a red(blue) line. When $t_2>t_1 $, $H^{(4)}_{1b}$ hosts a topologically nontrivial phase. Otherwise, it is trivial. (B) Along the high-symmetry points, we plot the band structure of $H^{(4)}_{1b}$. At the quarter filling, $H^{(4)}_{1b}$ is an insulator.  (C) RIXS process. We consider the RIXS transition from $\Gamma\to M$. An empty circle represents the hole in the valence band and the filled circles presents the created electron.}
	\label{fig:c4_band}
\end{figure}

In this supplemental material, we investigate another $C_4$ symmetric insulator $H^{(4)}_{1b}$\cite{hughes}. As in the quadrupolar insulator, we will suppress the spin index for the clarity. We will restore it back when we calculate the RIXS intensity. The real-space Hamiltonian is 
\begin{align} 
H^{(4)}_{1b} &= - \sum_{{\bf r} \in \mathbb{Z}^2} t_1  \left(c_{{\bf r}, 1}^\dagger c_{{\bf r}, 3} +c_{{\bf r},2}^\dagger c_{{\bf r}, 4} +c_{{\bf r}, 1}^\dagger c_{{\bf r}, 4} +c_{{\bf r}, 2}^\dagger c_{{\bf r}, 3} \right) + \text{h.c} \nonumber\\
&\quad-  \sum_{{\bf r} \in \mathbb{Z}^2} t_2 \left(c_{{\bf r}, 1}^\dagger c_{{\bf r}+{\bf e_x}, 3} +c_{{\bf r}+{\bf e_x},2}^\dagger c_{{\bf r}, 4} +c_{{\bf r}, 1}^\dagger c_{{\bf r}+{\bf e_y}, 4} +c_{{\bf r}+{\bf e_y}, 2}^\dagger c_{{\bf r}, 3} \right) + \text{h.c},
\end{align}
where $c_{\bm{r},\alpha}^\dagger$ creates an electron at ${\bf{r_\alpha}}$ with
\begin{align}
{\bf{r_1}} = {\bf{r}}+ (d,d),\quad {\bf{r_2}} = {\bf{r}}+ (-d,-d),\quad {\bf{r_3}} = {\bf{r}}+ (-d,d),\quad {\bf{r_4}} = {\bf{r}}+ (d,-d),
\end{align}
where $\bm{r}\in \mathbb{Z}^2$.  See [Fig.\ref{fig:c4_band}(A)] for the real-space hopping patterns of $H_{1b}^{(4)}$. 

By performing Fourier transformation, $c_{\bm{r},\alpha} = \sum_{k,\alpha} c_{\bm{k}, \alpha} e^{i \bm{k}\cdot \bm{r}}$,  the Hamiltonian is
\begin{align}
H_{4b}^{(4)} = \sum_{\bf{k}} c^\dagger_{{\bf{k}}}  h_{1b}^{(4)}({\bf k})c_{{\bf{k}}} 
\end{align}
where
\begin{align}
h_{1b}^{(4)}({\bf k}) = -\begin{pmatrix}
0&0&t_1 + t_2 e^{i k_x} &t_1 + t_2 e^{i k_y} \\
0&0&t_1 + t_2 e^{-i k_y} & t_1 + t_2 e^{-i k_x}\\
t_1 + t_2 e^{- i k_x}&t_1 + t_2 e^{i k_y}&0&0\\
t_1 + t_2 e^{-i k_y}&t_1 + t_2 e^{i k_x}&0&0
\end{pmatrix}. \label{C4-SYM}
\end{align}
We diagonalize the Hamiltonian via 
\begin{align}
H^{(1)}_{4b} &= \sum_{{\bf{k}},i=c/v, \;\eta=1,2} \varepsilon_{i,\eta}({\bf{k}}) \gamma_{i,\eta}^\dagger({\bf{k}}) \gamma_{i=c/v,\eta}({\bf{k}}),\\
\varepsilon_{i=c/v,1}({\bf{k}})&=\pm \sqrt{ t_1^2+t_2^2 + t_1 t_2 \left(\cos k_x + \cos k_y\right) + \sqrt{ \left(  t_1^2+t_2^2 +2 t_1 t_2 \cos k_x \right)\left(  t_1^2+t_2^2 +2 t_1 t_2 \cos k_y \right)}},\\
\varepsilon_{i=c/v,2}({\bf{k}})&=\pm \sqrt{ t_1^2+t_2^2 + t_1 t_2 \left(\cos k_x + \cos k_y\right) - \sqrt{ \left(  t_1^2+t_2^2 + 2t_1 t_2 \cos k_x \right)\left(  t_1^2+t_2^2 +2 t_1 t_2 \cos k_y \right)}},
\end{align}
with the correspoding four eigenstates, $\psi_{i=c/v,\eta=1}({\bf{k}})$ and $\psi_{i=c/v,\eta=2}({\bf{k}})$ respectively. See [Fig. \ref{fig:c4_band}(B)] for the band dispersions of $H_{1b}^{(4)}$. 

\subsection{Symmetry}
There are three symmetries in $H_{1b}^{(4)}$: $C_4$ rotation symmetry, and two mirror symmetries, and chiral symmetry. Although we discuss the mirror and chiral symmetries here, we do not impose them to be respected by the Hamiltonian, when we come to calculate the RIXS intensity. That is, we will assume that there are always small perturbations which break the symmetries explicitly.

First, let us discuss the $C_4$ rotation symmetry, which involves the matrix $\hat{\mathcal{C}}_4$. When it acts on the fermion, 
\begin{align}
C_4: \bm{c}_k \to \hat{\mathcal{C}}_4 \cdot \bm{c}_{r_4 [\bm{k}]} = 
\begin{pmatrix}
0&\sigma_0\\
\sigma_1&0\\
\end{pmatrix}
 \cdot \bm{c}_{r_4 [\bm{k}]},
\end{align}
where $r_{4} : (k_x,k_y) \rightarrow (-k_y ,k_x)$.

Similarly, the $M_{x,y}$ mirror symmetries are given by the corresponding matrices $\hat{{M}}_{x,y}$ where $x,y$ represents the mirror planes. 
\begin{align}
M_{x}: \bm{c}_k \to \hat{{M}}_{x} \cdot \bm{c}_{m_{x} [\bm{k}]} = 
\begin{pmatrix}
0&\sigma_0\\
\sigma_0&0\\
\end{pmatrix}
 \cdot \bm{c}_{m_x [\bm{k}]},\\
 M_{y}: \bm{c}_k \to \hat{{M}}_{y} \cdot \bm{c}_{m_{y} [\bm{k}]} = 
\begin{pmatrix}
0&\sigma_1\\
\sigma_1&0\\
\end{pmatrix}
 \cdot \bm{c}_{m_y[\bm{k}]},
\end{align}
where $m_x : (k_x,k_y) \rightarrow (-k_x ,k_y)$, $m_y : (k_x,k_y) \rightarrow (k_x ,-k_y)$. Lastly, the chiral symmetry is represented by $\hat{\Pi} \doteq \sigma^3 \otimes \sigma^0$.

\subsection{Topology}
For Eq.\eqref{C4-SYM} Hamiltonian, we summarize the $C_4$ rotation eigenvalues in [Table \ref{c4phase}], which are useful in the next section. These eigenvalues determine $\mathcal{C}$, which is either $0$ (trivial) or $1/2$ (topological). 

\begin{table}[h!]
\begin{tabular}{c|| c |c } 
$C_4$& $\Gamma({\bf{q}}=(0,0))$ & $M({\bf{q}}=(\pi,\pi))$  \\ 
\hline
$v$  &\text{sign}$(t_1 + t_2)$   &  \text{sign}$(t_1 + t_2)$ $ e^{2\pi i \mathcal{C} }$\\ 
\hline
$c$ &$-$\text{sign}$(t_1 + t_2)$   & $-$\text{sign}$(t_1 + t_2)$    $e^{2\pi i \mathcal{C} }$ \\
\hline
\end{tabular}\quad
\caption{ $C_4$ eigenvalue tables at high symmetric points. We summarize rotation eigenvalues of $(c/v,1)$ bands which is relevant band in the quarter filling. See the band dispersion [Fig. \ref{fig:c4_band}(B)].  When $|t_1|>|t_2|$, $\mathcal{C}=0$ which is the trivial phase. When $|t_1|<|t_2|$, $\mathcal{C}=\frac{1}{2}$ and $H_{1b}^{(4)}$ Eq.\eqref{C4-SYM} is in the topological phase. } 
\label{c4phase}
\end{table}

\subsection{RIXS intensity formula reflecting the band topology}
We derive the RIXS intensity formula of 2D C$_4$ symmetric higher-order topological insulator, for diagnosing the band topology. 
We consider the energy transfer $\omega = \Delta \varepsilon = \varepsilon_{c,1}(M) -\varepsilon_{v,1}(\Gamma)$ and 
and the momentum transfer ${\bf q}_n= (2n+1)(\pi,\pi)= M-\Gamma$ ($n\in\mathbb{Z}$). With these energy and momentum transfers, the RIXS quantum amplitude is written as
\begin{align}
\mathcal{A}_{fi} ({\bf{q}_n},\Delta \epsilon) = \mathcal{C}(\Delta\epsilon) \sum_{s=\{\uparrow,\downarrow\}} \psi_{c,1,s}^* (M)  \hat{\mathcal{M}}_{\bf{0+q_n,0}} \psi_{v,1,s} (\Gamma), 
\end{align}
with
\begin{align}
\hat{\mathcal{M}}_{{\bf{k+q},\bf{k}}}=
\begin{pmatrix}
e^{i (q_x+q_y) d}&0&0&0\\
0&e^{-i (q_x+q_y) d}&0&0\\
0&0&e^{i (-q_x+q_y) d}&0\\
0&0&0&e^{i (q_x-q_y) d}
\end{pmatrix}. 
\end{align}
Therefore, the RIXS intensity of the C$_4$ symmetric higher-order topological insulator is 
\begin{align}
\mathcal{I} ({\bf{q}}_n,\Delta\epsilon)  &= 2|\mathcal{C}(\Delta\epsilon)|^2   | \psi^\dagger_{c,1}(M) \hat{\mathcal{M}} _{0 + {\bf{q}_n} ,0} \psi_{v,1} (\Gamma)|^2,
\end{align}
where the overall factor 2 comes from spin degrees of freedom. 
Following the same calculation of QI model, the wave functions at the high symmetric points $\bm{k}_* = \Gamma, M$ are fixed by the $C_4$ symmetry (up to a complex phase), which gives
\begin{align}
\psi_{c/v,1} (\bm{k}_*=\Gamma, M)&\propto \frac{1}{2} \left( 1,r^2_{4,c/v}(\bm{k}_*),r_{4,c/v}(\bm{k}_*) ,r^3_{4,c/v}(\bm{k}_*) \right)
\end{align} 
where $r^4_{4,c/v}(\bm{k}_*)=1$. 
Inserting them into the RIXS quantum amplitude with the $C_4$ eigenvalues in [Table. \ref{c4phase}], we obtain
\begin{align}
\mathcal{I}({\bf{q}_n },\Delta \epsilon)  & =\frac{|\mathcal{C}(\Delta \epsilon)|^2}{2}\Big| \cos 2 (2n+1)\pi d -  e^{2\pi i \mathcal{C}}\Big|^2  \\
&=  2|\mathcal{C}(\Delta \epsilon)|^2\sin^4\Big( \big(  \left(2n+1\right )d +   \mathcal{C}\big) \pi \Big)=
\begin{cases}
\sin^4 (2n+1)\pi d & \text{for } |t_2|>|t_1| \; (\mathcal{C}=0),\\    
\cos^4 (2n+1)\pi d& \text{for } |t_2|<|t_1|\; (\mathcal{C}=\frac{1}{2}). 
\end{cases}
\label{C4-RIXS}
\end{align}

\subsection{Numerical simulation of RIXS intensity and fitting}
\begin{figure}[t!]
	\includegraphics[width=.7\textwidth]{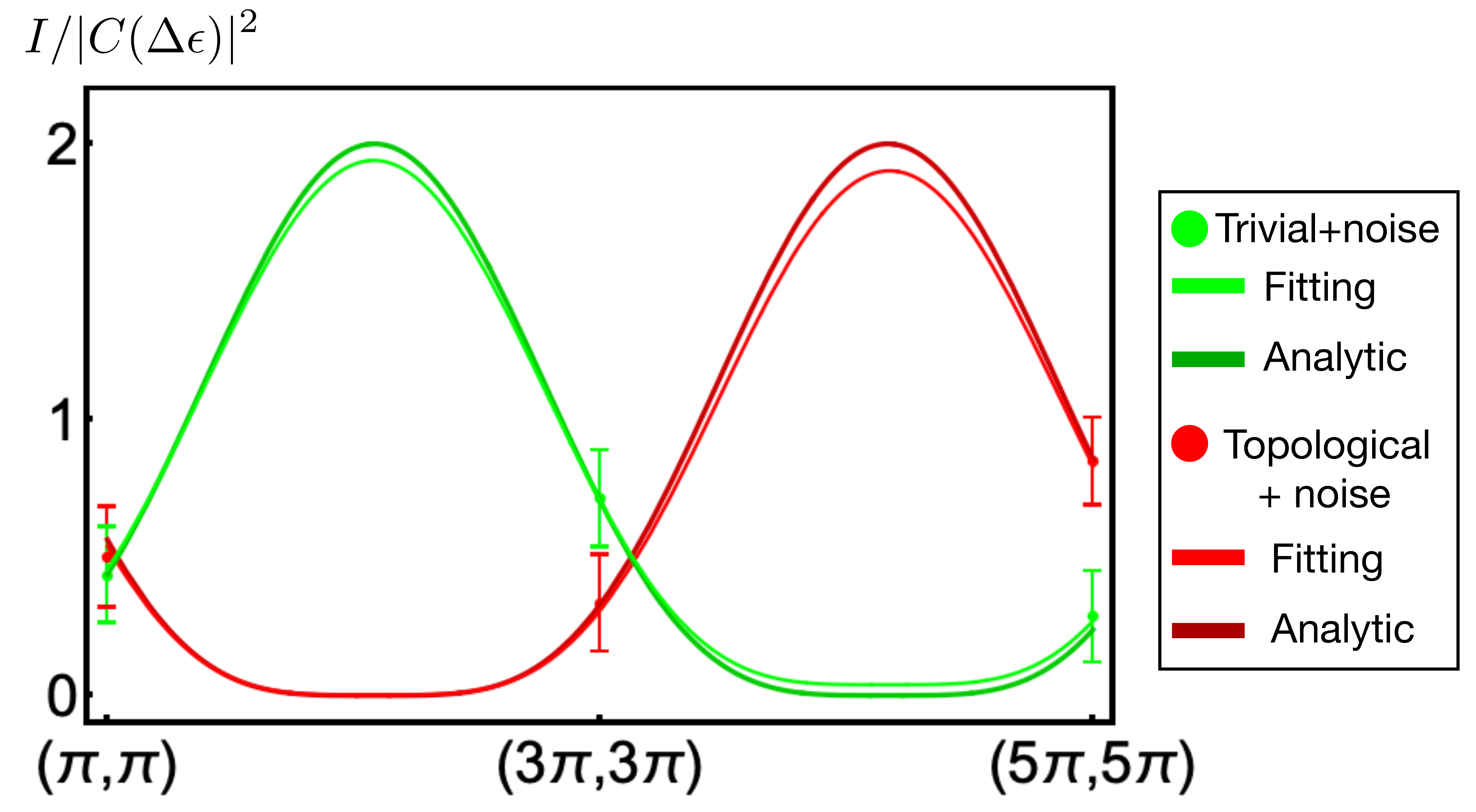}
	\caption{  RIXS intensity of  $H^{(4)}_{1b} $ model with momentum transfer $\bm{q}_n = (2n+1) (\pi,\pi) $ and energy transfer $\omega = \Delta\epsilon= \epsilon_{c,1}(M) - \epsilon_{v,1}(\Gamma)$. See the band dispersion [Fig. \ref{fig:c4_band}(B)]. }
	\label{fig:c42_RIXS}
\end{figure}

\paragraph{Simulation of the RIXS intensity:} The RIXS intensity data [Fig.\ref{fig:c42_RIXS}] with momentum transfer $\bm{q}_n=(2n+1)(\pi,\pi)$ are generated by adding the two contributions. 
\begin{align}
\mathcal{I}(\bm{q}_n, \omega) = \mathcal{I}_0 (\bm{q}_n, \omega) + \delta \mathcal{I}_{\text{random}}, 
\end{align}
where $\mathcal{I}_0 (\bm{q}_n, \omega)$ is the RIXS intensity from the insulator, and $\delta \mathcal{I}_{\text{random}}$ is the random noise. Essentially, $\mathcal{I}_0 (\bm{q}_n, \omega)$ is the analytic, perfect RIXS signal on the clean $H^{(1)}_{4b}$ model, which is obtained by applying Eq.(3) of the main text to the model. On the other hand, $\delta \mathcal{I}_{\text{random}}$ is the white noise drawn from the uniform distribution $[-0.3, 0.3] \times |\mathcal{C}(\Delta \epsilon)|^2$. For the topological case $\mathcal{C}=1/2$ (red circles in [Fig.\ref{fig:c42_RIXS}]), the parameters of the $H^{(1)}_{4b}$ model were $t_1 = 0.1$ and $t_2=1$. For the trivial case $\mathcal{C}=0$ (green circles in [Fig.\ref{fig:c42_RIXS}]), the parameters were $t_1 =1$ and $t_2 = 0.1$ (so that $\Delta \epsilon$ is the same for the both cases). We set $d=0.24$ for both the cases.

\paragraph{Fitting procedure:} We have attempted to fit the data with $\mathcal{I} (\bm{q}_n)/|C(\Delta\epsilon)|^2= A \sin^4\big((2n+1)\pi B +Q  \big)+C$, [Cf. Eq.\ref{C4-RIXS}]. Here $Q$ will be the diagnostics of the band topology, i.e. $\mathcal{C} = Q/\pi$ is the band index.

For the idealistic, perfect signal of the trivial case (without the noise $\delta \mathcal{I}_{\text{random}}$), we expect to find 
\begin{align}
A=2, B=0.24, C=0, Q=0,
\label{c4-ideal-triv}
\end{align} 
which is the darker green line in [Fig.\ref{fig:c42_RIXS}]. As the result of the fitting (the light green line in [Fig.\ref{fig:c42_RIXS}]), we found 
\begin{align}
A=1.90, B=0.23, C=0.04, Q=0.00,
\end{align} 
which are pretty close to the exact values (\ref{c4-ideal-triv}). In particular, this correctly determines $\mathcal{C} = Q/\pi = 0.00$.  

For the idealistic, perfect signal of topological case (without the noise $\delta \mathcal{I}_{\text{random}}$), we expect to find 
\begin{align}
A=2, B=0.24, C=0,Q=1.58,
\label{c4-ideal-top}
\end{align} 
which is the darker red line in [Fig.\ref{fig:c42_RIXS}]. 

As the result of the fitting (the lighter red line in [Fig.\ref{fig:c42_RIXS}]), we found 
\begin{align}
A=1.90, B=0.23, C=0.00, Q=1.57, 
\end{align} 
which are pretty close to the exact values (\ref{c4-ideal-top}). In particular, this correctly determines $\mathcal{C} = Q/\pi = 0.5$. 
 
\section{3D topological band insulator} \label{3dtbi-supp}
\subsection{Review of 3D TBI}
Here we will review the basics of the 3D topological band insulator (TBI) in a diamond lattice (see Ref.[\onlinecite{inv_TI}] for the details), which is consisted of two interpenetrating face-centered cubic lattices (FCC) with a displacement $\vec{d}= \frac{1}{8} (1,1,1)$. The tight-binding Hamiltonian is 
\begin{align}
H_{TBI} = t\sum_{\langle i,j \rangle} c^\dagger_{i}c_j +8 i \lambda_{SO} \sum_{\langle \langle  i,j\rangle \rangle}c^\dagger_i \bm{s} \cdot (\bm{d}^1_{ij}  \times \bm{d}^2_{ij}  ) c_{j}  + h.c.
\end{align}
where $t$ is a hopping parameter and $\bm{d}_{ij}^{1,2}$ are the two nearest neighbor bond vectors traverse between $i$ and $j$. Here, $\lambda_{SO}$ is the spin-orbit coupling strength.  
Under strain [111] direction, the hopping parameter along [111] direction $ t $ is changed by $t\to t+ \delta t$\cite{inv_TI}. Incorporating this and performing the Fourier transformation, $\bm{c}_{\bm{r}} = \sum_{\bm{k}} \bm{c}_{k} e^{i \bm{k}\cdot \bm{r}}$, we find 
\begin{align}
H_{TBI} = \sum_{\bm{k}} \bm{c}^\dagger_{\bm{k} } h_{TBI}({\bf{k}}) \bm{c}_{\bm{k}},
\end{align}
where $\bm{c}_{\bm{k}} =(c_{\bm{k},1,\uparrow},c_{\bm{k},1,\downarrow},c_{\bm{k},2,\uparrow},c_{\bm{k},2,\downarrow})$,
\begin{align}
h_{TBI}({\bf{k}}) =\sum_{a=1}^5  d_a({\bf{k}}) \Gamma^a,
\label{TBI-Hamiltonian}
\end{align}
and
\begin{align*}
x_1 &=  \frac{1}{2} (k_y+k_z), \quad x_2 = \frac{1}{2} (k_x+k_z), \quad x_3 = \frac{1}{2} (k_y+k_x),\\
d_1(k)&= t+ \delta t + t(\cos x_1 +\cos x_2 +\cos x_3 ),\\
d_2(k)&=  t(\sin x_1 +\sin x_2 +\sin x_3 ),\\
d_3(k)&=  \lambda_{SO} ( \sin x_2 - \sin x_3- \sin (x_2-x_1) + \sin (x_3-x_1)) ,\\
d_4(k)&=  \lambda_{SO} ( \sin x_3 - \sin x_1- \sin (x_3-x_2) + \sin (x_1-x_2)) ,\\
d_5(k)&=  \lambda_{SO} ( \sin x_1 - \sin x_2- \sin (x_1-x_3) + \sin (x_2-x_3)) ,\\
\Gamma^{a=1,2,3,4,5} &= ( \tau^1 \sigma^0,\tau^2 \sigma^0 ,\tau^3\sigma^1,\tau^3\sigma^2,\tau^3\sigma^3),
\end{align*}
and $\delta t$ is the change of the hopping parameter along the [111] direction. 

\subsection{Symmetry \& Topology}
The Hamiltonian $h_{TBI}({\bf{k}}) $ has the two crucial symmetries. 
The first is the inversion symmetry 
\begin{align}
\mathcal{I}: \bm{c}_k \to \hat{\mathcal{I}}\cdot \bm{c}_{-\bm{k}} = \tau_1  \sigma_0 \cdot \bm{c}_{-\bm{k}}.
\end{align}
The second symmetry is the (anti-unitary) time-reversal symmetry 
\begin{align}
\mathcal{T}: \bm{c}_k \to \hat{\mathcal{T}}\cdot \bm{c}_{-\bm{k}} = (i \tau_0 \sigma_2) K \cdot \bm{c}_{-\bm{k}}.
\end{align}
where K is a complex conjugate operator. 

The phase diagram of $H_{TBI}$ is simple: 
\begin{align}
\delta t >0 : \text{topological} \quad \delta t <0 : \text{trivial},
\end{align}
where the $\mathbb{Z}_2$ topological band index can be calculated as\cite{inv_TI,fu-kane-mele}
\begin{align}
(-1)^\nu &= \Pi _{\bm{k}_*\in TRIM} \mathcal{I}_v(\bm{k}_*)  =\Pi _{\bm{k}_*\in TRIM-\{\Gamma\} } \mathcal{I}_v(\Gamma)  \mathcal{I}_v(\bm{k}_*)\label{Index}  
\end{align}
where the time-reversal invariant momentum (TRIM) points are given as  
\begin{align}
\Gamma(0,0,0) ,X_1(2\pi,0,0), X_2(0,2\pi,0) ,X_3(0,0,2\pi),  L_1(\pi,\pi,\pi) ,L_2(\pi,-\pi,-\pi), L_3(-\pi,\pi,-\pi) , L_4(-\pi,-\pi,\pi),  
\end{align} 
and $\mathcal{I}_v(\bm{k}_*\in TRIM )  =\pm1$  is an inversion eigenvalue of the valence bands at the TRIM points. In the second equality of Eq.\eqref{Index}, we have used $\mathcal{I}_v(\Gamma)^8=1$. 

Let us present the inversion eigenvalues for a given value of the $\delta t$. When $\delta t>0$ (topological), the eigenvalues are  
\begin{align}
\begin{tabular}{ |c||c|c|c|c|c|c|c|c|c| } 
 \hline
$\bm{k}_*$ &     $\Gamma$ & $X_1$&$X_2$&$X_3$ &$ L_1$ &$L_2 $&$L_3$ &$L_4$  \\ 
 \hline
$\mathcal{I}_v(\bm{k}_*)$  &  $-1 $ & $-1$& $-1$& $-1$ &$1$ & $-1$&$-1$ &$ -1$ \\ 
 \hline
\end{tabular}
\end{align}
which gives $(-1)^7 = (-1)^\nu$ i.e. $\nu=1 $ mod 2. 
On the other hand, when $\delta t<0$ (trivial), 
\begin{align}
\begin{tabular}{ |c||c|c|c|c|c|c|c|c|c| } 
 \hline
$\bm{k}_*$ &     $\Gamma$ & $X_1$&$X_2$&$X_3$ &$ L_1$ &$L_2 $&$L_3$ &$L_4$  \\ 
 \hline
$\mathcal{I}_v(\bm{k}_*)$  &  $-1 $ & $1$& $1$& $1$ &$1$ & $-1$&$-1$ &$ -1$ \\ 
 \hline
\end{tabular}
\end{align}
which gives $(-1)^4 = (-1)^\nu$ i.e. $\nu=0 $ mod 2.

\subsection{Details of Proof for Eq.(8) of the main text}
Here we present the details for deriving Eq.(8) of the main text. We will use the fact that each of conduction and valence bands are two-fold degenerate at TRIM points because of the time reversal symmetry, and they have the same inversion eigenvalues. Our starting point is the non-spin-flip RIXS quantum amplitude,
\begin{align}
\mathcal{A}_{f}(\bm{q})  =& \mathcal{C}(\omega) \sum_{\bm{r} ,\alpha, s=\{ \uparrow,\downarrow\} } \langle f |c_{\bm{r},\alpha, s} c^\dagger_{\bm{r},\alpha ,s} e^{i \bm{q \cdot \bm{r}_\alpha}} | g\rangle,
\label{amplitude}
\end{align}
where $\alpha$ is the sublattice index, and $s$ is the spin. On this, we plug $c_{\bm{r},\alpha,s} = \sum_{\bm{k},\eta} U_{\alpha \eta}^\dagger \gamma_{\eta,s}(\bm{k}) e^{i \bm{k \cdot x}}$ , with
\begin{align}
{U}^\dagger = 
\begin{bmatrix}
\psi_{c,\mu}(\bm{k})|\psi_{c,\nu}(\bm{k})|\psi_{v,\mu}(\bm{k})| \psi_{v,\nu}(\bm{k})
\end{bmatrix}, 
\end{align}
gives
\begin{align}
\mathcal{A}_{f}(\bm{q})=& \mathcal{C}(\omega) \sum_{\bm{r},\alpha} \sum_{\bm{k,k}', \mu,\nu,s} \langle f|  U_{\alpha \mu}^{\dagger}(\bm{k}) U_{\nu \alpha}(\bm{k'}) \gamma_{\mu,s}(\bm{k})  \gamma_{\nu,s}^\dagger(\bm{k'}) e^{-i (\bm{k'-k+q)\cdot x}} e^{-i \bm{q\cdot d}_\alpha}  |g\rangle , \\
=& \mathcal{C}(\omega) \sum_{k,s} \langle f|  (U(\bm{k})  \hat{ \mathcal{M}}_ {\bm{k+q,k}}  U^\dagger(\bm{k+q}))_{\nu\mu}\gamma_{\mu,s}(\bm{k+q})  \gamma^\dagger_{\nu,s}(\bm{k}) |g \rangle ,
\label{a}
\end{align}
    where $\psi_{c/v,\mu}(\bm{k})$ is an eigenstate of $h_{TBI}(\bm{k})$ with a degeneracy index $\mu$ (due to the time-reversal symmetry), and $\hat{\mathcal{M}}_{\bm{q}}$ is
    \begin{align}
    \hat{\mathcal{M}}_{\bm{q}} &= 
    \begin{pmatrix}
    e^{i \bm{q}\cdot\bm{d}} &0&0&0\\
    0& e^{i \bm{q}\cdot\bm{d}} &0&0\\
    0&0& e^{-i \bm{q}\cdot\bm{d}} &0\\
    0&0& 0&e^{-i \bm{q}\cdot\bm{d}} 
    \end{pmatrix}
    =
    \cos\left( \bm{q}\cdot\bm{d}\right) 
    \begin{pmatrix}
    1 &0&0&0\\
    0& 1 &0&0\\
    0&0& 1 &0\\
    0&0& 0&1
    \end{pmatrix}
     + i \sin\left( \bm{q}\cdot\bm{d}\right) \begin{pmatrix}
    1 &0&0&0\\
    0& 1 &0&0\\
    0&0& -1 &0\\
    0&0& 0&-1
    \end{pmatrix}, \nonumber\\
    &=\cos\left( \bm{q}\cdot\bm{d}\right)  1_{4\times 4} + i\sin\left( \bm{q}\cdot\bm{d}\right) \hat{\Pi},
    \end{align}
    with $\vec{d} = \frac{1}{8}(1,1,1)$. Here $\hat{\Pi}$ can be written as $\Gamma^{12} = [\Gamma^1, \Gamma^2]/2i$.

The RIXS quantum amplitude is given by 
    \begin{align}
        \mathcal{A}_{f}^{\mu,\nu} ({\bf{q}}_n,\Delta \epsilon) &= \mathcal{C}(\Delta\epsilon)  \psi^\dagger_{c,\mu}({\bf k+q_n})\hat{\mathcal{M}}_{\bm{q}_n} \psi_{v,\nu}({ \bf{ k}}),\nonumber\\
        &= \mathcal{C}(\Delta\epsilon) \left(   \cos\left( {\bm {q}_n} \cdot\bm{d}\right)  \psi^\dagger_{c,\mu}(\bm{k+q_n}) \psi_{v,\nu}(\bm{k})+ i \sin\left( {\bm{q}_n}\cdot{\bm{d}}\right) \psi^\dagger_{c,\mu}(\bm{k+q_n})  \hat{\Pi}  \psi_{v,\nu}(\bm{k})\right). 
    \label{amplitude:3dTI}
    \end{align}

Using the symmetries, we would like to prove that the RIXS intensity is, 
%
    \begin{align}
    I(\bm{q}_n ,\Delta\epsilon)  &= \sum_{\mu,\nu} |\mathcal{A}_{f}^{\mu,\nu} ({\bf{q}}_n,\Delta \epsilon)|^2 \nonumber\\ 
&=2|\mathcal{C}(\Delta\epsilon)|^2 \sin^2 \left( {\bm{q}_n\cdot \bm{d}} + \frac{1}{4} \left( \mathcal{I}_v(\bm{q}_n) \mathcal{I}_v(\bm{k}) -1 \right) \pi   \right), \nonumber\\ 
&=
\begin{cases}
2|\mathcal{C}(\Delta\epsilon)|^2\sin^2 {\bm{q}_n\cdot \bm{d}} & \text{for } \mathcal{I}_v(\bm{q}_n) \mathcal{I}_v(\bm{k}) =1,\\    
2|\mathcal{C}(\Delta\epsilon)|^2\cos^2 {\bm{q}_n\cdot \bm{d}} & \text{for } \mathcal{I}_v(\bm{q}_n) \mathcal{I}_v(\bm{k})=-1. 
\end{cases}
    \label{3D_TI}
    \end{align}
which is the same as Eq.(8) of the main text. 


The first step of our proof is to show $\psi^\dagger_{\eta=c/v,\mu}(\bm{k})  \hat{\Pi} \psi_{\eta,\nu}(\bm{k})=0$ where $\bm{k}\in $TRIM, 
    \begin{align*}
    &\psi^\dagger_{\eta,\mu}(\bm{k})  \hat{\Pi} \psi_{\eta,\nu}(\bm{k}),\\
    =&\psi^\dagger_{\eta,\mu}(\bm{k})   \hat{\Pi} \hat{\mathcal{I}} \hat{\mathcal{I}} \psi_{\eta,\nu}(\bm{k}),\\
    =&\mathcal{I}_v(\bm{k})\psi^\dagger_{\eta,\mu}(\bm{k})  \hat{ \Pi} \hat{\mathcal{I} }\psi_{\eta,\nu}(\bm{k}), \\
    =& -\mathcal{I}_v(\bm{k})\psi^\dagger_{\eta,\mu}(\bm{k})  \hat{\mathcal{I}}\hat{ \Pi}  \psi_{\eta,\nu}(\bm{k}) ,\\
     =& -\psi^\dagger_{\eta,\mu}(\bm{k})  \hat{\Pi} \psi_{\eta,\nu}(\bm{k}),
    \end{align*}
 where we used $\{\hat{\mathcal{I}},\hat{\Pi} \}=0$. This implies that $\hat{\Pi} \psi_{c/v,\mu}(\bm{k})$ is orthogonal to $\psi_{c/v,\nu}({\bm{k}})$. Thence, $\hat{\Pi} \psi_{c/v,\mu}(\bm{k})$ belongs to the space spanned by $\psi_{v/c,\nu}({\bm{k}})$. Similarly, if $\hat{\mathcal{I}} \psi_{v,\mu}(\bm{k}) = \mathcal{I}_{v}(\bm{k})  \psi_{v,\mu}(\bm{k}) $, then
\begin{align}
\hat{\Pi} \hat{\mathcal{I}} \psi_{v,\mu}(\bm{k}) = \mathcal{I}_{v}(\bm{k})  \hat{\Pi} \psi_{v,\mu}(\bm{k}) \Rightarrow  \hat{\mathcal{I}} \left( \hat{\Pi} \psi_{v,\mu}(\bm{k}) \right) = - \mathcal{I}_{v}(\bm{k})  \left( \hat{\Pi} \psi_{v,\mu}(\bm{k}) \right),
\end{align}
which means that $\mathcal{I}_{c}(\bm{k})=-\mathcal{I}_{v}(\bm{k})$. 



To proceed, we choose the convenient orthnormal basis at $\bm{k}, \bm{k}+\bm{q}_n \in$ TRIM as follows: 
\begin{align}
\{\psi_{v,1}(\bm{k}),\psi_{v,2}(\bm{k}),\psi_{c,1}(\bm{k}),\psi_{c,2}(\bm{k})  \},\nonumber
\end{align} 
where $\psi_{c, \mu}(\bm{k}) = \hat{\Pi} \psi_{v, \mu}(\bm{k})$. Similary, we define  
\begin{align}
\{\psi_{v,1}(\bm{k}+\bm{q}_n),\psi_{v,2}(\bm{k}+\bm{q}_n),\psi_{c,1}(\bm{k}+\bm{q}_n),\psi_{c,2}(\bm{k}+\bm{q}_n)  \} ,\nonumber
\end{align} 
such that $\psi^\dagger_{c,\mu}(\bm{k+q_n})\psi_{v,\nu\neq\mu}(\bm{k})=0$ and $\psi^\dagger_{c,\mu}(\bm{k+q_n})\psi_{c,\nu\neq\mu}(\bm{k})=0$. \textit{This choice allows us to forget about the degeneracy index $(\mu, \nu)$ because the amplitude is diagonal in this index. We used this convention in the proof of our main text, where we suppressed the degeneracy index.} With this, Eq.\eqref{amplitude:3dTI} becomes 
    \begin{align}
        \mathcal{A}_{f}^{\mu\nu} ({\bf{q}}_n,\Delta \epsilon) &= \delta^{\mu\nu} \mathcal{C}(\Delta\epsilon) \left(   \cos\left( {\bm {q}_n} \cdot\bm{d}\right)  \psi^\dagger_{c,\mu}(\bm{k+q_n}) \psi_{v,\mu}(\bm{k})+ i \sin\left( {\bm{q}_n}\cdot{\bm{d}}\right) \psi^\dagger_{c,\mu}(\bm{k+q_n})  \hat{\Pi}  \psi_{v,\mu}(\bm{k})\right). 
    \label{amplitude:3dTI-2}
    \end{align}

We now look for the symmetry constraints on the terms in (\ref{amplitude:3dTI}). For example, 
    \begin{align}
    \psi^\dagger_{c,\mu}(\bm{k+q_n}) \psi_{v,\mu}(\bm{k})=\psi^\dagger_{c,\mu}(\bm{k+q_n}) \hat{\mathcal{I}}^\dagger  \hat{\mathcal{I}}  \psi_{v,\mu}(\bm{k})=-\mathcal{I}_v(\bm{k+q_n}) \mathcal{I}_v(\bm{k}) \psi^\dagger_{c,\mu}(\bm{k+q_n}) \psi_{v,\mu}(\bm{k}). 
    \end{align}
Therefore, 
\begin{align}
\text{if}\quad  \mathcal{I}_v(\bm{k+q_n}) \mathcal{I}_v(\bm{k}) =1 \Rightarrow  \psi^\dagger_{c,\mu}(\bm{k+q_n})\psi_{v,\mu}(\bm{k}) =0.
\end{align}
At the same time, we automatically find $|\psi^\dagger_{c,\mu}(\bm{k+q_n}) \hat{\Pi} \psi_{v,\mu}(\bm{k})|=1$ since $\psi^\dagger_{v,\mu}(\bm{k}) \hat{\Pi} \psi_{v,\mu}(\bm{k})=0$ and  $\psi^\dagger_{v,\mu}(\bm{k+q_n})\psi_{c/v,\nu\neq\mu}(\bm{k})=0$. Hence, Eq.\eqref{amplitude:3dTI-2} becomes 
    \begin{align}
        \mathcal{A}_{f}^{\mu\nu} ({\bf{q}}_n,\Delta \epsilon) &= \delta^{\mu\nu} \mathcal{C}(\Delta\epsilon)  i \sin\left( {\bm{q}_n}\cdot{\bm{d}}\right) 
    \label{amplitude:3dTI-3}
    \end{align}
which leads to $\mathcal{I} = 2 |\mathcal{C}(\Delta\epsilon)|^2 \sin^2 ({\bm{q}_n}\cdot{\bm{d}})$.

Similarly, 
    \begin{align}
    \psi^\dagger_{c,\mu}(\bm{k+q_n}) \hat{\Pi} \psi_{v,\mu}(\bm{k})=\psi^\dagger_{c,\mu}(\bm{k+q_n}) \hat{\mathcal{I}}^\dagger  \hat{\mathcal{I}} \hat{\Pi}  \psi_{v,\mu}(\bm{k})=\mathcal{I}_v(\bm{k+q_n}) \mathcal{I}_v(\bm{k}) \psi^\dagger_{c,\mu}(\bm{k+q_n}) \hat{\Pi}  \psi_{v,\mu}(\bm{k}), 
    \end{align}
so
\begin{align}
\text{if}\quad  \mathcal{I}_v(\bm{k+q_n}) \mathcal{I}_v(\bm{k}) =-1 \Rightarrow  \psi^\dagger_{c,\mu}(\bm{k+q_n})\hat{\Pi} \psi_{v,\mu}(\bm{k}) =0. 
\end{align}
This enforces $\psi^\dagger_{c,\mu}(\bm{k+q_n}) \psi_{v,\mu}(\bm{k}) =1$. Hence, Eq.\eqref{amplitude:3dTI-2} becomes 
    \begin{align}
        \mathcal{A}_{f}^{\mu\nu} ({\bf{q}}_n,\Delta \epsilon) &= \delta^{\mu\nu} \mathcal{C}(\Delta\epsilon) \cos\left( \bm{q}_n\cdot\bm{d}\right)  , 
    \label{amplitude:3dTI-3}
    \end{align}
which leads to $\mathcal{I} = 2 |\mathcal{C}(\Delta\epsilon)|^2 \cos^2 ({\bm{q}_n}\cdot{\bm{d}})$. This completes our proof.

\subsection{Numerical Details for [Fig.2(C)] in the main text}
Let us provide some numerical details for the plot [Fig.2(C)] in the main text. 

\paragraph{Simulation of the RIXS intensity:} 
The RIXS intensity data (the filled circles in [Fig.2(C)]) with momentum transfer $\bm{q}_n = (2\pi,0,0) + n(2\pi,2\pi,2\pi)$ (which connects $\Gamma \to X(2\pi,0,0)$) are generated by adding the two contributions. 
\begin{align}
\mathcal{I}(\bm{q}_n, \omega) = \mathcal{I}_0 (\bm{q}_n, \omega) + \delta \mathcal{I}_{\text{random}}, 
\end{align}
where $\mathcal{I}_0 (\bm{q}_n, \omega)$ is the RIXS intensity obtained by applying Eq.(3) of the main text to the 3D topological insulator model\cite{inv_TI,fu-kane-mele}, and $\delta \mathcal{I}_{\text{random}}$ is the random noise. $\delta \mathcal{I}_{\text{random}}$ is drawn from the uniform distribution $[-0.3, 0.3] \times |\mathcal{C}(\Delta \epsilon)|^2$. Hence, $\mathcal{I}_0 (\bm{q}_n, \omega)$ is the idealistic, perfect RIXS signal of the clean 3D topological band insulator model.  Red circles correspond to the $\mathcal{I}_v(\Gamma)\mathcal{I}_v(X)=-1$ case and green circles correspond to the $\mathcal{I}_v(\Gamma)\mathcal{I}_v(X)=1$ case. 

\paragraph{Fitting procedure:} 
We have attempted to fit the data with $\mathcal{I} (\bm{q}_n)/|\mathcal{C}(\omega)|^2  = A \sin^2\big( \bm{q}_n\cdot\bm{d} + B \big)+C$ [Cf. Eq.\ref{3D_TI}]. Note that $B$ is the diagnostics of the band topology, because we expect $B = \frac{1}{4} \left( \mathcal{I}_v(\bm{q}_n) \mathcal{I}_v(\bm{k}) -1 \right) \pi$ if the fitting works well enough. 

For the ideal, clean signal of $\mathcal{I}_v(\Gamma)\mathcal{I}_v(X)=1$ case (without the noise $\delta \mathcal{I}_{\text{random}}$), we expect to find 
\begin{align}
A=2.00, B=0.00, C=0.00, \label{TBI-E} 
\end{align} 
which is the dotted green line in [Fig.2(C)]. As the result of the fitting (the solid green line in [Fig.2(C)]), we found 
\begin{align}
A=1.96, B=0.01, C=0.00,
\end{align}
which agrees excellently with the exact values Eq.\eqref{TBI-E}. In particular, from $B$, we can correctly infer $\mathcal{I}_v(\bm{q}_n) \mathcal{I}_v(\bm{k}) = 4B/\pi+1 \approx 1$. 

For the perfect, ideal signal of $\mathcal{I}_v(\Gamma)\mathcal{I}_v(X)=-1$ case (without the noise $\delta \mathcal{I}_{\text{random}}$), we expect to find 
\begin{align}
A=2.00, B=1.58, C=0.00, \label{TBI-E2}
\end{align} 
which is the dotted red line in [Fig.2(C)]. As the result of the fitting (the solid red line in [Fig.2(C)]), we found 
\begin{align}
A=1.96, B=1.58, C=0.00,
\end{align} 
which are pretty close to the exact result Eq.\eqref{TBI-E2}. In particular, from $B \approx \pi/2$, we can correctly infer $\mathcal{I}_v(\bm{q}_n) \mathcal{I}_v(\bm{k}) \approx -1$. 

\subsection{Demonstration}
In this section, we demonstrate how the RIXS data can be used to read off topological band index. Suppose a series of RIXS intensity that connects $\Gamma$  and other TRIM points $(X,L_1, L_2)$ by $\bm{q}_{i=1,2,3}$ is obtained. 
    For example, 
    \begin{align}
    I_{\Gamma \to X(2\pi,0,0)}(\bm{ q_1} ,\Delta \epsilon_1 )  &= |\mathcal{C}(\Delta \epsilon_1 )|^2 \sum_\mu |\psi^\dagger_{c,\mu}(X)\hat{\mathcal{M}}_{\bm{ q_1}}\psi_{v,\mu}(\Gamma) |^2  \propto \sin^2 \left(  \bm{ q_1} \cdot\bm{ d}\right)  ,\\
    I_{\Gamma \to L_1(\pi,\pi,\pi) }(\bm{ q_2} ,\Delta \epsilon_2 )  &= |\mathcal{C}(\Delta \epsilon_2 )|^2 \sum_\mu |\psi^\dagger_{c,\mu}(L_1)\hat{\mathcal{M}}_{\bm{ q_2}}\psi_{v,\mu}(\Gamma) |^2  \propto \cos^2 \left(  \bm{ q_2} \cdot\bm{ d}\right)  ,\\
    I_{\Gamma \to L_2(\pi,-\pi,\pi)}(\bm{ q_3} ,\Delta \epsilon_3 )  &= |\mathcal{C}(\Delta \epsilon_3 )|^2\sum_\mu  |\psi^\dagger_{c,\mu}(L_2)\hat{\mathcal{M}}_{\bm{ q_3}}\psi_{v,\mu}(\Gamma) |^2  \propto \sin^2 \left(  \bm{ q_3} \cdot\bm{ d}\right) .
    %
    %
    \end{align}
    From these results, we can reconstruct the inversion eigenvalues at $\bm{k}\in$ TRIM,  see [Table.\ref{relativeEV}]. 
    \begin{align}
    \begin{tabular}{ |c||c|c|c|c|c|c|c|c|c| } 
     \hline
     \bf{k} &     $\Gamma$ & $X_1$&$X_2$&$X_3$ &$ L_1$ &$L_2 $&$L_3$ &$L_4$  \\ 
     \hline
    $\mathcal{I}_v(\bm{k}) $   &  $\mathcal{I}_v(\Gamma)$ & $\mathcal{I}_v(\Gamma)$& $\mathcal{I}_v(\Gamma)$& $\mathcal{I}_v(\Gamma)$ &-$\mathcal{I}_v(\Gamma)$ & $\mathcal{I}_v(\Gamma)$&$\mathcal{I}_v(\Gamma)$ &$ \mathcal{I}_v(\Gamma)$ \\ 
     \hline
    \end{tabular}
    \label{relativeEV}
    \end{align}
    From the eigenvalues, we can also infer the $\mathbb{Z}_2 $ topological index i.e. 
    \begin{align}
    (-1)^\nu=-(\mathcal{I}_v(\Gamma))^8  = -1\quad  \Rightarrow \nu \equiv 1 \;\text{mod 2}.
    \end{align}
    This actually corresponds to $\delta t>0$ case of $h_{TBI}(\bm{k})$. 

\begin{figure}[t!]
	\includegraphics[width=.4\textwidth]{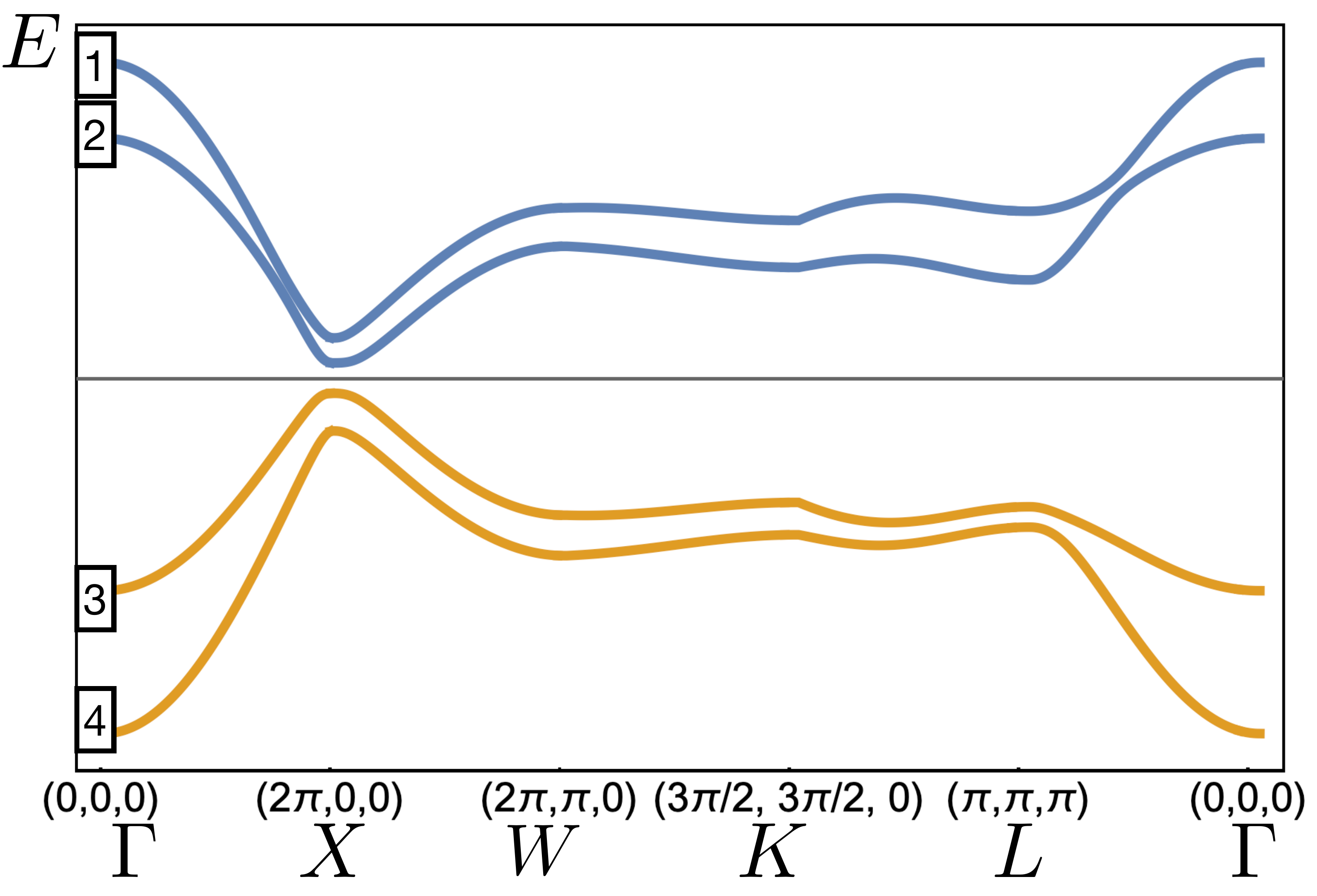}
		\caption{The band dispersion of $h_{TBI,8}(\bm{k}) $ with $(t_1,t_2, \delta t_1 ,\delta t_2 ,\lambda_{SO,1}, \lambda_{SO,2}, e_1  ,e_2,e_3) = (0.5,0.6,       -0.2,  0.25,        0.3,      0.3, 0.12,  -0.15,0.12)$ along the high-symmetry cuts in the Brillouin zone.  }
	\label{fig:3DTI_perturbed_band}
\end{figure}

Note that the above protocol can be generalized to non-minimal models. For example, as a concrete illustration, we will consider the double copies of Eq.\eqref{TBI-Hamiltonian} with different Hamiltonian parameters. 
\begin{align}
h_{TBI,8}(\bm{k}) = 
\begin{pmatrix}
h_{TBI}(t_1,\delta t_1 ,\lambda_{SO,1} ; \bm{k}) + e_1 \mathbb{I}_{4\times 4} &e_3 \mathbb{I}_{4\times 4}\\
e_3 \mathbb{I}_{4\times 4}& h_{TBI}(t_2,\delta t_2 ,\lambda_{SO,2} ; \bm{k}) +e_2 \mathbb{I}_{4\times 4}
\end{pmatrix},
\end{align}
Here, $h_{TBI}(t_{i},\delta t_{i} ,\lambda_{SO,{i}} ; \bm{k})$ $(i=1,2)$ is one of two copies of bloch Hamiltonian \eqref{TBI-Hamiltonian} with a parameter set $( t_{i}, \delta t_{i}, \lambda_{SO,{i}})$ and $e_{a=1,2} \mathbb{I}_{4\times 4}$ is for the constant energy shifts for each bands. Here, $e_{3}$ is the hybridizations between the two copies. The band dispersion of $h_{TBI,8}(\bm{k})$ is plotted on [Fig.\ref{fig:3DTI_perturbed_band}]. 

We note that this model shares the same structure factor $\hat{\mathcal{M}}_{\bm{q}}$ as before. I.e., $\hat{\mathcal{M}}_{\bm{q},8} = \hat{\mathcal{M}}_{\bm{q}} \oplus \hat{\mathcal{M}}_{\bm{q}}$. We can explicitly write the RIXS intensity and quantum amplitudes out in terms of the Bloch functions:
    \begin{align}
        \mathcal{A}_{i \to f}^{\mu,\nu} ({\bm{q}}_n,\Delta \epsilon) &= \mathcal{C}(\Delta\epsilon)  \psi^\dagger_{f,\mu}({\bm {k+q_n}})\hat{\mathcal{M}}_{\bm{q}_n,8} \psi_{i,\nu}({ \bm{ k}}),\nonumber\\
        &= \mathcal{C}(\Delta\epsilon) \left(   \cos\left( {\bm {q}_n} \cdot\bm{d}\right)  \psi^\dagger_{f,\mu}(\bm{k+q_n}) \psi_{i,\nu}(\bm{k})+ i \sin\left( {\bm{q}_n}\cdot{\bm{d}}\right) \psi^\dagger_{f,\mu}(\bm{k+q_n})  \hat{\Pi}_8  \psi_{i,\nu}(\bm{k})\right). 
    \label{amplitude:3dTI-two-copy}
    \end{align}
where $\mu$ and $\nu$ are degeneracy indices for each bands. 
Finally, the inversion symmetry is $\hat{\mathcal{I} }_8 = \Gamma^1 \oplus \Gamma^1 $ which implies that $\{ \hat{\mathcal{I}}_8 , \hat{\Pi}_8 \}=0$ still holds. Since all the algebra straightforwardly follows even in this case, we can easily show that 
    \begin{align}
    I(\bm{q}_n ,\Delta\epsilon)  
&=
\begin{cases}
2|\mathcal{C}(\Delta\epsilon)|^2\sin^2 {\bm{q}_n\cdot \bm{d}} & \text{for } \mathcal{I}_f(\bm{k}+\bm{q}_n) \mathcal{I}_i(\bm{k}) =1,\\    
2|\mathcal{C}(\Delta\epsilon)|^2\cos^2 {\bm{q}_n\cdot \bm{d}} & \text{for } \mathcal{I}_f(\bm{k}+\bm{q}_n) \mathcal{I}_i(\bm{k})=-1,  
\end{cases}
    \label{8bandTI}
    \end{align}
for $\bm{q}_n$  that connects TRIM points $\bm{k}$ and $\bm{k}+\bm{q}_n$. E.g. $\bm{q}_n = 2\pi \left(n+1  ,   n , n  \right)$ as the momentum connecting $\Gamma$ to $X$. 

Below, we will work with the parameters,
\begin{align*}
(t_1,t_2, \delta t_1 ,\delta t_2 ,\lambda_{SO,1}, \lambda_{SO,2}, e_1  ,e_2,e_3) = (0.5,0.6,       -0.2,  0.25,        0.3,      0.3, 0.12,  -0.15,0.12),
\end{align*}
which are arbitrarily chosen. We will label each two-fold degenerate bands as in [Fig. \ref{fig:3DTI_perturbed_band}] and summarize their inversion eigenvalues at TRIM points on [Table.\ref{relativeEV_perturbed}]. 

    \begin{align}
    \begin{tabular}{ |c||c|c|c|c|c|c|c|c|c| } 
     \hline
     \bf{k} &                               $\Gamma$ & $X_1$&$X_2$&$X_3$ &$ L_1$ &$L_2 $&$L_3$ &$L_4$  \\ 
     \hline
    $\mathcal{I}_1(\bm{k}) $   &  $1$ & $-1$& $-1$& $-1$  &$-1$ & $1$&$1$ &$ 1$ \\ \hline
    $\mathcal{I}_2(\bm{k}) $   &  $1$ & $1$& $1$& $1$     & $-1$ & $1$&$1$ &$ 1$ \\ \hline
    $\mathcal{I}_3(\bm{k}) $   &  $-1$ & $1$& $1$& $1$    &$1$ & $-1$&$-1$ &$ -1$ \\ \hline
    $\mathcal{I}_4(\bm{k}) $   &  $-1$ & $-1$& $-1$& $-1$  &$1$ & $-1$&$-1$ &$ -1$ \\ 
     \hline
    \end{tabular}
    \label{relativeEV_perturbed}
    \end{align}

\begin{figure}[h!]
	\includegraphics[width=.8\textwidth]{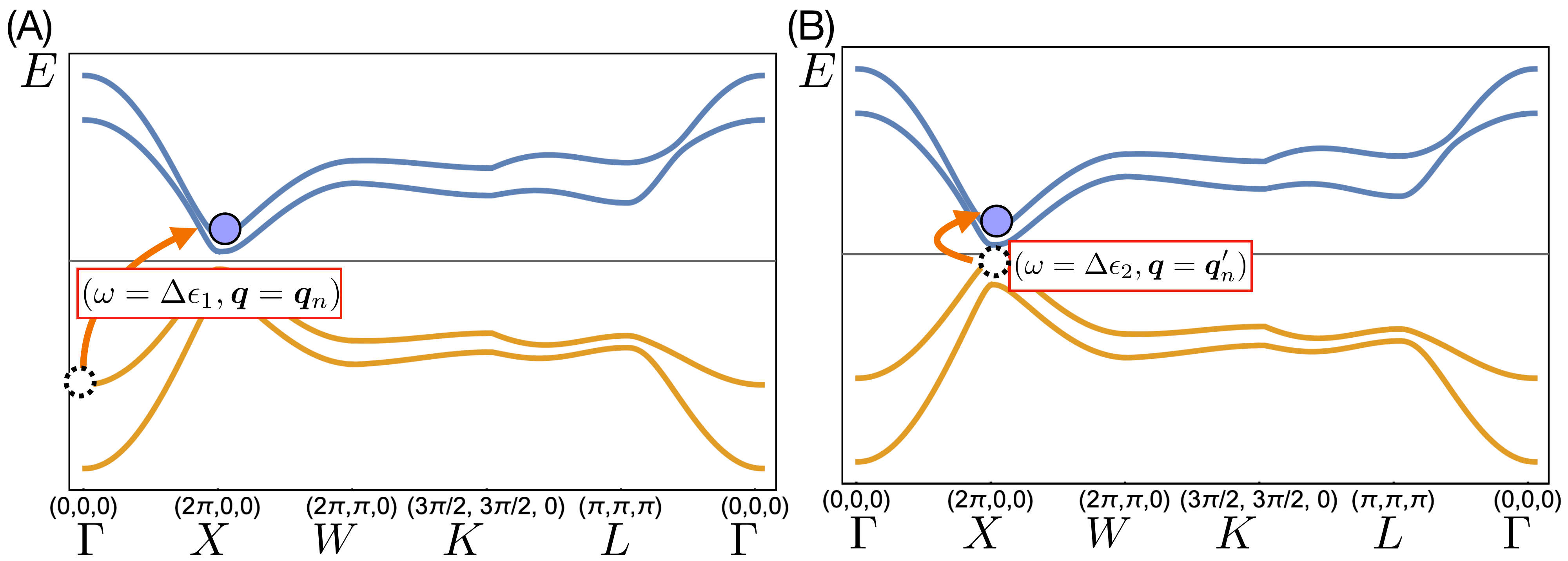}
		\caption{The two RIXS transitions with appropriate energy-momentum transfers that we consider. (A) With energy-momentum transfer $(\Delta\epsilon_1 = \epsilon_2(X) -\epsilon_3(\Gamma) , \bm{q}_n = 2\pi(n+1,n,n) )$ $(n\in \mathbb{Z})$, this RIXS intensity is determined by the inversion eigenvalues $\mathcal{I}_2(X) $ and $\mathcal{I}_3(\Gamma) $. See Eq.\eqref{8bandTI}. (B)  With energy-momentum transfer $(\Delta\epsilon_2 = \epsilon_2(X) -\epsilon_3(X) , \bm{q}'_n = 2\pi(n,n,n) )$ $(n\in \mathbb{Z})$, the RIXS intensity of this transition is determined by the $\mathcal{I}_2(X) $ and $\mathcal{I}_3(X) $. See Eq.\eqref{8bandTI}.}
	\label{fig:3DTI_cTI_transition}
\end{figure}

Next, we demonstrate our protocol to read off the product of the inversion eigenvalues from the momentum oscillation of the RIXS intensity. From this, we will determine the topological $\mathbb{Z}_2$ band index at the end. As a concrete illustration, we will take the state at $X$ of the $2$-band as the reference and will attempt to read off the band topology of the $3$-band. For this, we consider the two example transitions: (A) $\psi_{3,\mu}(\Gamma) \to \psi_{2,\mu}(X)$, which we can choose degeneracy index $\mu$ as before,  by momentum transfer $\bm{q}_n = 2\pi \left(n+1  ,   n , n  \right)$ $(n\in \mathbb{Z})$  and (B) $\psi_{3,\mu}(X)\to \psi_{2,\mu}(X)$ by momentum transfer $\bm{q}'_n = 2\pi \left(n  ,  n ,n  \right)$ [with $n\in \mathbb{Z}$], see [Fig. \ref{fig:3DTI_cTI_transition},(A) and (B)]. 

We have numerically simulated those two RIXS transitions with noises taken from a random uniform distribution, $[-0.2,0.2]\times |\mathcal{C}(\Delta\epsilon)|^2$ to mimic experimental situation. With the data, we try to fit $\mathcal{I} (\bm{q}_n)/|\mathcal{C}(\Delta\epsilon)|^2  = A \sin^2\big( \bm{q}_n\cdot\bm{d} + B \big)+C$ [Cf. Eq.\ref{3D_TI}] for (A). As before, $B$ is the diagnostics of the band topology, $B = \frac{1}{4} \left( \mathcal{I}_3(\Gamma) \mathcal{I}_2(X) +1 \right) \pi$ where $\mathcal{I}_a(\bm{k})$ is an inversion eigenvalue of the $a$-band at $\bm{k}$. 

\begin{figure}[t!]
	\includegraphics[width=.8\textwidth]{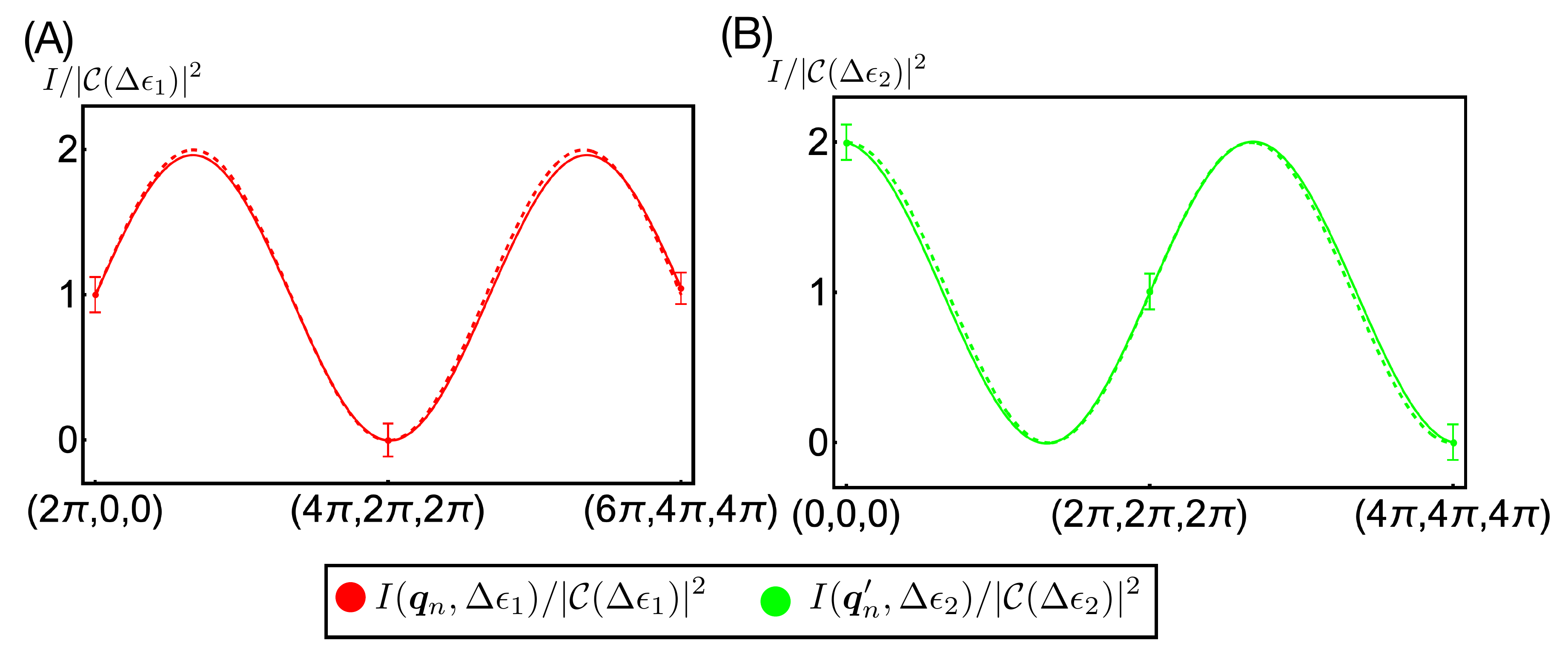}
		\caption{The two RIXS intensities of $h_{TBI,8}(\bm{k})$ with appropriate energy and momentum transfers.  Red (A) and green (B) dotted lines represent ideal results without noises and red (A) and green (B) solid lines are the result of the fitting from simulated data. For detailed procedure, see from \eqref{pTIGX-ideal} to \eqref{pTIXX-fit}. }
	\label{fig:3DTI_cTI_oscillation}
\end{figure}

For the ideal limit of (A), clean signal of $\mathcal{I}_3(\Gamma)\mathcal{I}_2(X)=-1$ case (without the noise), we expect to find 
\begin{align}
A=2.00, B=0.00, C=0.00,  
\label{pTIGX-ideal}
\end{align} 
which is the dotted red line in [Fig. \ref{fig:3DTI_cTI_oscillation} (A)]. As the result of the fitting (the solid red line in [Fig.\ref{fig:3DTI_cTI_oscillation} (A)]), we found 
\begin{align}
A=1.99, B=0.01, C=0.00,
\end{align}
which agrees well with ideal result. 
In particular, from $B$, we can correctly infer $\mathcal{I}_3(\Gamma) \mathcal{I}_2(X) = 4B/\pi-1 \approx -1$. 

For the ideal limit of (B), ideal signal of $\mathcal{I}_3(X)\mathcal{I}_2(X)=1$ case (without the noise), we expect to find 
\begin{align}
A=2.00, B=1.58, C=0.00,
\end{align} 
which is the dotted green line in [Fig.\ref{fig:3DTI_cTI_transition} (B)]. As the result of the fitting (the solid green line in [Fig.\ref{fig:3DTI_cTI_transition} (B)]), we found 
\begin{align}
A=2.00, B=1.58, C=0.00,
\label{pTIXX-fit}
\end{align} 
which also agrees with ideal case. In particular, from $B \approx \pi/2$, we can correctly infer $\mathcal{I}_3(X) \mathcal{I}_2(X) \approx 1$. 

This can be repeated for all the TRIM points of the $3$-band, which allows us to determine $\mathcal{I}_3(X) \mathcal{I}_2(\bf{k}_*)$ with $\bm{k}_* \in TRIM$. Then, we can determine the band-wise $\mathbb{Z}_2$ invariant $\nu_0 (\text{3-band})$ of the $3$-band via 
\begin{align}
(-1)^{\nu_0 (\text{3-band})}= \prod_{\bm{k}_* \in TRIM} \mathcal{I}_3(\bm{k}_*) =  \prod_{\bm{k}_* \in TRIM} [\mathcal{I}_3(\bm{k}_*)\mathcal{I}_2(X)], 
\end{align}
since $\mathcal{I}_2(X)^8 = 1$. There are two remarks. The same can be repeated for the other filled band, i.e., $4$-band, to completely determine the $\mathbb{Z}_2$ invariant. Next, in the above, we have selected the $X$ point of the $2$-band as the reference, but (if there is any preference) one can choose other high-symmetry point with other band as the reference. 

\section{3D chiral hinge insulator  }
\subsection{Review of 3D chiral hinge insulator  }
Here, we will review the basics of 3D chiral hinge insulator (CHI) by following  Ref.[\onlinecite{CHI}]. Its momentum-space Hamiltonian is  
%
\begin{align}
H_{CHI}&= \sum_{\bm{k}} \psi^\dagger(\bm{k}) h_{CHI}(\bm{k}) \psi(\bm{k}),\nonumber\\
h_{CHI}(\bm{k}) &=\left( M+ t \sum_{i} \cos k_i \right) \tau_1  \sigma_0  + \Delta_1 \sum_{i=x,y,z} \sin k_i \tau_3 \sigma_i  + \Delta_2 \left(\cos k_x - \cos k_y \right) \tau_2  \sigma_0, \label{CHI-M}
\end{align}
where
$
\psi^T(\bm{k}) = (c_{\alpha =0, \uparrow}(\bm{k}),c_{\alpha =0,\downarrow}(\bm{k}),c_{\alpha =1, \uparrow}(\bm{k}),c_{\alpha =1,\downarrow}(\bm{k}))
$. 
Here, $\Delta_2$ term is a $C_{4}^z$-symmetry-breaking term by external effect such as $[1\bar{1}0]$ strain. 
$\alpha=0,1$  is sublattice index, and and $s=\uparrow,\downarrow$ is the spin index.  
Also ${\bm{r}_\alpha}=\bm{r} + (-1)^\alpha \bm{d} =\bm{r} - (-1)^\alpha (0,0,d/2)$ with ${\bm{r}} \in \mathbb{Z}^3$ and $0<d<\frac{1}{2}$ without loss of generality. 
Each eigenstates are two-fold degenerate protected by the symmetries of the system, which we will explain in the next section. 

\subsection{Symmetry \& Topology} 
The remaining symmetries of the CHI Hamiltonian are $\hat{I}\hat{T}$ and $\hat{C}^z_4 \hat{I}$. Each are represented as 
\begin{align}
\hat{C}^z_4=\tau_0  e^{-i \frac{\pi}{4} \sigma_z} , \quad  \hat{I} = \tau_1 \sigma_0 ,\quad \hat{T}=\tau_0  \sigma_y K,
\end{align}
when acting on the fermion fields. Here, $\hat{I}$ is an inversion operator and $\hat{T}$ is time-reversal operator with the complex conjugation operator $K$. 
They satisfy 
\begin{align}
(\hat{C}^z_4 \hat{T})^4 = -1, \quad (\hat{C}^z_4 \hat{I})^4 = -1, \quad [\hat{C}^z_4 \hat{I} , \hat{I} \hat{T} ] = 0, \quad \{\hat{C}^z_4 \hat{I} , \hat{\Pi} \} = 0, 
\label{algebra}
\end{align}
where $\hat{\Pi}=\tau_3    \sigma_0 $ is a chiral-symmetry operator. As in the other models, we will in general assume that the chiral symmetry is explicitly broken by perturbations to the CHI Hamiltonian. 

The topology of the CHI Hamiltoinan can be expressed as the inversion eigenvalues of some high-symmetry points in momentum space, namely the Roto-Inversion Symmetric Points(RISP) such that $\left(C^z_{4}I \right)\bm{k}= \bm{k}$: 
\begin{align}
RISP= \{ ( 0,0, 0),( 0,0, \pi),( \pi,\pi, 0),( \pi,\pi, \pi)\}.
\end{align}
Note that because of Eq.\eqref{algebra}, $\hat{ C_{4}^{z}} \hat{I} $ eigenvalues of the two-fold degenerate eigenstates at RISP has to be complex conjugate pairs, $\{ \mathcal{I}_v(\bm{k}) e^{i\pi/4},\mathcal{I}_v(\bm{k}) e^{-i\pi/4}\}$ with $\mathcal{I}_v(\bm{k})=\pm1$ is an inversion eigenvalue of the band at $\bm{k}$. 

The topological band index is given by the modified Fu-Kane formula \cite{CHI} 
\begin{align}
(-1)^\nu = \Pi _{\bm{k}\in RISP} \mathcal{I}_v(\bm{k})= \Pi _{\bm{k}\in  RISP-\{ \Gamma\} } \mathcal{I}_v(\bm{k})\mathcal{I}_{v}(\Gamma).
\label{z2index}
\end{align}
Here $\mathcal{I}_v(\bm{k})$ is an inversion eigenvalue of {\it one} of the two filled band at $\bm{k}$ . In the second equality, we used $\mathcal{I}^4_{v}(\Gamma)=1$. When $\Delta_2$ is finite,  $H_{CHI}$ becomes topological and hosts a chiral hinge state for $1<|M/t|<3$. Otherwise, it is trivial.

\subsection{RIXS intensity of 3D chiral hinge insulator}
We present the RIXS intensity of 3D chiral hinge insulator, which can be used to diagnose the band topology. 
As in the 3D topological band insulator case, we consider the RIXS intensity from a reference point, say $\Gamma(0,0,0)$, to the other RISP $\bm{k}_*$. That is, we select the momentum transfer: $\bm{q}_n = \bm{k}_* + \sum_{i=1}^{3} n_i \bm{G}_i$ with $\bm{G}_i$ being a primitive reciprocal vector. 
The RIXS intensity can be expanded as 
\begin{align}
\frac{I(\bm{q}_n ,\Delta \epsilon) }{|\mathcal{C}(\Delta \epsilon)|^2} &= \sum_{\mu,\nu} |\psi^\dagger_{c,\mu}(\bm{k}_* )  \hat{\mathcal{M}}_{{\bf q_n}} \psi_{v,\nu} (\Gamma)|^2\nonumber,\\
&= \sum_{\mu,\nu} |\cos(\bm{q}_n\cdot\bm{d})   \psi^\dagger_{c,\mu}(\bm{k}_* )  \psi_{v,\nu} (\Gamma)+i\sin(\bm{q}_n\cdot\bm{d})    \psi^\dagger_{c,\mu}(\bm{k}_* )    \hat{\Pi} \psi_{v,\nu} (\Gamma)   |^2,
\end{align}
where
\begin{align}
\hat{\mathcal{M}}_{\bm{q}_n} =\cos(\bm{q}_n\cdot\bm{d}) + i\sin(\bm{q}_n\cdot\bm{d})   \hat{\Pi},
\end{align}
and $\mu,\nu$ are degeneracy indices. 


We will show how the inversion eigenvalues $\mathcal{I}_v(\bm{k}_*) \mathcal{I}_v(\Gamma)$ (the input data for determining the $\mathbb{Z}_2$ band topology Eq.\eqref{z2index}) constrains the RIXS intensity $I(\bm{q}_n ,\Delta \epsilon)$.  
We first note that the roto-inversion $\hat{C}^z_{4}\hat{I}$ eigenvalues of $\{ \psi_{v,{\mu}}(\bm{k}_*),\psi_{v,\bar{\mu}\neq\mu} (\Gamma)\}$ can be ordered to be $\{\mathcal{I}_v(\bm{k}_*) e^{\pm i \frac{\pi}{4} }, \mathcal{I}_v(\Gamma) e^{\mp i \frac{\pi}{4} }\}$ from  \ref{algebra}.This immediately implies 
\begin{align}
\psi^\dagger_{c,\mu}(\bm{k}_*)\psi_{v,\bar{\mu}\neq\mu}(\Gamma) = 0 
\end{align} 
where $\psi_{c,\mu} (\bm{k}_*)= \hat{\Pi} \psi_{v,\mu}(\bm{k}_*)$. This is because
\begin{align}
\psi^\dagger_{c,{\mu}}(\bm{k}_*)\psi_{v,\bar{\mu}}(\Gamma) &= \psi^\dagger_{c,{\mu}}(\bm{k}_*)(\hat{C}^z_4 \hat{I})^\dagger(\hat{C}^z_4 \hat{I}) \psi_{v,\bar{\mu}}(\Gamma) ,\nonumber\\
&=-\mathcal{I}_v(\bm{k}_*) e^{\pm i \pi/4}  \mathcal{I}_v(\Gamma) e^{\pm i \pi /4} \psi^\dagger_{c,{\mu}}(\bm{k}_*)\psi_{v,\bar{\mu}}(\Gamma) ,\nonumber\\
&= \mp i \mathcal{I}_v(\bm{k}_*) \mathcal{I}_v(\Gamma)  \psi^\dagger_{c,{\mu}}(\bm{k}_*)\psi_{v,\bar{\mu}}(\Gamma) ,
\end{align}
which gives $\psi^\dagger_{c,{\mu}}(\bm{k}_*)\psi_{v,\bar{\mu}}(\Gamma)=0$. 
In the second line, we used $\mathcal{I}_{v}(\bm{k}_*) = -\mathcal{I}_{c}(\bm{k}_*)$ from $\{ \hat{C}^z_{4} \hat{{I}},\hat{\Pi} \}=0$ for $\bm{k}_* \in$ TRIM. 
Similarly, 
\begin{align}
\psi^\dagger_{c,{\mu}}(\bm{k}_*)\hat{\Pi}\psi_{v,\bar{\mu}}(\Gamma) &= \psi^\dagger_{c,{\mu}}(\bm{k}_*)(\hat{C}^z_4 \hat{I})^\dagger(\hat{C}^z_4 \hat{I}) \hat{\Pi}\psi_{v,\bar{\mu}}(\Gamma) ,\nonumber\\
&=\mathcal{I}_v(\bm{k}_*) e^{\pm i \pi/4}  \mathcal{I}_v(\Gamma) e^{\pm i \pi /4} \psi^\dagger_{c,{\mu}}(\bm{k}_*)\hat{\Pi}\psi_{v,\bar{\mu}}(\Gamma) ,\nonumber\\
&= \pm i \mathcal{I}_v(\bm{k}_*) \mathcal{I}_v(\Gamma)  \psi^\dagger_{c,{\mu}}(\bm{k}_*)\hat{\Pi}\psi_{v,\bar{\mu}}(\Gamma) .
\end{align}
so, $\psi^\dagger_{c,{\mu}}(\bm{k}_*)\hat{\Pi}\psi_{v,\bar{\mu}}(\Gamma) =0$.
Therefore, the RIXS intensity for this case is diagonal in the degeneracy index, and is reduced to
\begin{align}
\frac{I(\bm{q}_n ,\Delta \epsilon) }{|\mathcal{C}(\Delta \epsilon)|^2} &=  \sum_{\mu} |\cos(\bm{q}_n\cdot\bm{d})   \psi^\dagger_{c,\mu}(\bm{k}_* )  \psi_{v,\mu} (\Gamma)+    i\sin(\bm{q}_n\cdot\bm{d})    \psi^\dagger_{c,\mu}(\bm{k}_* ) \hat{\Pi}    \psi_{v,\mu} (\Gamma)   |^2. 
\label{3DTI_RIXS}
\end{align}
For this case, we can next prove that 
\begin{align}
\begin{cases}
|\psi^\dagger_{c,\mu}(\bm{k}_*)\psi_{v,\mu}(\Gamma)|^2 =1,\quad |\psi^\dagger_{c,\mu}(\bm{k}_*)\hat{\Pi} \psi_{v,\mu}(\Gamma)|^2 =0 \quad \text{for} \quad \mathcal{I}_v(\bm{k}_*) \mathcal{I}_v(\Gamma) =-1,\\
|\psi^\dagger_{c,\mu}(\bm{k}_*)\psi_{v,\mu}(\Gamma)|^2 =0,\quad|\psi^\dagger_{c,\mu}(\bm{k}_*)\hat{\Pi} \psi_{v,\mu}(\Gamma)|^2 =1 \quad \text{for} \quad \mathcal{I}_v(\bm{k}_*) \mathcal{I}_v(\Gamma)=1 ,
\end{cases}
\end{align}
which gives
\begin{align}
\frac{I(\bm{q}_n ,\Delta \epsilon) }{|\mathcal{C}(\Delta \epsilon)|^2} &= 
 \sum_{\mu} \left( \cos^2({\bm{q_n}\cdot{\bm{d}}}) |  \psi^\dagger_{c,\mu}(\bm{k}_* )  \psi_{v,\mu} (\Gamma)|^2  +  \sin^2({\bm{q_n}\cdot{\bm{d}}})  |   \psi^\dagger_{c,\mu}(\bm{k}_* ) \hat{\Pi}    \psi_{v,\mu} (\Gamma)|^2    \right) ,\\
&=2\sin^2\left({\bm{q}_n\cdot{\bm{d}}} + \frac{1}{4} \left( \mathcal{I}_v(\bm{k}_*) \mathcal{I}_v(\Gamma) -1 \right)\pi \right)=
\begin{cases}
2\cos^2({\bm{q}_n\cdot{\bm{d}}})   \quad  \text{for} \quad \mathcal{I}_v(\bm{k}_*) \mathcal{I}_v(\Gamma) =-1\\
2\sin^2({\bm{q}_n\cdot{\bm{d}}})  \quad  \text{for} \quad \mathcal{I}_v(\bm{k}_*) \mathcal{I}_v(\Gamma)=1 ,
\label{Int_chiral}
\end{cases}
\end{align}

To prove this, we consider $\psi^\dagger_{c,{\mu}}(\bm{k}_*)\psi_{v,\mu}(\Gamma)$, where we insert an identity, $(\hat{C}^z_4 \hat{I})^\dagger(\hat{C}^z_4 \hat{I})$ 
\begin{align}
\psi^\dagger_{c,{\mu}}(\bm{k}_*)\psi_{v,\mu}(\Gamma) &= \psi^\dagger_{c,{\mu}}(\bm{k}_*)(\hat{C}^z_4 \hat{I})^\dagger(\hat{C}^z_4 \hat{I})  \psi_{v,\mu}(\Gamma) ,\nonumber\\
&=-\mathcal{I}_v(\bm{k}_*) e^{\pm i \pi/4}  \mathcal{I}_v(\Gamma) e^{\mp i \pi /4} \psi^\dagger_{c,{\mu}}(\bm{k}_*)\psi_{v,\mu}(\Gamma) ,\nonumber\\
&= - \mathcal{I}_v(\bm{k}_*) \mathcal{I}_v(\Gamma) \psi^\dagger_{c,{\mu}}(\bm{k}_*)\psi_{v,\mu}(\Gamma) .
\end{align}
Therefore, if $\mathcal{I}_v(\bm{k}_*) \mathcal{I}_v(\Gamma) =1$, then $\psi^\dagger_{c,{\mu}}(\bm{k}_*)\psi_{v,\mu}(\Gamma) =0$. At the same time, we have $\psi^\dagger_{c,{\mu}}(\bm{k}_*)\psi_{v,\bar{\mu}}(\Gamma) =0$, $\psi^\dagger_{c,{\mu}}(\bm{k}_*)\hat{\Pi} \psi_{v,\bar{\mu}}(\Gamma) =0$ for $\bar{\mu} \neq \mu$. They together imply $|\psi^\dagger_{c,{\mu}}(\bm{k}_*)\hat{\Pi}\psi_{v,\mu}(\Gamma)| =1$. For $\mathcal{I}_v(\bm{k}_*) \mathcal{I}_v(\Gamma)=-1$ case, one can similary show that $\psi^\dagger_{c,{\mu}}(\bm{k}_*) \hat{\Pi} \psi_{v,\mu}(\Gamma) =0$ and $|\psi^\dagger_{c,{\mu}}(\bm{k}_*) \psi_{v,\bar{\mu}}(\Gamma)| =1$. This completes our proof.

\subsection{Numerical simulation of RIXS intensity and fitting}
In this section, we numerically demonstrate how we can read off the topological band index from RIXS intensities. As a concrete example, we will consider the CHI model Eq.\eqref{CHI-M}. In the model, we will focus on determining the product of the inversion eigenvalues of the two RISP(\ref{z2index}). 
At $M/t = -3$, the band gap is inverted at the RISP $Z$ and $\mathcal{I}_v(Z)$ flips its sign. 
This results in the change $\mathbb{Z}_2$ index because of Eq.\eqref{z2index} from $\nu=0$ mod 2 for $M/t<-3$ to $\nu=1$ mod 2 for $M/t>-3$.   
We will attempt to diagnose the transition via the RIXS intensities.

\begin{figure}[t!]
	\includegraphics[width=.7\textwidth]{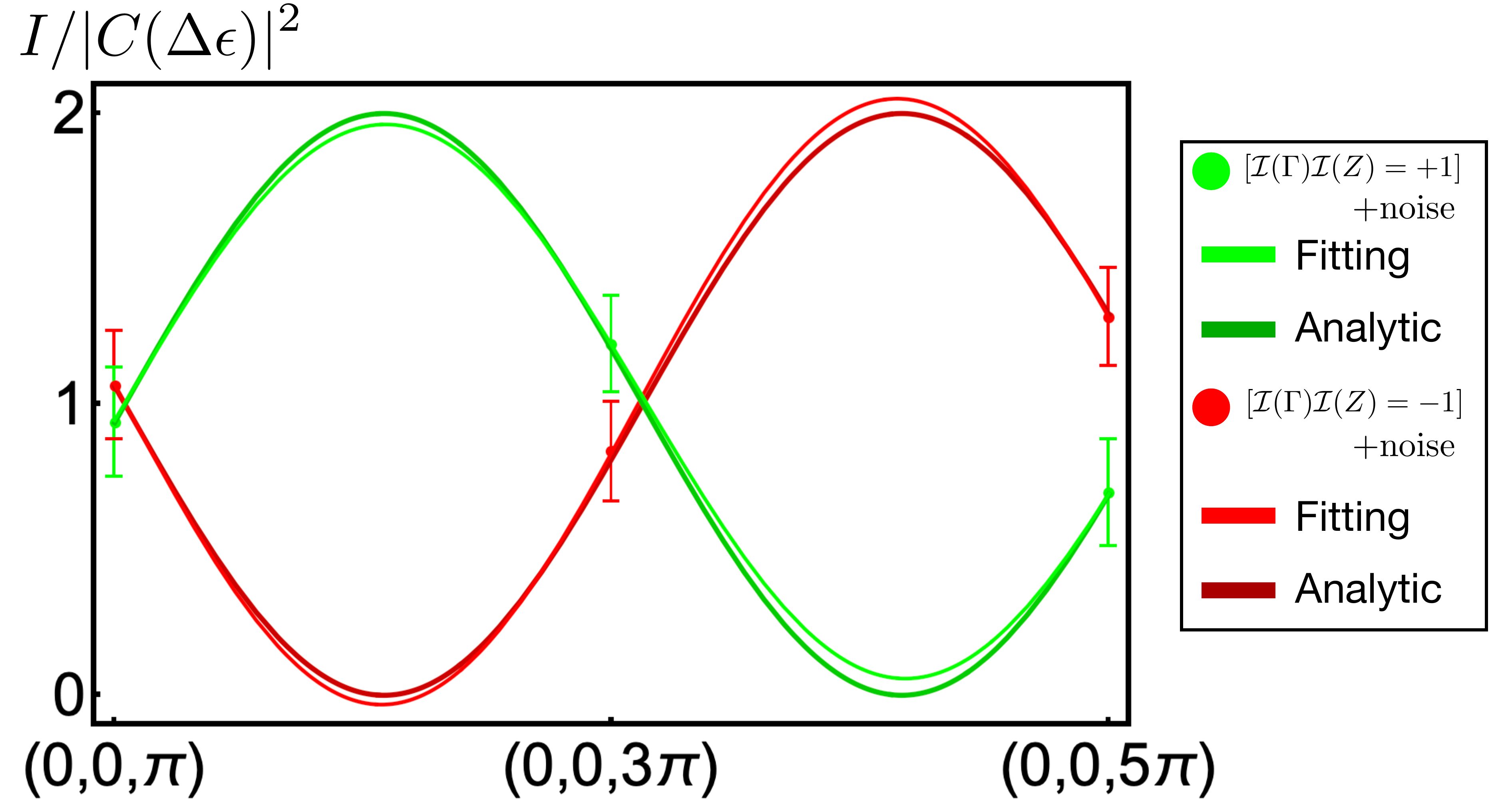}
	\caption{  RIXS intensity of  $H_{CHI} $ model with momentum transfer $\bm{q}_n = (2n+1) (0,0,\pi) $ and energy transfer $\omega = \Delta\epsilon= \epsilon_{c}(Z) - \epsilon_{v}(\Gamma)$. }
	\label{fig:chiral_RIXS}
\end{figure}

\paragraph{Simulation of the RIXS intensity:} The RIXS intensity data [Fig.\ref{fig:chiral_RIXS}] with momentum transfer $\bm{q}_n=(2n+1)(0,0,\pi)$ and energy transfer $\omega = \Delta \epsilon = \epsilon_c(Z) - \epsilon_v(\Gamma)$ are generated by adding the two contributions  
\begin{align}
\mathcal{I}(\bm{q}_n, \omega) = \mathcal{I}_0 (\bm{q}_n, \omega) + \delta \mathcal{I}_{\text{random}}, 
\end{align}
where $\mathcal{I}_0 (\bm{q}_n, \omega)$ is the RIXS intensity from the insulator, and $\delta \mathcal{I}_{\text{random}}$ is the random noise. Essentially, $\mathcal{I}_0 (\bm{q}_n, \omega)$ is the analytic, perfect RIXS signal on the clean $H_{CHI}$ model, which is obtained by applying Eq.(3) of the main text to the model. On the other hand, $\delta \mathcal{I}_{\text{random}}$ is the white noise drawn from the uniform distribution $[-0.3, 0.3] \times |\mathcal{C}(\Delta \epsilon)|^2$.

For the topological case $\nu=1$ mod 2, i.e., $\mathcal{I}(\Gamma)\mathcal{I}(Z) = +1$, (red circles in [Fig.\ref{fig:chiral_RIXS}]), the parameter of the $H_{CHI}$ model is $M/t=-2.9$. 
For the trivial case $\nu=0$ mod 2,  i.e., $\mathcal{I}(\Gamma)\mathcal{I}(Z) = -1$, (green circles in [Fig.\ref{fig:chiral_RIXS}]), the parameter is $M/t=-3.1$ (so that $\Delta \epsilon$ is the same for the both cases). We set $d=0.24$ for both the cases.

\paragraph{Fitting procedure:} We have attempted to fit the data with $\mathcal{I} (\bm{q}_n)/|C(\Delta\epsilon)|^2= A \sin^2\big((2n+1)\pi B +I  \big)+C$, [Cf. Eq.\ref{Int_chiral}]. Here $I\in[-\pi,0]$ will be the diagnostics of the band topology, i.e. $I=  \frac{\pi}{4} \left( \mathcal{I}_v(Z) \mathcal{I}_v(\Gamma) -1 \right)$.

For the idealistic, perfect signal of the trivial case (without the noise $\delta \mathcal{I}_{\text{random}}$), we expect to find 
\begin{align}
A=2, B=0.24, C=0, I=0,
\label{chiral-ideal-triv}
\end{align} 
which is the darker green line in [Fig.\ref{fig:chiral_RIXS}]. As the result of the fitting (the light green line in [Fig.\ref{fig:chiral_RIXS}]), we found 
\begin{align}
A=1.90, B=0.23, C=0.05, I=0.00,
\end{align} 
which are pretty close to the exact values (\ref{chiral-ideal-triv}). In particular, this correctly determines $\mathcal{I}_v(Z) \mathcal{I}_v(\Gamma) = 1.00$.  

For the idealistic, perfect signal of topological case (without the noise $\delta \mathcal{I}_{\text{random}}$), we expect to find 
\begin{align}
A=2, B=0.24, C=0,I=-1.58,
\label{chiral-ideal-top}
\end{align} 
which is the darker red line in [Fig.\ref{fig:chiral_RIXS}]. 
As the result of the fitting (the lighter red line in [Fig.\ref{fig:chiral_RIXS}]), we found 
\begin{align}
A=2.08, B=0.24, C=-0.03, I=-1.57, 
\end{align} 
which are pretty close to the exact values (\ref{chiral-ideal-top}). In particular, this correctly determines $\mathcal{I}_v(Z) \mathcal{I}_v(\Gamma) = -0.99$. 
Therefore, one can read off $\mathcal{I}_v(\Gamma)\mathcal{I}_v(Z)$ from this RIXS intensity, from which one can infer the topological band index.